\documentclass[]{pasj01}
\usepackage{graphicx}

\begin{document} 
%\Received{}%{yyyy/mm/dd}
%\Accepted{}%{yyyy/mm/dd}
%\Published{yyyy/mm/dd}

\title{Great Optically Luminous Dropout Research Using Subaru HSC (GOLDRUSH). I. \\ 
UV Luminosity Functions at $z \sim 4-7$\\ 
Derived with the Half-Million Dropouts \\
on the 100 deg$^2$ Sky
\thanks{Based on data collected at the Subaru Telescope 
and retrieved from the HSC data archive system, 
which is operated by the Subaru Telescope 
and Astronomy Data Center at National Astronomical Observatory of Japan.}
}

%%% begin:list of authors
% Do NOT capitalize all letters in "textsc".
\author{
Yoshiaki \textsc{Ono},\altaffilmark{1} 
Masami \textsc{Ouchi},\altaffilmark{1,2} 
Yuichi \textsc{Harikane},\altaffilmark{1,3} 
Jun \textsc{Toshikawa},\altaffilmark{1} 
Michael \textsc{Rauch},\altaffilmark{4} 
Suraphong \textsc{Yuma},\altaffilmark{5} 
Marcin \textsc{Sawicki},\altaffilmark{6} 
Takatoshi \textsc{Shibuya},\altaffilmark{1} 
Kazuhiro \textsc{Shimasaku},\altaffilmark{7,8} 
Masamune \textsc{Oguri},\altaffilmark{2,3,8} 
Chris \textsc{Willott},\altaffilmark{9} 
Mohammad \textsc{Akhlaghi},\altaffilmark{10} 
Masayuki \textsc{Akiyama},\altaffilmark{11} 
Jean \textsc{Coupon},\altaffilmark{12} 
Nobunari \textsc{Kashikawa},\altaffilmark{13,14} 
Yutaka \textsc{Komiyama},\altaffilmark{13,14} 
Akira \textsc{Konno},\altaffilmark{1,7} 
Lihwai \textsc{Lin},\altaffilmark{15} 
Yoshiki \textsc{Matsuoka},\altaffilmark{16} 
Satoshi \textsc{Miyazaki},\altaffilmark{13,14} 
Tohru \textsc{Nagao},\altaffilmark{16} 
Kimihiko \textsc{Nakajima},\altaffilmark{17,$\dagger$} 
John \textsc{Silverman},\altaffilmark{2} 
Masayuki \textsc{Tanaka},\altaffilmark{13} 
Yoshiaki \textsc{Taniguchi},\altaffilmark{18} 
and 
Shiang-Yu \textsc{Wang}\altaffilmark{15} 
}
%et al.%

\altaffiltext{1}{
Institute for Cosmic Ray Research, The University of Tokyo, Kashiwa, Chiba 277-8582, Japan
}
\altaffiltext{2}{
Kavli Institute for the Physics and Mathematics of the Universe (WPI), The University of Tokyo, 5-1-5 Kashiwanoha, Kashiwa, Chiba 277-8583, Japan 
}
\altaffiltext{3}{
Department of Physics, Graduate School of Science, The University of Tokyo, 7-3-1 Hongo, Bunkyo-ku, Tokyo, 113-0033, Japan 
}
\altaffiltext{4}{
Carnegie Observatories, 813 Santa Barbara Street, Pasadena, CA 91101, USA 
}
\altaffiltext{5}{
Department of Physics, Faculty of Science, Mahidol University, Bangkok 10400, Thailand 
}
\altaffiltext{6}{
Saint Mary's University, Department of Astronomy \& Physics, 923 Robie Street, Halifax, B3J 3Z4, Canada
}
\altaffiltext{7}{
Department of Astronomy, Graduate School of Science, The University of Tokyo, 7-3-1 Hongo, Bunkyo-ku, Tokyo 113-0033, Japan
}
\altaffiltext{8}{
Research Center for the Early Universe, Graduate School of Science, The University of Tokyo, 7-3-1 Hongo, Bunkyo-ku, Tokyo 113-0033, Japan
}
\altaffiltext{9}{
Herzberg Astronomy and Astrophysics, National Research Council, 5071 West Saanich Road, Victoria, V9E 2E7, Canada
}
\altaffiltext{10}{
CRAL, Observatoire de Lyon, CNRS, Universite Lyon 1, 9 avenue Ch. Andre, 69561 Saint Genis-Laval Cedex, France
}
\altaffiltext{11}{
Astronomical Institute, Tohoku University, Aramaki, Aoba-ku, Sendai, 980-8578
}
\altaffiltext{12}{
Department of Astronomy, University of Geneva, ch. d’\'Ecogia 16, 1290 Versoix, Switzerland 
}
\altaffiltext{13}{
National Astronomical Observatory of Japan, Mitaka, Tokyo 181-8588, Japan
}
\altaffiltext{14}{
Depertment of Astronomical, School of Physical Sciences, SOKENDAI (The Graduate University for Advanced Studies), Mitaka, Tokyo 181-8588, Japan 
}
\altaffiltext{15}{
Institute of Astronomy \& Astrophysics, Academia Sinica, Taipei 10617, Taiwan
}
\altaffiltext{16}{
Research Center for Space and Cosmic Evolution, Ehime University, Bunkyo-cho 2-5, Matsuyama, 790-8577
}
\altaffiltext{17}{
European Southern Observatory, Karl-Schwarzschild-Str. 2, D-85748, Garching bei Munchen, Germany 
}
\altaffiltext{18}{
The Open University of Japan, 2-11, Wakaba, Mihama-ku, Chiba, Chiba 261-8586, Japan
}
\altaffiltext{$\dagger$}{
JSPS Overseas Research Fellow 
}

\email{ono@icrr.u-tokyo.ac.jp}

%\author{B-Firstname \textsc{B-Familyname},\altaffilmark{2}}
%\altaffiltext{2}{B-Address of Institute}

%\author{C-Firstname \textsc{C-Familyname}\altaffilmark{3}}
%\altaffiltext{3}{C-Address of Institute}
%%% end:list of authors

%% `\KeyWords{}' always has to be placed before `\maketitle'.
\KeyWords{
galaxies: formation ---
galaxies: evolution ---
galaxies: high-redshift 
} 
%Do NOT move this preamble from here!

\maketitle

\begin{abstract}
We study the UV luminosity functions (LFs) at $z\sim 4$, $5$, $6,$ and $7$
based on the deep large-area optical images taken 
by the Hyper Suprime-Cam (HSC) Subaru strategic program (SSP).
On the 100 deg$^2$ sky of the HSC SSP data available to date,
we make enormous samples consisting of a total of 
579,565 dropout candidates at $z\sim 4-7$ by
the standard color selection technique, 
358 out of which are spectroscopically 
confirmed by our follow-up spectroscopy and other studies.
We obtain UV LFs at $z \sim 4-7$ that span a very wide UV luminosity range 
of $\sim 0.002$ -- $100 \, L_{\rm UV}^\ast$ ($-26 < M_{\rm UV} < -14$ mag) 
by combining LFs from our program and the ultra-deep 
\textit{Hubble Space Telescope} legacy surveys.
We derive three parameters of the best-fit Schechter function,
$\phi^\ast$, $M_{\rm UV}^\ast$, and $\alpha$,
of the UV LFs in the magnitude range where the AGN contribution
is negligible, and find that $\alpha$ and $\phi^\ast$ decrease from $z\sim 4$ to $7$
with no significant evolution of $M_{\rm UV}^\ast$.
Because our HSC SSP data bridge the LFs of galaxies and AGNs
with great statistical accuracy, 
we carefully investigate the bright end  of the galaxy UV LFs that are estimated 
by the subtraction of the AGN contribution 
either aided with spectroscopy or the best-fit AGN UV LFs. 
We find that the bright end of the galaxy UV LFs cannot be explained 
by the Schechter function fits at $> 2 \sigma$ significance, 
and require either double power-law functions 
or modified Schechter functions 
that consider a magnification bias due to gravitational lensing.
\end{abstract}

%%%%%%%%%%%%%%%%%%%%%%%%%%%%%%%%%%%%%%%%%%%%%%%%%%%%%%%%%%%%%%%%%
%%%%%%%%%%%%%%%%%%%%%%%%%%%%%%%%%%%%%%%%%%%%%%%%%%%%%%%%%%%%%%%%%
\section{Introduction} \label{sec:introduction}
%%%%%%%%%%%%%%%%%%%%%%%%%%%%%%%%%%%%%%%%%%%%%%%%%%%%%%%%%%%%%%%%%
%%%%%%%%%%%%%%%%%%%%%%%%%%%%%%%%%%%%%%%%%%%%%%%%%%%%%%%%%%%%%%%%%

One of the important observables to study the formation and evolution of galaxies 
is the galaxy luminosity function (LF), 
which is the measure of the number of galaxies per unit volume as a function of luminosity. 
The form of the LF in the rest-frame UV is of significant interest,  
since it is closely related to ongoing star formation 
and contains key information about the physical processes that shape galaxies.

Great progress has been made in determining the faint end of the UV LFs 
(see the recent review of \cite{2016ARA&A..54..761S}). 
Analyses of sources in deep blank fields including the \textit{Hubble} Ultra Deep field (HUDF) 
have resulted in identifying $z \sim 4-10$ galaxy candidates 
down to $\sim -17$ mag 
(\cite{2013ApJ...763L...7E}; \cite{2013ApJ...768..196S}; \cite{2013MNRAS.432.2696M}; 
\cite{2015ApJ...803...34B}; \cite{2015ApJ...810...71F}). 
Recently,  
it becomes possible to probe even fainter sources 
with the \textit{Hubble} Frontier Fields (HFF) project, 
which takes advantage of 
the gravitational lens magnification effects of galaxy clusters 
(\cite{2015ApJ...799...12I}; \cite{2015ApJ...814...69A}; \cite{2016ApJ...819..114K}; 
\cite{2016ApJ...823L..40C}; \cite{2016MNRAS.459.3812M}; \cite{2017ApJ...835..113L}; 
\cite{2017arXiv170204867I}). 
They have investigated the shape of the UV LF down to $\sim -14$ mag, 
at around which 
many cosmological hydrodynamic simulations of galaxy formation 
predict a flattening 
(e.g., 
\cite{2011ApJ...729...99M}; \cite{2012ApJ...753...16K}; \cite{2013ApJ...776...34K}; \cite{2013ApJ...766...94J}; \cite{2014MNRAS.442.2560W}; 
\cite{2015ApJ...807L..12O}; \cite{2016MNRAS.462..235L}; \cite{2016ApJ...825L..17G}; \cite{2016MNRAS.463.1462O}; \cite{2017MNRAS.464.1633F}),  
although 
it has been pointed out that 
the analyses for the lensing fields 
would be significantly affected by systematic errors 
such as the one from 
the assumed size distribution of faint galaxies \citep{2017ApJ...843...41B}  
and the constructed magnification maps \citep{2017ApJ...843..129B}.

Together with studying the faint end of the UV LFs, 
it is important to investigate their bright-end shapes. 
Previous studies have shown that 
the UV LF of low-$z$ galaxies has an exponential cutoff 
(e.g., \cite{2012MNRAS.420.1239L}; \cite{2014MNRAS.439.1245K}), 
which is thought to be caused by several different mechanisms 
such as  
heating from an active galactic nucleus (AGN; \cite{2004MNRAS.347.1093B}; 
\cite{2004ApJ...608...62S}; \cite{2004ApJ...600..580G}; \cite{2006MNRAS.365...11C}; \cite{2006MNRAS.370..645B}), 
inefficiency of gas cooling in high-mass dark matter haloes 
(e.g., \cite{1977ApJ...215..483B}; \cite{1977MNRAS.179..541R}; \cite{1977ApJ...211..638S}; 
\cite{2003ApJ...599...38B}),  
and dust attenuation, 
which becomes substantial for the most luminous galaxies 
(e.g., \cite{1996ApJ...457..645W}; \cite{2000ApJ...544..218A}; \cite{2005ApJ...619L..59M}). 
However, 
at very high redshifts where typical dark matter halo masses are small, 
these processes may be ineffective yet (e.g., \cite{2008ApJ...686..230B}).  
Interestingly, recent studies by 
\citet{2015MNRAS.452.1817B} and \citet{2017MNRAS.466.3612B} 
using a $1.7$ deg$^2$ imaging survey 
have claimed an overabundance of galaxies at the bright end of the $z \geq 6$ LF 
over 
the best-fit Schechter function. 
It may indicate 
different astrophysical conditions in high-$z$ and low-$z$ galaxies. 
Another possible explanation for the 
overabundance at the bright end 
is contribution of light from AGNs. 
At a lower redshift of $z \sim 3$, 
around the peak of the quasar number density, 
there is evidence that the UV LF at the absolute UV magnitude $M_{\rm UV} \lesssim -24$ mag 
has a significant contribution from  
faint quasars \citep{2013ApJ...774...28B}. 
Gravitational lensing magnification bias also needs to be considered 
\citep{2011Natur.469..181W,2011ApJ...742...15T,2015ApJ...805...79M,2015MNRAS.450.1224B}. 
It is also possible that merger systems are blended at ground-based resolution 
and appear as bright extended objects  
\citep{2017MNRAS.466.3612B}. 
Due to the small number densities of these luminous galaxies, 
previous studies lack information on 
the most luminous $z \gtrsim 4$ galaxies with $M_{\rm UV} \lesssim -23$ mag 
(e.g., \cite{2004ApJ...611..660O}; \cite{2005PASJ...57..447S}; \cite{2006ApJ...642..653S}; 
\cite{2006ApJ...653..988Y}; \cite{2007MNRAS.376.1557I}; \cite{2009MNRAS.395.2196M}; 
\cite{2009ApJ...706.1136O}; \cite{2010A&A...511A..20C}; \cite{2010A&A...523A..74V}; 
\cite{2013AJ....145....4W}; \cite{2015MNRAS.452.1817B}; \cite{2017MNRAS.466.3612B}; 
\cite{2017arXiv170604613S}). 
To study 
a possible deviation from the commonly used Schechter functional form, 
it is necessary to construct 
a sample of rare luminous high-$z$ galaxies down to very low space densities 
based on wider multi-wavelength deep imaging surveys. 
In addition, spectroscopic redshifts for a subsample are vital 
to estimate the contaminant fraction for LF calculation.

In this study, 
we present results from our systematic search for 
very luminous galaxies at $z \sim 4-7$ 
based on 
wide and deep optical Hyper Suprime-Cam 
(HSC; \cite{2012SPIE.8446E..0ZM}; see also 
\cite{miyazaki2017}; 
\cite{komiyama2017}; 
\cite{furusawa2017}; 
\cite{kawanomoto2017}) 
images obtained by the Subaru Strategic Program 
(HSC SSP; \cite{2017arXiv170405858A}). 
With a large field of view of about $1.8$ deg$^2$ and excellent sensitivity, 
HSC is one of the best ground-based instruments 
for searching for intrinsically luminous but apparently faint rare sources such as luminous high-$z$ galaxies. 
The HSC SSP survey was 
awarded 300 nights of Subaru observing time over 5 years from 2014. 
The survey consists of three layers: Wide (W), Deep (D), and UltraDeep (UD).  
The W layer will cover $1400$ deg$^2$ 
with five broadband filters of $g$, $r$, $i$, $z$, and $y$ 
down to $5\sigma$ limits of about $26$ mag ($24-25$ mag) in $gri$ ($zy$). 
The D (UD) layers will cover $27$ ($3.5$) deg$^2$ 
with the five broadband filters 
down to $5\sigma$ limits of about $27$ ($28$) mag in $gri$ and $25-26$ ($26-27$) mag in $zy$. 
The D (UD) layers will also be observed with 
three narrowband filters of NB387 (NB101), NB816, and NB921. 
Public versions of the reduced HSC SSP images 
and source catalogs are available to the community 
on the HSC SSP website.\footnote{http://hsc.mtk.nao.ac.jp/ssp/}  
This wide-field deep survey will enable us 
to cover an unprecedentedly large cosmic volume at $z \gtrsim 4$ 
and 
to identify a large number of very rare bright sources that reside at the bright end of the UV LF, 
which has been poorly explored by previous high-$z$ galaxy studies. 
The present paper is 
one in a series of papers from twin continuing programs 
devoted to scientific results on high-$z$ galaxies 
based on the HSC SSP survey data products. 
One program is 
Great Optically Luminous Dropout Research Using Subaru HSC (GOLDRUSH). 
This program provides  
precise determinations of the the very bright end of the galaxy UV LFs at $z \sim 4-7$, 
which are presented in this paper, 
robust clustering measurements of luminous galaxy candidates at $z \sim 4-6$ 
\citep{2017arXiv170406535H}, 
and construction of a sizable sample of $z \sim 4$ galaxy protocluster candidates 
\citep{toshikawa2017}. 
The other program is 
Systematic Identification of LAEs for Visible Exploration and Reionization Research Using Subaru HSC 
(SILVERRUSH; 
\cite{2017arXiv170407455O}; \cite{2017arXiv170408140S}; \cite{2017arXiv170500733S}; \cite{2017arXiv170501222K}; 
R. Higuchi et al. in preparation). 
Data products from these programs such as catalogs of dropouts and LAEs 
will be provided on our project webpage at 
http:\slash\slash{}cos.icrr.u-tokyo.ac.jp\slash{}rush.html.

This paper is organized as follows. 
In the next section, we describe our HSC SSP data and spectroscopic follow-up observations. 
The sample selection and analyses for measuring UV LF are described in Section \ref{sec:sample_selection}. 
We show the results of our UV LF measurements 
and discuss the shapes of the UV LFs in Section \ref{sec:results_and_discussion}.  
A summary is presented in Section \ref{sec:summary}. 
Throughout this paper, 
we use magnitudes in the AB system \citep{1983ApJ...266..713O} 
and 
assume a flat universe with 
$\Omega_{\rm m} = 0.3$, 
$\Omega_\Lambda = 0.7$, 
and $H_0 = 70$ km s$^{-1}$ Mpc$^{-1}$.

%%%%%%%%%%%%%%%%%%%%%%%%%%%%%%%%%%%%%%%
\begin{longtable}{ccccccccc}
  \caption{HSC SSP data used in this study. 
  (1) Field name. (2) Right ascension. (3) Declination. (4) Effective area in deg$^2$. 
  (5)--(9) $5\sigma$ limiting magnitude measured with $1\farcs5$ diameter circular apertures in $g$, $r$, $i$, $z$, and $y$. 
  }\label{tab:HSCdata}
  \hline              
  Field & R.A. & Decl. & Area & $g$ & $r$ & $i$ & $z$ & $y$ \\
   & (J2000) & (J2000) & (deg$^2$) & (ABmag) & (ABmag) & (ABmag) & (ABmag) & (ABmag) \\
   (1) & (2) & (3) & (4) & (5) & (6) & (7) & (8) & (9) \\   
\endhead
  \hline
\endfoot
  \hline
\endlastfoot
  \hline
  \multicolumn{9}{c}{UltraDeep (UD)} \\
  UD-SXDS & 02:18:00.00 & $-$05:00:00.00 & $1.1$ & $27.15$ & $26.68$ & $26.53$ & $25.96$ & $25.15$ \\
  UD-COSMOS & 10:00:28.60 & 02:12:21.00 & $1.3$ & $27.13$ & $26.84$ & $26.46$ & $26.10$ & $25.28$ \\
  \hline
  \multicolumn{9}{c}{Deep (D)} \\
  D-XMM-LSS & 02:16:51.57 & $-$03:43:08.43 & $2.4$ & $26.73$ & $26.30$ & $25.88$ & $25.42$ & $24.40$ \\
  D-COSMOS & 10:00:59.50 & 02:13:53.06 & $6.5$ & $26.56$ & $26.19$ & $26.04$ & $25.58$ & $24.76$ \\
  D-ELAIS-N1 & 16:10:00.00 & 54:17:51.07  & $3.3$ & $26.77$ & $26.13$ & $25.87$ & $25.16$ & $24.25$ \\
  D-DEEP2-3 & 23:30:22.22 & $-$00:44:37.69  & $5.5$ & $26.69$ & $26.25$ & $25.96$ & $25.29$ & $24.56$ \\
  \hline
  \multicolumn{9}{c}{Wide (W)} \\
  W-XMM & 02:16:51.57 & $-$03:43:08.43 & $28.5$ & $26.43$ & $25.93$ & $25.71$ & $25.00$ & $24.25$ \\
  W-GAMA09H & 09:05:11.11 & 00:44:37.69  & $12.4$ & $26.35$ & $25.88$ & $25.65$ & $25.07$ & $24.45$ \\
  W-WIDE12H & 11:57:02.22 & 00:44:37.69 & $15.2$ & $26.38$ & $25.95$ & $25.82$ & $25.15$ & $24.23$ \\
  W-GAMA15H & 14:31:06.67& $-$00:44:37.69  & $16.6$ & $26.39$ & $25.96$ & $25.81$ & $25.11$ & $24.31$ \\
  W-HECTOMAP & 16:08:08.14 & 43:53:03.47  & $4.8$ & $26.47$ & $26.04$ & $25.82$ & $25.09$ & $24.07$ \\
  W-VVDS & 22:37:02.22 & 00:44:37.69  & $5.1$ & $26.31$ & $25.87$ & $25.74$ & $24.98$ & $24.23$ \\
  \hline
  Total & --- & --- & $102.7$ & --- & --- & --- & --- & --- \\
  \hline
\end{longtable}
%%%%%%%%%%%%%%%%%%%%%%%%%%%%%%%%%%%%%%%

%%%%%%%%%%%%%%%%%%%%%%%%%%%%%%%%%%%%%%%
{\footnotesize
\begin{longtable}{llcl}
  \caption{The selection criteria for our source catalog construction.  
  }\label{tab:HSCflag}
  \hline              
  Parameter & Value & Band & Comment \\
\endhead
  \hline
\endfoot
  \hline
\endlastfoot
  \hline
\verb|detect_is_primary| & True & --- & Object is a primary one with no deblended children. \\
\verb|flags_pixel_edge| & False & $grizy$ & Locate within images \\
\verb|flags_pixel_interpolated_center| & False & $grizy$ & None of the central $3 \times 3$ pixels of an object is interpolated. \\
\verb|flags_pixel_saturated_center| & False & $grizy$ &  None of the central $3 \times 3$ pixels of an object is saturated. \\
\verb|flags_pixel_cr_center| & False & $grizy$ & None of the central $3 \times 3$ pixels of an object is masked as cosmic ray. \\
\verb|flags_pixel_bad| & False & $grizy$ & None of the pixels in the footprint of an object is labelled as bad. \\
\verb|flags_pixel_bright_object_any| & False & $grizy$ & None of the pixels in the footprint of an object is close to bright sources. \\
\verb|centroid_sdss_flags| & False & $ri$ for $g$-drop & Object centroid measurement has no problem. \\
 & False & $iz$ for $r$-drop &  \\
 & False & $zy$ for $i$-drop &  \\
 & False & $y$ for $z$-drop &   \\
\verb|cmodel_flux_flags| & False & $gri$ for $g$-drop & Cmodel flux measurement has no problem. \\
& False & $riz$ for $r$-drop &  \\
& False & $izy$ for $i$-drop &   \\
& False & $zy$ for $z$-drop &   \\
\verb|merge_peak| & True & $ri$ for $g$-drop & Detected in $r$ and $i$ \\
& False/True & $g$/$iz$ for $r$-drop &  Undetected in $g$ and detected in $r$ and $i$ \\
& False/True & $gr$/$zy$ for $i$-drop & Undetected in $g$ and $r$, and detected in $z$ and $y$  \\
& False/True & $gri$/$y$ for $z$-drop & Undetected in $g$, $r$ and $i$, and detected in $y$ \\
\verb|blendedness_abs_flux| & $<0.2$ & $ri$ for $g$-drop & The target photometry is not significantly affected by neighbors.\\
& $<0.2$ & $iz$ for $r$-drop &  \\
& $<0.2$ & $zy$ for $i$-drop &  \\
& $<0.2$ & $y$ for $z$-drop &  \\
\hline
\end{longtable}
}
%%%%%%%%%%%%%%%%%%%%%%%%%%%%%%%%%%%%%%%

%%%%%%%%%%%%%%%%%%%%%%%%%%%%%%%%%%%%%%%%%%%%%%%%%%%%%%%%%%%%%%%%%
%%%%%%%%%%%%%%%%%%%%%%%%%%%%%%%%%%%%%%%%%%%%%%%%%%%%%%%%%%%%%%%%%
\section{Data} \label{sec:data}
%%%%%%%%%%%%%%%%%%%%%%%%%%%%%%%%%%%%%%%%%%%%%%%%%%%%%%%%%%%%%%%%%
%%%%%%%%%%%%%%%%%%%%%%%%%%%%%%%%%%%%%%%%%%%%%%%%%%%%%%%%%%%%%%%%%

%%%%%%%%%%%%%%%%%%%%%%%%%%%%%%%%%%%%%%%%%%%%%%%%%%%%%%%%%%%%%%%%%
\subsection{Imaging Data} \label{sec:imaging_data}
%%%%%%%%%%%%%%%%%%%%%%%%%%%%%%%%%%%%%%%%%%%%%%%%%%%%%%%%%%%%%%%%%

In this study, 
we use early data products of the HSC SSP that are obtained in 2014--2016 
\citep{2017arXiv170208449A}. 
Specifically, 
we use the internal data release of S16A, 
where additional data taken in 2016 January -- April have been merged with 
the version of Public Data Release 1. 
The HSC images were reduced with 
version 4.0.2 of the HSC pipeline, hscPipe \citep{2017arXiv170506766B},
which uses codes from the Large Synoptic Survey Telescope (LSST) software pipeline 
\citep{2008arXiv0805.2366I,2010SPIE.7740E..15A,2015arXiv151207914J}. 
The HSC pipeline performs 
CCD-by-CCD reduction, calibration for astrometry, and photometric zero point determination. 
The pipeline then conducts  
mosaic-stacking that combines reduced CCD images into a large stacked image, 
and 
creates source catalogs by detecting and measuring sources on the stacked images. 
The HSC astrometry and photometry are calibrated with 
the Pan-STARRS $3\pi$ catalog 
\citep{2012ApJ...750...99T,2012ApJ...756..158S,2013ApJS..205...20M}. 
Full details of the HSC observations, data reduction,  
and object detection and photometric catalog creation 
are provided in \citet{2017arXiv170208449A}. 
In this study, we estimate total magnitudes and colors of sources 
by using the cmodel magnitude,  
which is a weighted combination of exponential and de Vaucouleurs fits 
to the light profile of each object 
(\cite{2004AJ....128..502A}; \cite{2017arXiv170506766B}). 
The source colors are measured through forced photometry. 
We correct all the magnitudes for Galactic extinction 
by using the dust map of \citet{1998ApJ...500..525S}.

The current HSC SSP survey data cover 
$6$ distinct areas on the sky in the W layer, 
$4$ areas in the D layer, 
and $2$ areas in the UD layer. 
To obtain uniform data sets, 
we mask regions which are affected by bright source halos \citep{2017arXiv170500622C}.  
We also mask regions where exposure times are relatively short 
by using the hscPipe parameter \verb|countinputs| $N_{\rm c}$, 
which denotes the number of exposures at a source position for a given filter.   
For the W-layer data, regions where $N_{\rm c} \geq (3, \, 3, \, 5, \, 5, \, 5)$ for $(g, \, r, \, i, \, z, \, y)$ are used.  
For the D-layer data, regions where $N_{\rm c} \geq (3, \, 3, \, 5, \, 5, \, 5)$ for $(g, \, r, \, i, \, z, \, y)$ are used.  
For the UD-COSMOS data, regions where $N_{\rm c} \geq (17, \, 16, \, 27, \, 47, \, 62)$ for $(g, \, r, \, i, \, z, \, y)$ are used.   
For the UD-SXDS data, regions where $N_{\rm c} \geq (13, \, 13, \, 27, \, 42, \, 38)$ for $(g, \, r, \, i, \, z, \, y)$ are used.   
After the masks are applied, 
the total effective area is about $100$ deg$^2$. 
Thanks to the large volumes that we probe, 
the influence of cosmic variance on the shape of the estimated LF 
is expected to be small \citep{2008ApJ...676..767T}. 
Table \ref{tab:HSCdata} summarizes 
the effective areas 
and the $5\sigma$ limiting magnitudes of our data.

First, we select isolated or cleanly deblended sources 
from the detected source catalog 
available on the database \citep{takata2017} 
that is provided by the HSC SSP survey team. 
We then require that 
none of the pixels in their footprint are interpolated, 
none of the central $3 \times 3$ pixels are saturated, 
none of the central $3 \times 3$ pixels are affected by cosmic rays, 
and there are no bad pixels in their footprint.   
We also require that 
there are no problems in measuring cmodel fluxes 
in $gri$ images for $g$-dropouts, 
in $riz$ images for $r$-dropouts, 
in $izy$ images for $i$-dropouts, 
and
in $zy$ images for $z$-dropouts.  
In addition, 
we remove sources 
if there are any problems in measuring their centroid positions 
in $ri$ images for $g$-dropouts, 
in $iz$ images for $r$-dropouts, 
in $zy$ images for $i$-dropouts, 
and 
in $y$ images for $z$-dropouts.  
The selection criteria for our source catalog construction are listed in Table \ref{tab:HSCflag}.

%%%%%%%%%%%%%%%%%%%%%%%%%%%%%%%%%%%%%%
\begin{figure*}
 \begin{center}
  \includegraphics[width=14cm]{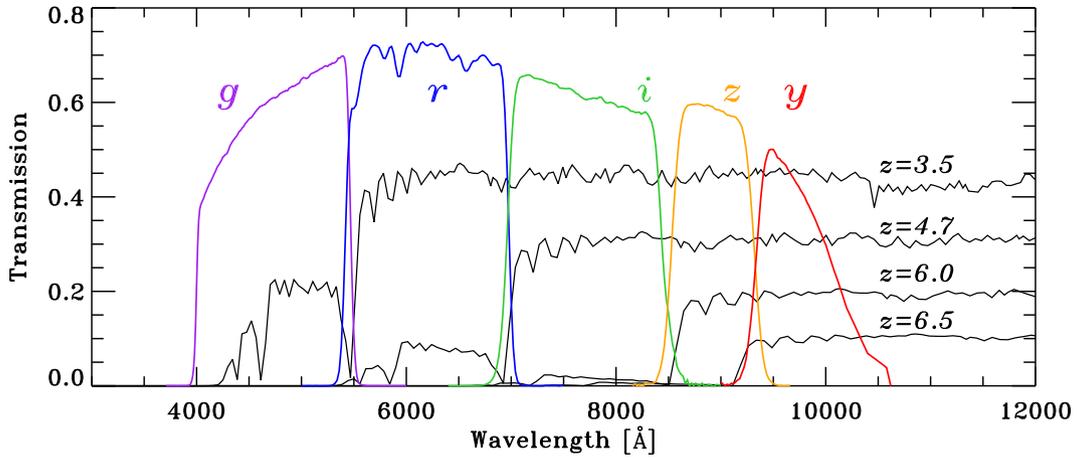} 
 \end{center}
\caption{
Transmissions of the five HSC broadband filters used in this study 
(purple: $g$, blue: $r$, green: $i$, orange: $z$, and red: $y$) 
together with four spectra of star-forming galaxies 
at $z = 3.5$, $4.7$, $6.0$, and $6.5$  
from the \citet{2003MNRAS.344.1000B} library (black lines). 
}\label{fig:filters_SEDs}
\end{figure*}
%%%%%%%%%%%%%%%%%%%%%%%%%%%%%%%%%%%%%%%

%%%%%%%%%%%%%%%%%%%%%%%%%%%%%%%%%%%%%%%%%%%%%%%%%%%%%%%%%%%%%%%%%
\subsection{Spectroscopic Data} \label{sec:spectroscopic_data}
%%%%%%%%%%%%%%%%%%%%%%%%%%%%%%%%%%%%%%%%%%%%%%%%%%%%%%%%%%%%%%%%%

We carried out spectroscopic follow-up observations 
for sources in our catalogs 
with the Faint Object Camera and Spectrograph (FOCAS; \cite{2002PASJ...54..819K}) 
on the Subaru Telescope 
on 2015 September 7 (S15B-188S, PI: Y. Ono), 
December 2, 4 (S15B-059, PI: S. Yuma), 
and 12 (S16A-211S, PI: Y. Ono),  
and 
with the Low Dispersion Survey Spectrograph 3 (LDSS3) 
on the Magellan II Clay telescope 
in 2015 November 
(PI: M. Rauch). 
Our sources were filler targets 
in the FOCAS observations of S15B-059 and the LDSS observations. 
In the FOCAS observations, 
we used the $300$ line mm$^{-1}$ grism and the VPH900 grism 
with the SO58 order-cut filter. 
The spectroscopic observations were made 
in the long slit mode or multi-object slit mode. 
Slit widths were 0$\farcs$8. 
The integration times were 2,000--6,000 sec. 
Flux calibration was carried out 
with spectra of the spectroscopic standard stars G191B2B, Feige 34, and GD153. 
In the LDSS3 observations, 
the VPH RED grism and the OG590 filter were used. 
The spectroscopic observations were made in the long slit mode. 
Slit widths were $1\farcs0$. 
The integration times were 3,600--5,400 sec. 
Flux calibration was carried out 
with spectra of the spectroscopic standard star LTT 9239. 
Note that we have also been awarded observing time 
with the Gemini Multi-Object Spectrographs (GMOS; \cite{2004PASP..116..425H}) 
on the Gemini South telescope (PI: M. Sawicki), 
but at the time of writing this paper, 
no useful data had yet been obtained from this program.

In addition to the observations described above, 
we include results of our observations with 
the Inamori Magellan Areal Camera and Spectrograph (IMACS; \cite{2011PASP..123..288D})
on the Magellan I Baade telescope in 2007 -- 2011 (PI: M. Ouchi). 
The IMACS observations were carried out on 
2007 November 11--14, 
2008 November 29--30, 
December 1--2,  
December 18--20,  
2009 October 11--13, 
2010 February 8--9, 
July 9--10,  
and 
2011 January 3--4. 
In these observations, 
main targets were 
high-$z$ Ly$\alpha$ emitter (LAE) candidates 
found in the deep Subaru Suprime-Cam narrowband images 
obtained in the SXDS \citep{2008ApJS..176..301O,2010ApJ...723..869O} 
and COSMOS fields \citep{2007ApJS..172..523M,2009ApJ...696..546S}, 
and 
high-$z$ dropout galaxy candidates 
selected from the deep broadband images 
in these two fields 
\citep{2008ApJS..176....1F,2007ApJS..172...99C} 
were also observed as mask fillers. 
The data are reduced with the Carnegie Observatories System for MultiObject Spectroscopy 
({\sc cosmos}) pipeline.\footnote{http://code.obs.carnegiescience.edu/cosmos}
Details of the IMACS observations and data reduction will be presented elsewhere.

%%%%%%%%%%%%%%%%%%%%%%%%%%%%%%%%%%%%%%%
\begin{table}
  \tbl{
  Number of Sources in our $z \sim 4$, $z \sim 5$, $z \sim 6$, and $z \sim 7$ 
  Galaxy Candidate Samples. 
  }{%
  \begin{tabular}{ccccc}
      \hline
  Field & $z\sim4$ & $z\sim5$ & $z\sim6$ & $z\sim7$  \\
   & \# & \# & \# & \# \\
      \hline
  \multicolumn{5}{c}{UltraDeep (UD)} \\
  UD-SXDS & $9916$ & $1209$ & $36$ & ---$^\dagger$ \\
  UD-COSMOS & $10644$ & $1990$ & $50$ & ---$^\dagger$ \\
  \hline
  \multicolumn{5}{c}{Deep (D)} \\
  D-XMM-LSS & $6730$ & $711$ & $6$ & $0$ \\
  D-COSMOS & $45767$ & $6282$ & $64$ & $4$ \\
  D-ELAIS-N1 & $19631$ & $612$ & $15$ & $1$ \\
  D-DEEP2-3 & $35963$ & $1498$ & $47$ & $5$ \\
  \hline
  \multicolumn{5}{c}{Wide (W)} \\
  W-XMM & $113582$ & $6371$ & $81$ & $7$ \\
  W-GAMA09H & $44670$ & $5989$ & $98$ & $16$ \\
  W-WIDE12H & $94544$ & $5243$ & $36$ & $8$ \\
  W-GAMA15H & $104224$ & $6457$ & $73$ & $14$ \\
  W-HECTOMAP & $30663$ & $1082$ & $11$ & $7$ \\
  W-VVDS & $23677$ & $1500$ & $20$ & $11$ \\
  \hline
  Total & $540011$ & $38944$ & $537$ & 73 \\
  \hline
    \end{tabular}}\label{tab:dropout_samples}
\begin{tabnote}
$^\dagger$ 
Our $z \sim 7$ dropout search focuses on the W and D layers. 
See Section 3.1 for details. 
\end{tabnote}
\end{table}
%%%%%%%%%%%%%%%%%%%%%%%%%%%%%%%%%%%%%%%

%%%%%%%%%%%%%%%%%%%%%%%%%%%%%%%%%%%%%%%
{\footnotesize
\begin{longtable}{ccccccccc}
  \caption{
  Spectroscopically identified galaxies and AGNs in our dropout samples. 
  (1) Object ID. (2) Right ascension. (3) Declination. 
  (4) Spectroscopic redshift. 
  (5) Apparent magnitude. 
  (6) UV absolute magnitude. 
  (7) The dropout sample in which the source is selected: 
  1 = $g$-dropout, 2 = $r$-dropout, 3 = $i$-dropout, and 4 = $z$-dropout. 
  (8) Galaxy/AGN flag (1 = galaxy; 2 = AGN). 
  (9) Reference of spectroscopic redshift. 
  S08 = \citet{2008ApJ...675.1076S}, 
  O08 = \citet{2008ApJS..176..301O}, 
  W10 = \citet{2010AJ....140..546W}, 
  C12 = \citet{2012MNRAS.422.1425C}, 
  Mas12 = \citet{2012ApJ...755..169M}, 
  M12 = \citet{2012ApJ...760..128M}, 
  W13 = \citet{2013AJ....145....4W}, 
  L13 = \citet{2013A&A...559A..14L}, 
  K15 = \citet{2015ApJ...798...28K}, 
  Kr15 = \citet{2015ApJS..218...15K}, 
  W16 = \citet{2016ApJ...819...24W}, 
  T16 = \citet{2016ApJ...826..114T}, 
  Mo16 = \citet{2016ApJS..225...27M}, 
  M16 = \citet{2016ApJ...828...26M}, 
  P17 = \citet{2017A&A...597A..79P}, 
  T17 = \citet{2017A&A...600A.110T}, 
  Y17 = \citet{2017AJ....153..184Y}, 
  Mas17 = \citet{2017ApJ...841..111M}, 
  M17 = \citet{2017arXiv170405854M}, 
  S17 = \citet{2017arXiv170500733S}, 
  and 
  H17 = R. Higuchi et al. in preparation.  
  }\label{tab:spectroscopy_identifications}
  \hline
  ID & R.A. & Decl. & $z_{\rm spec}$ & $m$ & $M_{\rm UV}$ & Sample & Flag & Reference \\
   & (J2000) & (J2000) &  & (mag) & (mag) &  &  & \\
   (1) & (2) & (3) & (4) & (5) & (6) & (7) & (8) & (9) \\   
\hline
\endhead
  \hline
\endfoot
  \hline
\endlastfoot
  \hline
    \multicolumn{9}{c}{Galaxies} \\
HSC J015949--035945 & 01:59:49.36 & $-$03:59:45.24 & $5.770$ & $24.0$ & $-22.7$ & 3 & 1 & M17 \\
HSC J021033--052304 & 02:10:33.82 & $-$05:23:04.28 & $5.890$ & $23.7$ & $-23.0$ & 3 & 1 & M16 \\
HSC J021041--055917 & 02:10:41.28 & $-$05:59:17.87 & $5.820$ & $24.3$ & $-22.4$ & 3 & 1 & M16 \\
HSC J021545--055529 & 02:15:45.20 & $-$05:55:29.03 & $5.740$ & $23.9$ & $-22.7$ & 3 & 1 & M16 \\
HSC J021551--050938 & 02:15:51.21 & $-$05:09:38.49 & $4.848$ & $24.5$ & $-21.8$ & 2 & 1 & This Study \\
HSC J021624--045516 & 02:16:24.70 & $-$04:55:16.55 & $5.706$ & $25.7$ & $-20.9$ & 2 & 1 & H17 \\
HSC J021640--050129 & 02:16:40.67 & $-$05:01:29.43 & $3.699$ & $24.4$ & $-21.5$ & 1 & 1 & O08 \\
HSC J021654--050216 & 02:16:54.14 & $-$05:02:16.50 & $4.284$ & $25.7$ & $-20.5$ & 1 & 1 & This Study \\
HSC J021658--053419 & 02:16:58.03 & $-$05:34:19.17 & $3.790$ & $25.4$ & $-20.5$ & 1 & 1 & S08 \\
HSC J021704--045215 & 02:17:04.17 & $-$04:52:15.69 & $4.826$ & $24.9$ & $-21.4$ & 2 & 1 & This Study \\
HSC J021708--043301 & 02:17:08.19 & $-$04:33:01.45 & $5.005$ & $25.2$ & $-21.2$ & 2 & 1 & This Study \\
HSC J021711--050806 & 02:17:11.20 & $-$05:08:06.44 & $4.084$ & $22.8$ & $-23.3$ & 1 & 1 & Mo16 \\
HSC J021712--051041 & 02:17:12.45 & $-$05:10:41.44 & $4.371$ & $24.9$ & $-21.3$ & 1 & 1 & Mo16 \\
HSC J021714--044510 & 02:17:14.01 & $-$04:45:10.77 & $3.988$ & $25.6$ & $-20.5$ & 1 & 1 & This Study \\
HSC J021714--052516 & 02:17:14.71 & $-$05:25:16.07 & $3.729$ & $25.6$ & $-20.3$ & 1 & 1 & O08 \\
HSC J021715--044418 & 02:17:15.29 & $-$04:44:18.21 & $3.980$ & $25.2$ & $-20.8$ & 1 & 1 & This Study \\
HSC J021715--044751 & 02:17:15.98 & $-$04:47:51.54 & $3.700$ & $24.6$ & $-21.3$ & 1 & 1 & This Study \\
HSC J021716--044336 & 02:17:16.54 & $-$04:43:36.90 & $4.808$ & $25.0$ & $-21.3$ & 2 & 1 & This Study \\
HSC J021718--044945 & 02:17:18.33 & $-$04:49:45.29 & $4.356$ & $25.4$ & $-20.8$ & 1 & 1 & This Study \\
HSC J021719--044853 & 02:17:19.13 & $-$04:48:53.46 & $4.239$ & $25.8$ & $-20.3$ & 1 & 1 & This Study \\
HSC J021721--050046 & 02:17:21.96 & $-$05:00:46.83 & $3.666$ & $24.8$ & $-21.0$ & 1 & 1 & This Study \\
HSC J021722--053059 & 02:17:22.01 & $-$05:30:59.58 & $3.798$ & $24.6$ & $-21.3$ & 1 & 1 & This Study \\
HSC J021723--044315 & 02:17:23.74 & $-$04:43:15.81 & $4.034$ & $24.5$ & $-21.6$ & 1 & 1 & This Study \\
HSC J021726--051839 & 02:17:26.07 & $-$05:18:39.38 & $4.077$ & $24.5$ & $-21.5$ & 1 & 1 & This Study \\
HSC J021727--044202 & 02:17:27.49 & $-$04:42:02.31 & $3.971$ & $24.0$ & $-22.0$ & 1 & 1 & This Study \\
HSC J021727--044413 & 02:17:27.70 & $-$04:44:13.81 & $3.683$ & $25.7$ & $-20.2$ & 1 & 1 & O08 \\
HSC J021734--050514 & 02:17:34.38 & $-$05:05:14.53 & $3.986$ & $23.3$ & $-22.7$ & 1 & 1 & This Study \\
HSC J021734--044558 & 02:17:34.57 & $-$04:45:58.95 & $5.702$ & $25.4$ & $-21.2$ & 2 & 1 & H17 \\
HSC J021735--051032 & 02:17:35.33 & $-$05:10:32.42 & $6.120$ & $24.8$ & $-21.9$ & 3 & 1 & C12 \\
HSC J021736--043334 & 02:17:36.06 & $-$04:33:34.87 & $4.293$ & $25.9$ & $-20.3$ & 1 & 1 & This Study \\
HSC J021736--044549 & 02:17:36.39 & $-$04:45:49.10 & $4.019$ & $24.9$ & $-21.1$ & 1 & 1 & This Study \\
HSC J021737--044650 & 02:17:37.85 & $-$04:46:50.35 & $3.899$ & $24.9$ & $-21.0$ & 1 & 1 & This Study \\
HSC J021738--052057 & 02:17:38.93 & $-$05:20:57.91 & $4.110$ & $26.2$ & $-19.9$ & 1 & 1 & S08 \\
HSC J021739--044832 & 02:17:39.78 & $-$04:48:32.82 & $3.889$ & $25.3$ & $-20.7$ & 1 & 1 & This Study \\
HSC J021739--053253 & 02:17:39.89 & $-$05:32:53.68 & $4.487$ & $24.5$ & $-21.7$ & 2 & 1 & This Study \\
HSC J021740--045103 & 02:17:40.34 & $-$04:51:03.55 & $4.545$ & $25.4$ & $-20.8$ & 1 & 1 & This Study \\
HSC J021742--043608 & 02:17:42.73 & $-$04:36:08.80 & $4.397$ & $25.6$ & $-20.6$ & 1 & 1 & This Study \\
HSC J021744--050642 & 02:17:44.56 & $-$05:06:42.41 & $4.549$ & $24.6$ & $-21.7$ & 1 & 1 & This Study \\
HSC J021744--044401 & 02:17:44.63 & $-$04:44:01.64 & $4.377$ & $26.0$ & $-20.2$ & 1 & 1 & This Study \\
HSC J021745--052735 & 02:17:45.37 & $-$05:27:35.55 & $3.648$ & $24.5$ & $-21.3$ & 1 & 1 & O08 \\
HSC J021746--051553 & 02:17:46.01 & $-$05:15:53.54 & $4.037$ & $24.8$ & $-21.2$ & 1 & 1 & This Study \\
HSC J021746--045045 & 02:17:46.14 & $-$04:50:45.67 & $4.039$ & $25.1$ & $-20.9$ & 1 & 1 & This Study \\
HSC J021747--045229 & 02:17:47.71 & $-$04:52:29.85 & $3.648$ & $23.8$ & $-22.0$ & 1 & 1 & O08 \\
HSC J021748--044935 & 02:17:48.97 & $-$04:49:35.64 & $4.221$ & $26.3$ & $-19.8$ & 1 & 1 & This Study \\
HSC J021749--043753 & 02:17:49.54 & $-$04:37:53.04 & $4.212$ & $25.4$ & $-20.8$ & 1 & 1 & This Study \\
HSC J021750--043230 & 02:17:50.86 & $-$04:32:30.52 & $4.095$ & $26.1$ & $-20.0$ & 1 & 1 & This Study \\
HSC J021751--052637 & 02:17:51.00 & $-$05:26:37.54 & $4.920$ & $24.2$ & $-22.2$ & 2 & 1 & This Study \\
HSC J021751--045627 & 02:17:51.01 & $-$04:56:27.44 & $3.701$ & $25.7$ & $-20.2$ & 1 & 1 & This Study \\
HSC J021752--053120 & 02:17:52.06 & $-$05:31:20.93 & $4.109$ & $24.9$ & $-21.2$ & 1 & 1 & This Study \\
HSC J021752--050700 & 02:17:52.78 & $-$05:07:00.15 & $3.700$ & $25.9$ & $-19.9$ & 1 & 1 & This Study \\
HSC J021754--050913 & 02:17:54.87 & $-$05:09:13.80 & $3.704$ & $26.2$ & $-19.7$ & 1 & 1 & This Study \\
HSC J021755--043203 & 02:17:55.30 & $-$04:32:03.70 & $4.626$ & $24.5$ & $-21.7$ & 1 & 1 & This Study \\
HSC J021756--053352 & 02:17:56.82 & $-$05:33:52.97 & $3.986$ & $25.1$ & $-21.0$ & 1 & 1 & This Study \\
HSC J021758--052135 & 02:17:58.22 & $-$05:21:35.54 & $4.646$ & $25.7$ & $-20.6$ & 1 & 1 & This Study \\
HSC J021758--043417 & 02:17:58.42 & $-$04:34:17.68 & $4.082$ & $24.6$ & $-21.5$ & 1 & 1 & This Study \\
HSC J021759--052507 & 02:17:59.47 & $-$05:25:07.57 & $3.820$ & $26.1$ & $-19.9$ & 1 & 1 & S08 \\
HSC J021800--052410 & 02:18:00.11 & $-$05:24:10.46 & $4.470$ & $25.7$ & $-20.5$ & 1 & 1 & S08 \\
HSC J021803--022029 & 02:18:03.42 & $-$02:20:29.73 & $5.900$ & $24.0$ & $-22.7$ & 3 & 1 & M17 \\
HSC J021807--052048 & 02:18:07.74 & $-$05:20:48.03 & $3.638$ & $25.0$ & $-20.8$ & 1 & 1 & O08 \\
HSC J021808--044845 & 02:18:08.12 & $-$04:48:45.54 & $4.852$ & $25.1$ & $-21.3$ & 2 & 1 & This Study \\
HSC J021813--051505 & 02:18:13.31 & $-$05:15:05.25 & $4.270$ & $25.5$ & $-20.7$ & 1 & 1 & S08 \\
HSC J021813--051841 & 02:18:13.53 & $-$05:18:41.03 & $3.572$ & $23.9$ & $-21.9$ & 1 & 1 & This Study \\
HSC J021813--043057 & 02:18:13.56 & $-$04:30:57.07 & $3.668$ & $25.6$ & $-20.3$ & 1 & 1 & O08 \\
HSC J021813--051840 & 02:18:13.79 & $-$05:18:40.91 & $5.013$ & $24.2$ & $-22.2$ & 2 & 1 & This Study \\
HSC J021814--043904 & 02:18:14.15 & $-$04:39:04.39 & $4.469$ & $24.9$ & $-21.3$ & 1 & 1 & This Study \\
HSC J021817--051027 & 02:18:17.08 & $-$05:10:27.32 & $3.702$ & $25.6$ & $-20.3$ & 1 & 1 & This Study \\
HSC J021823--051121 & 02:18:23.76 & $-$05:11:21.32 & $3.681$ & $25.9$ & $-20.0$ & 1 & 1 & This Study \\
HSC J021825--044429 & 02:18:25.06 & $-$04:44:29.20 & $4.023$ & $25.2$ & $-20.8$ & 1 & 1 & This Study \\
HSC J021826--051003 & 02:18:26.22 & $-$05:10:03.47 & $3.699$ & $25.0$ & $-20.9$ & 1 & 1 & O08 \\
HSC J021827--051947 & 02:18:27.71 & $-$05:19:47.18 & $3.677$ & $25.8$ & $-20.1$ & 1 & 1 & O08 \\
HSC J021828--042956 & 02:18:28.94 & $-$04:29:56.05 & $4.344$ & $26.0$ & $-20.1$ & 1 & 1 & This Study \\
HSC J021830--052808 & 02:18:30.79 & $-$05:28:08.55 & $3.579$ & $23.8$ & $-22.1$ & 1 & 1 & This Study \\
HSC J021832--043827 & 02:18:32.62 & $-$04:38:27.68 & $4.212$ & $25.9$ & $-20.2$ & 1 & 1 & This Study \\
HSC J021835--053550 & 02:18:35.18 & $-$05:35:50.45 & $3.671$ & $25.1$ & $-20.7$ & 1 & 1 & O08 \\
HSC J021835--042321 & 02:18:35.94 & $-$04:23:21.63 & $5.755$ & $24.9$ & $-21.7$ & 2 & 1 & S17 \\
HSC J021836--043906 & 02:18:36.29 & $-$04:39:06.85 & $3.797$ & $25.3$ & $-20.6$ & 1 & 1 & This Study \\
HSC J021837--044603 & 02:18:37.70 & $-$04:46:03.47 & $4.635$ & $24.7$ & $-21.5$ & 1 & 1 & This Study \\
HSC J021838--052023 & 02:18:38.50 & $-$05:20:23.05 & $3.678$ & $25.1$ & $-20.7$ & 1 & 1 & This Study \\
HSC J021838--050943 & 02:18:38.90 & $-$05:09:43.94 & $6.190$ & $25.0$ & $-21.8$ & 3 & 1 & C12 \\
HSC J021842--052340 & 02:18:42.06 & $-$05:23:40.24 & $4.722$ & $23.8$ & $-22.5$ & 2 & 1 & This Study \\
HSC J021845--052718 & 02:18:45.54 & $-$05:27:18.72 & $3.676$ & $25.3$ & $-20.6$ & 1 & 1 & This Study \\
HSC J021845--044139 & 02:18:45.56 & $-$04:41:39.19 & $4.016$ & $25.1$ & $-20.9$ & 1 & 1 & This Study \\
HSC J021848--043755 & 02:18:48.15 & $-$04:37:55.06 & $3.659$ & $24.9$ & $-20.9$ & 1 & 1 & This Study \\
HSC J021848--050224 & 02:18:48.77 & $-$05:02:24.22 & $4.599$ & $26.1$ & $-20.1$ & 1 & 1 & This Study \\
HSC J021851--052228 & 02:18:51.24 & $-$05:22:28.39 & $3.671$ & $24.6$ & $-21.3$ & 1 & 1 & This Study \\
HSC J021851--043022 & 02:18:51.31 & $-$04:30:22.86 & $4.353$ & $25.6$ & $-20.6$ & 1 & 1 & This Study \\
HSC J021852--043832 & 02:18:52.25 & $-$04:38:32.69 & $3.769$ & $26.1$ & $-19.9$ & 1 & 1 & This Study \\
HSC J021852--053008 & 02:18:52.93 & $-$05:30:08.02 & $3.819$ & $23.7$ & $-22.3$ & 1 & 1 & This Study \\
HSC J021853--044628 & 02:18:53.62 & $-$04:46:28.14 & $3.894$ & $25.6$ & $-20.3$ & 1 & 1 & This Study \\
HSC J021856--044556 & 02:18:56.66 & $-$04:45:56.37 & $4.125$ & $24.8$ & $-21.3$ & 1 & 1 & This Study \\
HSC J021911--045646 & 02:19:11.65 & $-$04:56:46.72 & $5.009$ & $24.8$ & $-21.6$ & 2 & 1 & This Study \\
HSC J021915--045511 & 02:19:15.37 & $-$04:55:11.87 & $3.672$ & $25.4$ & $-20.5$ & 1 & 1 & O08 \\
HSC J021917--050739 & 02:19:17.32 & $-$05:07:39.26 & $3.693$ & $25.0$ & $-20.9$ & 1 & 1 & O08 \\
HSC J021921--045712 & 02:19:21.19 & $-$04:57:12.51 & $4.790$ & $24.9$ & $-21.5$ & 2 & 1 & This Study \\
HSC J021930--050915 &  02:19:30.57  &  $-$05:09:15.86  & $4.580$ & $22.8$ & $-23.5$ & 2 & 1 & This Study \\
HSC J021938--045405 & 02:19:38.21 & $-$04:54:05.18 & $4.914$ & $25.1$ & $-21.3$ & 2 & 1 & This Study \\
HSC J021942--045525 & 02:19:42.18 & $-$04:55:25.51 & $4.226$ & $25.3$ & $-20.8$ & 1 & 1 & This Study \\
HSC J021944--045055 & 02:19:44.67 & $-$04:50:55.56 & $3.669$ & $25.5$ & $-20.4$ & 1 & 1 & O08 \\
HSC J021948--045606 & 02:19:48.60 & $-$04:56:06.79 & $5.074$ & $25.2$ & $-21.2$ & 2 & 1 & This Study \\
HSC J021950--050845 & 02:19:50.63 & $-$05:08:45.75 & $4.316$ & $24.3$ & $-21.9$ & 1 & 1 & This Study \\
HSC J022001--050446 & 02:20:01.52 & $-$05:04:46.36 & $5.237$ & $25.1$ & $-21.3$ & 2 & 1 & This Study \\
HSC J022006--050413 & 02:20:06.67 & $-$05:04:13.60 & $4.242$ & $25.7$ & $-20.4$ & 1 & 1 & This Study \\
HSC J022424--041931 & 02:24:24.65 & $-$04:19:31.64 & $3.555$ & $25.5$ & $-20.3$ & 1 & 1 & T16 \\
HSC J022429--042143 & 02:24:29.52 & $-$04:21:43.05 & $3.550$ & $25.6$ & $-20.2$ & 1 & 1 & T16 \\
HSC J022444--041935 & 02:24:44.62 & $-$04:19:35.53 & $3.463$ & $24.8$ & $-21.0$ & 1 & 1 & T16 \\
HSC J022511--041620 & 02:25:11.52 & $-$04:16:20.44 & $4.276$ & $25.5$ & $-20.6$ & 1 & 1 & T16 \\
HSC J022517--041402 & 02:25:17.28 & $-$04:14:02.38 & $3.738$ & $24.8$ & $-21.1$ & 1 & 1 & T16 \\
HSC J022520--042219 & 02:25:20.56 & $-$04:22:19.22 & $3.886$ & $24.1$ & $-21.9$ & 1 & 1 & L13 \\
HSC J022530--041515 & 02:25:30.08 & $-$04:15:15.78 & $3.827$ & $24.6$ & $-21.4$ & 1 & 1 & T16 \\
HSC J022533--041445 & 02:25:33.01 & $-$04:14:45.29 & $3.766$ & $24.8$ & $-21.1$ & 1 & 1 & T16 \\
HSC J022533--041541 & 02:25:33.69 & $-$04:15:41.51 & $3.699$ & $23.6$ & $-22.3$ & 1 & 1 & L13 \\
HSC J022533--042236 & 02:25:33.89 & $-$04:22:36.90 & $4.259$ & $24.8$ & $-21.4$ & 1 & 1 & L13 \\
HSC J022539--041420 & 02:25:39.70 & $-$04:14:20.79 & $3.754$ & $24.6$ & $-21.3$ & 1 & 1 & T16 \\
HSC J022541--041606 & 02:25:41.75 & $-$04:16:06.74 & $3.843$ & $25.0$ & $-21.0$ & 1 & 1 & T16 \\
HSC J022545--043737 & 02:25:45.60 & $-$04:37:37.76 & $3.417$ & $23.9$ & $-21.8$ & 1 & 1 & L13 \\
HSC J022549--042215 & 02:25:49.89 & $-$04:22:15.00 & $3.865$ & $24.6$ & $-21.4$ & 1 & 1 & L13 \\
HSC J022607--042617 & 02:26:07.71 & $-$04:26:17.51 & $4.636$ & $24.4$ & $-21.9$ & 2 & 1 & L13 \\
HSC J022611--041921 & 02:26:11.54 & $-$04:19:21.60 & $3.755$ & $25.3$ & $-20.6$ & 1 & 1 & T16 \\
HSC J022619--042225 & 02:26:19.92 & $-$04:22:25.65 & $3.309$ & $24.2$ & $-21.5$ & 1 & 1 & L13 \\
HSC J022626--043219 & 02:26:26.45 & $-$04:32:19.94 & $3.428$ & $24.6$ & $-21.1$ & 1 & 1 & L13 \\
HSC J022632--042749 & 02:26:32.79 & $-$04:27:49.34 & $3.251$ & $24.0$ & $-21.6$ & 1 & 1 & L13 \\
HSC J022636--043701 & 02:26:36.93 & $-$04:37:01.94 & $3.450$ & $24.6$ & $-21.2$ & 1 & 1 & L13 \\
HSC J022644--042147 & 02:26:44.46 & $-$04:21:47.41 & $3.395$ & $24.3$ & $-21.4$ & 1 & 1 & L13 \\
HSC J022646--041822 & 02:26:46.51 & $-$04:18:22.49 & $3.438$ & $24.4$ & $-21.4$ & 1 & 1 & L13 \\
HSC J022658--041019 & 02:26:58.54 & $-$04:10:19.61 & $3.406$ & $23.6$ & $-22.1$ & 1 & 1 & L13 \\
HSC J022659--041832 & 02:26:59.60 & $-$04:18:32.95 & $3.873$ & $23.7$ & $-22.3$ & 1 & 1 & L13 \\
HSC J022717--042824 & 02:27:17.06 & $-$04:28:24.67 & $3.703$ & $23.5$ & $-22.4$ & 1 & 1 & L13 \\
HSC J022723--041841 & 02:27:23.07 & $-$04:18:41.83 & $3.289$ & $24.2$ & $-21.5$ & 1 & 1 & L13 \\
HSC J022727--042226 & 02:27:27.97 & $-$04:22:26.11 & $3.521$ & $23.4$ & $-22.4$ & 1 & 1 & L13 \\
HSC J022728--042302 & 02:27:28.81 & $-$04:23:02.80 & $3.512$ & $24.6$ & $-21.2$ & 1 & 1 & L13 \\
HSC J022746--041527 & 02:27:46.84 & $-$04:15:27.77 & $3.672$ & $23.5$ & $-22.4$ & 1 & 1 & L13 \\
HSC J022754--042637 & 02:27:54.45 & $-$04:26:37.88 & $3.835$ & $22.4$ & $-23.6$ & 1 & 1 & L13 \\
HSC J084818+004509 &  08:48:18.34  &  00:45:09.33  &  $5.80$ & $23.7$ & $-22.9$ & 3 & 1 & This Study \\
HSC J084021+010311 &  08:40:21.29  &  01:03:11.41  & $5.61$ & $24.1$ & $-22.5$ & 3 & 1 & This Study \\
HSC J085723+014254 & 08:57:23.95 & 01:42:54.56 & $5.820$ & $23.9$ & $-22.8$ & 3 & 1 & M16 \\
HSC J090704+002624 &  09:07:04.05  &  00:26:24.79  &  $5.96$ & $24.3$ & $-22.4$ & 3 & 1 & This Study \\
HSC J095820+021658 & 09:58:20.93 & 02:16:58.61 & $4.916$ & $25.4$ & $-21.0$ & 2 & 1 & M12 \\
HSC J095830+020630 & 09:58:30.62 & 02:06:30.91 & $4.891$ & $25.5$ & $-20.9$ & 2 & 1 & M12 \\
HSC J095835+020454 & 09:58:35.24 & 02:04:54.92 & $4.092$ & $25.2$ & $-20.8$ & 1 & 1 & M12 \\
HSC J095839+020519 & 09:58:39.44 & 02:05:19.89 & $4.093$ & $24.9$ & $-21.1$ & 1 & 1 & M12 \\
HSC J095842+021523 & 09:58:42.78 & 02:15:23.95 & $3.933$ & $24.7$ & $-21.3$ & 1 & 1 & M12 \\
HSC J095847+020700 & 09:58:47.48 & 02:07:00.79 & $4.155$ & $25.8$ & $-20.3$ & 1 & 1 & M12 \\
HSC J095901+020303 & 09:59:01.50 & 02:03:03.19 & $3.855$ & $24.7$ & $-21.3$ & 1 & 1 & M12 \\
HSC J095902+020902 & 09:59:02.11 & 02:09:02.51 & $5.305$ & $24.1$ & $-22.4$ & 2 & 1 & M12 \\
HSC J095902+021742 & 09:59:02.18 & 02:17:42.57 & $4.158$ & $26.0$ & $-20.1$ & 1 & 1 & M12 \\
HSC J095904+021843 & 09:59:04.31 & 02:18:43.36 & $4.802$ & $24.8$ & $-21.5$ & 2 & 1 & M12 \\
HSC J095907+020721 & 09:59:07.26 & 02:07:21.31 & $5.181$ & $24.1$ & $-22.4$ & 2 & 1 & M12 \\
HSC J095910+024623 & 09:59:10.84 & 02:46:23.54 & $3.515$ & $23.5$ & $-22.3$ & 1 & 1 & Mas17 \\
HSC J095910+021700 & 09:59:10.96 & 02:17:00.92 & $4.248$ & $24.5$ & $-21.6$ & 1 & 1 & This Study \\
HSC J095916+020714 & 09:59:16.24 & 02:07:14.84 & $3.867$ & $24.3$ & $-21.7$ & 1 & 1 & M12 \\
HSC J095928+015342 & 09:59:28.34 & 01:53:42.03 & $4.719$ & $24.6$ & $-21.7$ & 2 & 1 & M12 \\
HSC J095929+022950 & 09:59:29.35 & 02:29:50.15 & $4.840$ & $25.7$ & $-20.6$ & 2 & 1 & M12 \\
HSC J095932+014205 & 09:59:32.47 & 01:42:05.98 & $4.483$ & $24.6$ & $-21.6$ & 1 & 1 & M12 \\
HSC J095934+024015 & 09:59:34.78 & 02:40:15.40 & $4.097$ & $25.8$ & $-20.2$ & 1 & 1 & M12 \\
HSC J095936+024309 & 09:59:36.29 & 02:43:09.72 & $4.420$ & $24.7$ & $-21.5$ & 1 & 1 & M12 \\
HSC J095936+020218 & 09:59:36.62 & 02:02:18.24 & $4.720$ & $25.2$ & $-21.1$ & 2 & 1 & M12 \\
HSC J095944+013817 & 09:59:44.36 & 01:38:17.10 & $4.285$ & $24.9$ & $-21.3$ & 1 & 1 & M12 \\
HSC J095945+020325 & 09:59:45.91 & 02:03:25.78 & $4.150$ & $24.6$ & $-21.5$ & 1 & 1 & M12 \\
HSC J095946+020642 & 09:59:46.02 & 02:06:42.47 & $4.894$ & $24.7$ & $-21.7$ & 2 & 1 & M12 \\
HSC J095946+014840 & 09:59:46.36 & 01:48:40.48 & $4.653$ & $25.1$ & $-21.2$ & 1 & 1 & M12 \\
HSC J095946+024215 & 09:59:46.70 & 02:42:15.85 & $4.849$ & $25.4$ & $-20.9$ & 2 & 1 & M12 \\
HSC J095947+022232 & 09:59:47.07 & 02:22:32.84 & $4.414$ & $25.1$ & $-21.0$ & 1 & 1 & M12 \\
HSC J095948+022720 & 09:59:48.52 & 02:27:20.37 & $4.582$ & $25.3$ & $-21.0$ & 1 & 1 & M12 \\
HSC J095952+022424 & 09:59:52.83 & 02:24:24.33 & $4.185$ & $25.2$ & $-20.9$ & 1 & 1 & M12 \\
HSC J095953+023411 & 09:59:53.99 & 02:34:11.79 & $4.953$ & $24.7$ & $-21.7$ & 2 & 1 & M12 \\
HSC J095954+021516 & 09:59:54.52 & 02:15:16.50 & $5.688$ & $25.9$ & $-20.7$ & 2 & 1 & M12 \\
HSC J095954+021039 & 09:59:54.77 & 02:10:39.26 & $5.662$ & $25.6$ & $-21.0$ & 2 & 1 & M12 \\
HSC J095955+023808 & 09:59:55.17 & 02:38:08.18 & $4.541$ & $24.7$ & $-21.5$ & 2 & 1 & M12 \\
HSC J095956+023557 & 09:59:56.02 & 02:35:57.82 & $3.919$ & $24.8$ & $-21.2$ & 1 & 1 & M12 \\
HSC J095956+021227 & 09:59:56.54 & 02:12:27.12 & $5.983$ & $24.6$ & $-22.1$ & 3 & 1 & W13 \\
HSC J095957+023113 & 09:59:57.26 & 02:31:13.02 & $4.542$ & $24.9$ & $-21.3$ & 1 & 1 & M12 \\
HSC J100000+015956 & 10:00:00.82 & 01:59:56.61 & $5.655$ & $24.9$ & $-21.6$ & 3 & 1 & M12 \\
HSC J100001+022750 & 10:00:01.49 & 02:27:50.06 & $4.085$ & $25.1$ & $-20.9$ & 1 & 1 & M12 \\
HSC J100002+022103 & 10:00:02.09 & 02:21:03.24 & $4.270$ & $25.1$ & $-21.0$ & 1 & 1 & M12 \\
HSC J100002+022523 & 10:00:02.32 & 02:25:23.98 & $5.053$ & $24.7$ & $-21.8$ & 2 & 1 & M12 \\
HSC J100004+023735 & 10:00:04.06 & 02:37:35.77 & $5.158$ & $24.3$ & $-22.2$ & 2 & 1 & M12 \\
HSC J100004+020845 & 10:00:04.17 & 02:08:45.68 & $4.840$ & $25.1$ & $-21.2$ & 2 & 1 & M12 \\
HSC J100005+020312 & 10:00:05.11 & 02:03:12.23 & $5.240$ & $25.1$ & $-21.4$ & 2 & 1 & M12 \\
HSC J100007+023414 & 10:00:07.36 & 02:34:14.42 & $4.550$ & $24.9$ & $-21.4$ & 1 & 1 & M12 \\
HSC J100008+021136 & 10:00:08.78 & 02:11:36.46 & $5.866$ & $24.9$ & $-21.8$ & 3 & 1 & M12 \\
HSC J100016+022005 & 10:00:16.25 & 02:20:05.07 & $4.301$ & $24.3$ & $-21.9$ & 1 & 1 & T17 \\
HSC J100016+022149 & 10:00:16.58 & 02:21:49.73 & $3.433$ & $24.8$ & $-20.9$ & 1 & 1 & T17 \\
HSC J100016+022117 & 10:00:16.95 & 02:21:17.18 & $3.617$ & $24.5$ & $-21.4$ & 1 & 1 & Kr15 \\
HSC J100017+015807 & 10:00:17.68 & 01:58:07.18 & $4.994$ & $24.9$ & $-21.5$ & 2 & 1 & M12 \\
HSC J100018+022840 & 10:00:18.01 & 02:28:40.11 & $3.374$ & $24.8$ & $-20.9$ & 1 & 1 & T17 \\
HSC J100018+021247 & 10:00:18.42 & 02:12:47.00 & $4.386$ & $25.0$ & $-21.2$ & 1 & 1 & M12 \\
HSC J100018+022814 & 10:00:18.83 & 02:28:14.16 & $4.756$ & $24.7$ & $-21.6$ & 2 & 1 & M12 \\
HSC J100019+021539 & 10:00:19.74 & 02:15:39.73 & $4.065$ & $25.1$ & $-20.9$ & 1 & 1 & T17 \\
HSC J100022+024103 & 10:00:22.51 & 02:41:03.25 & $5.661$ & $25.3$ & $-21.3$ & 2 & 1 & M12 \\
HSC J100023+023244 & 10:00:23.35 & 02:32:44.91 & $3.614$ & $24.6$ & $-21.2$ & 1 & 1 & T17 \\
HSC J100023+020304 & 10:00:23.36 & 02:03:04.36 & $4.518$ & $24.5$ & $-21.7$ & 1 & 1 & M12 \\
HSC J100023+021520 & 10:00:23.44 & 02:15:20.22 & $3.558$ & $24.6$ & $-21.2$ & 1 & 1 & T17 \\
HSC J100024+023309 & 10:00:24.05 & 02:33:09.48 & $3.357$ & $24.9$ & $-20.9$ & 1 & 1 & T17 \\
HSC J100024+023136 & 10:00:24.11 & 02:31:36.58 & $4.016$ & $24.7$ & $-21.4$ & 1 & 1 & M12 \\
HSC J100024+022911 & 10:00:24.21 & 02:29:11.02 & $3.836$ & $25.5$ & $-20.5$ & 1 & 1 & T17 \\
HSC J100025+022224 & 10:00:25.73 & 02:22:24.18 & $3.422$ & $24.5$ & $-21.2$ & 1 & 1 & Kr15 \\
HSC J100028+022327 & 10:00:28.26 & 02:23:27.93 & $3.685$ & $25.3$ & $-20.6$ & 1 & 1 & T17 \\
HSC J100030+015143 & 10:00:30.68 & 01:51:43.57 & $4.278$ & $24.4$ & $-21.7$ & 1 & 1 & M12 \\
HSC J100032+022909 & 10:00:32.42 & 02:29:09.61 & $4.815$ & $25.1$ & $-21.2$ & 2 & 1 & T17 \\
HSC J100032+021528 & 10:00:32.61 & 02:15:28.34 & $4.450$ & $24.8$ & $-21.4$ & 1 & 1 & M12 \\
HSC J100032+022856 & 10:00:32.84 & 02:28:56.95 & $4.007$ & $25.0$ & $-21.1$ & 1 & 1 & T17 \\
HSC J100033+015023 & 10:00:33.54 & 01:50:23.89 & $4.375$ & $23.8$ & $-22.4$ & 1 & 1 & This Study \\
HSC J100034+015921 & 10:00:34.30 & 01:59:21.23 & $4.466$ & $23.8$ & $-22.5$ & 1 & 1 & M12 \\
HSC J100034+022655 & 10:00:34.31 & 02:26:55.43 & $3.332$ & $25.4$ & $-20.3$ & 1 & 1 & Kr15 \\
HSC J100034+021524 & 10:00:34.33 & 02:15:24.53 & $5.121$ & $24.9$ & $-21.5$ & 2 & 1 & M12 \\
HSC J100034+013616 & 10:00:34.62 & 01:36:16.43 & $4.902$ & $25.1$ & $-21.2$ & 2 & 1 & M12 \\
HSC J100035+023310 & 10:00:35.08 & 02:33:10.48 & $3.414$ & $24.8$ & $-21.0$ & 1 & 1 & T17 \\
HSC J100035+022729 & 10:00:35.82 & 02:27:29.43 & $3.398$ & $24.6$ & $-21.1$ & 1 & 1 & T17 \\
HSC J100037+022540 & 10:00:37.73 & 02:25:40.32 & $3.989$ & $24.2$ & $-21.8$ & 1 & 1 & T17 \\
HSC J100038+015534 & 10:00:38.91 & 01:55:34.32 & $4.325$ & $25.5$ & $-20.7$ & 1 & 1 & M12 \\
HSC J100039+021444 & 10:00:39.01 & 02:14:44.94 & $4.385$ & $24.9$ & $-21.2$ & 1 & 1 & T17 \\
HSC J100039+020414 & 10:00:39.36 & 02:04:14.14 & $4.107$ & $24.8$ & $-21.3$ & 1 & 1 & M12 \\
HSC J100041+022637 & 10:00:41.08 & 02:26:37.31 & $4.867$ & $25.5$ & $-20.8$ & 2 & 1 & M12 \\
HSC J100041+021714 & 10:00:41.17 & 02:17:14.12 & $4.578$ & $24.9$ & $-21.4$ & 2 & 1 & T17 \\
HSC J100041+022817 & 10:00:41.43 & 02:28:17.22 & $3.709$ & $24.9$ & $-21.0$ & 1 & 1 & T17 \\
HSC J100042+022230 & 10:00:42.59 & 02:22:30.79 & $4.386$ & $24.6$ & $-21.6$ & 1 & 1 & T17 \\
HSC J100042+021811 & 10:00:42.64 & 02:18:11.44 & $4.588$ & $24.3$ & $-21.9$ & 2 & 1 & T17 \\
HSC J100043+022241 & 10:00:43.23 & 02:22:41.93 & $4.930$ & $24.4$ & $-22.0$ & 2 & 1 & M12 \\
HSC J100044+022244 & 10:00:44.85 & 02:22:44.77 & $3.740$ & $24.4$ & $-21.5$ & 1 & 1 & T17 \\
HSC J100047+022856 & 10:00:47.01 & 02:28:56.99 & $4.032$ & $25.0$ & $-21.0$ & 1 & 1 & T17 \\
HSC J100047+021802 & 10:00:47.67 & 02:18:02.08 & $4.586$ & $24.4$ & $-21.9$ & 1 & 1 & M12 \\
HSC J100047+023243 & 10:00:47.93 & 02:32:43.19 & $3.990$ & $24.4$ & $-21.6$ & 1 & 1 & T17 \\
HSC J100048+022224 & 10:00:48.45 & 02:22:24.71 & $3.475$ & $24.1$ & $-21.7$ & 1 & 1 & T17 \\
HSC J100049+021543 & 10:00:49.22 & 02:15:43.79 & $3.898$ & $24.4$ & $-21.5$ & 1 & 1 & T17 \\
HSC J100055+021309 & 10:00:55.44 & 02:13:09.19 & $4.872$ & $25.9$ & $-20.4$ & 2 & 1 & M12 \\
HSC J100056+015746 & 10:00:56.04 & 01:57:46.38 & $4.740$ & $25.4$ & $-20.9$ & 2 & 1 & M12 \\
HSC J100101+020531 & 10:01:01.05 & 02:05:31.51 & $4.938$ & $24.9$ & $-21.4$ & 2 & 1 & M12 \\
HSC J100101+022358 & 10:01:01.20 & 02:23:58.15 & $4.930$ & $24.9$ & $-21.5$ & 2 & 1 & M12 \\
HSC J100102+013526 & 10:01:02.70 & 01:35:26.56 & $4.324$ & $25.2$ & $-21.0$ & 1 & 1 & M12 \\
HSC J100105+020920 & 10:01:05.28 & 02:09:20.56 & $4.110$ & $25.7$ & $-20.4$ & 1 & 1 & M12 \\
HSC J100105+015502 & 10:01:05.35 & 01:55:02.46 & $3.772$ & $24.2$ & $-21.7$ & 1 & 1 & M12 \\
HSC J100105+020948 & 10:01:05.96 & 02:09:48.65 & $4.562$ & $24.6$ & $-21.6$ & 1 & 1 & M12 \\
HSC J100109+021513 & 10:01:09.72 & 02:15:13.45 & $5.712$ & $25.9$ & $-20.7$ & 2 & 1 & M12 \\
HSC J100109+020430 & 10:01:09.87 & 02:04:30.12 & $4.217$ & $25.4$ & $-20.8$ & 1 & 1 & M12 \\
HSC J100110+021956 & 10:01:10.14 & 02:19:56.29 & $4.534$ & $24.4$ & $-21.8$ & 1 & 1 & M12 \\
HSC J100110+020729 & 10:01:10.99 & 02:07:29.47 & $4.057$ & $26.0$ & $-20.0$ & 1 & 1 & M12 \\
HSC J100111+023805 & 10:01:11.35 & 02:38:05.14 & $4.802$ & $24.9$ & $-21.4$ & 2 & 1 & M12 \\
HSC J100114+021842 & 10:01:14.25 & 02:18:42.46 & $4.584$ & $25.6$ & $-20.7$ & 1 & 1 & M12 \\
HSC J100116+021030 & 10:01:16.94 & 02:10:30.55 & $4.658$ & $24.8$ & $-21.5$ & 2 & 1 & M12 \\
HSC J100117+015719 & 10:01:17.11 & 01:57:19.23 & $4.488$ & $24.4$ & $-21.8$ & 2 & 1 & M12 \\
HSC J100119+023022 & 10:01:19.11 & 02:30:22.88 & $4.375$ & $24.7$ & $-21.5$ & 1 & 1 & M12 \\
HSC J100119+021150 & 10:01:19.69 & 02:11:50.41 & $3.788$ & $24.6$ & $-21.4$ & 1 & 1 & M12 \\
HSC J100120+022408 & 10:01:20.61 & 02:24:08.53 & $5.249$ & $24.8$ & $-21.7$ & 2 & 1 & M12 \\
HSC J100121+021621 & 10:01:21.90 & 02:16:21.83 & $4.301$ & $24.2$ & $-21.9$ & 1 & 1 & M12 \\
HSC J100122+015907 & 10:01:22.25 & 01:59:07.24 & $3.813$ & $24.6$ & $-21.4$ & 1 & 1 & M12 \\
HSC J100122+022502 & 10:01:22.60 & 02:25:02.95 & $4.530$ & $25.0$ & $-21.3$ & 1 & 1 & M12 \\
HSC J100123+015600 & 10:01:23.84 & 01:56:00.46 & $5.726$ & $25.9$ & $-20.8$ & 2 & 1 & M12 \\
HSC J100125+020508 & 10:01:25.06 & 02:05:08.31 & $5.032$ & $24.7$ & $-21.7$ & 2 & 1 & M12 \\
HSC J100126+014526 & 10:01:26.66 & 01:45:26.30 & $4.527$ & $23.8$ & $-22.5$ & 1 & 1 & M12 \\
HSC J100136+015517 & 10:01:36.81 & 01:55:17.02 & $4.432$ & $24.9$ & $-21.3$ & 1 & 1 & M12 \\
HSC J100145+015712 & 10:01:45.12 & 01:57:12.36 & $4.909$ & $24.9$ & $-21.5$ & 2 & 1 & M12 \\
HSC J100147+015505 & 10:01:47.05 & 01:55:05.59 & $4.020$ & $25.4$ & $-20.6$ & 1 & 1 & M12 \\
HSC J100148+015727 & 10:01:48.56 & 01:57:27.89 & $4.919$ & $25.0$ & $-21.4$ & 2 & 1 & M12 \\
HSC J100151+014729 & 10:01:51.47 & 01:47:29.56 & $4.357$ & $25.1$ & $-21.1$ & 1 & 1 & This Study \\
HSC J100154+023226 & 10:01:54.14 & 02:32:26.77 & $4.268$ & $25.2$ & $-20.9$ & 1 & 1 & M12 \\
HSC J100155+015803 & 10:01:55.03 & 01:58:03.38 & $4.173$ & $24.2$ & $-21.9$ & 1 & 1 & M12 \\
HSC J100159+015612 & 10:01:59.47 & 01:56:12.95 & $4.441$ & $24.6$ & $-21.6$ & 1 & 1 & M12 \\
HSC J100216+023438 & 10:02:16.25 & 02:34:38.56 & $5.657$ & $25.2$ & $-21.4$ & 2 & 1 & This Study \\
HSC J100218+021940 & 10:02:18.85 & 02:19:40.04 & $4.663$ & $24.5$ & $-21.8$ & 2 & 1 & This Study \\
HSC J100219+015736 & 10:02:19.12 & 01:57:36.65 & $4.098$ & $24.0$ & $-22.0$ & 1 & 1 & M12 \\
HSC J100223+015351 & 10:02:23.12 & 01:53:51.29 & $3.807$ & $24.6$ & $-21.4$ & 1 & 1 & M12 \\
HSC J100226+021750 & 10:02:26.84 & 02:17:50.25 & $5.045$ & $23.9$ & $-22.5$ & 2 & 1 & This Study \\
HSC J100233+023539 & 10:02:33.86 & 02:35:39.50 & $4.487$ & $25.5$ & $-20.7$ & 1 & 1 & This Study \\
HSC J100234+022922 & 10:02:34.16 & 02:29:22.39 & $4.515$ & $24.9$ & $-21.3$ & 1 & 1 & This Study \\
HSC J100242+015649 & 10:02:42.92 & 01:56:49.69 & $4.004$ & $25.2$ & $-20.8$ & 1 & 1 & M12 \\
HSC J100245+015535 & 10:02:45.66 & 01:55:35.91 & $5.633$ & $25.4$ & $-21.2$ & 2 & 1 & M12 \\
HSC J100332+024552 &  10:03:32.98  &  02:45:52.97   & $5.957$ & $25.2$ & $-21.5$ & 3 & 1 & This Study \\
HSC J162833+431210 & 16:28:33.02 & 43:12:10.56 & $6.030$ & $23.9$ & $-22.8$ & 3 & 1 & M17 \\
HSC J163026+431558 & 16:30:26.36 & 43:15:58.60 & $6.020$ & $24.0$ & $-22.7$ & 3 & 1 & M17 \\
HSC J232558+002557 & 23:25:58.43 & 00:25:57.53 & $5.703$ & $25.3$ & $-21.3$ & 2 & 1 & S17 \\
\hline
    \multicolumn{9}{c}{AGNs} \\
HSC J020258--025153 & 02:02:58.21 & $-$02:51:53.59 & $6.030$ & $23.1$ & $-23.6$ & 3 & 2 & M17 \\
HSC J020402--034319 & 02:04:02.55 & $-$03:43:19.67 & $3.820$ & $20.7$ & $-25.2$ & 1 & 2 & P15 \\
HSC J020423--051323 & 02:04:23.83 & $-$05:13:23.40 & $3.768$ & $20.8$ & $-25.1$ & 1 & 2 & P15 \\
HSC J020429--031257 & 02:04:29.27 & $-$03:12:57.41 & $3.615$ & $20.4$ & $-25.4$ & 1 & 2 & P15 \\
HSC J020611--025537 & 02:06:11.20 & $-$02:55:37.82 & $6.030$ & $21.7$ & $-25.0$ & 3 & 2 & M17 \\
HSC J020630--032847 & 02:06:30.47 & $-$03:28:47.13 & $3.527$ & $20.7$ & $-25.1$ & 1 & 2 & P15 \\
HSC J021013--045620 & 02:10:13.19 & $-$04:56:20.79 & $6.438$ & $22.3$ & $-24.5$ & 3 & 2 & W10 \\
HSC J021131--042126 & 02:11:31.07 & $-$04:21:26.74 & $3.875$ & $21.0$ & $-25.0$ & 1 & 2 & P15 \\
HSC J021527--060359 & 02:15:27.28 & $-$06:03:59.82 & $4.065$ & $21.2$ & $-24.8$ & 1 & 2 & P15 \\
HSC J021712--054109 & 02:17:12.98 & $-$05:41:09.66 & $4.563$ & $21.5$ & $-24.7$ & 1 & 2 & W16 \\
HSC J021831--044354 & 02:18:31.37 & $-$04:43:54.41 & $3.724$ & $22.0$ & $-23.9$ & 1 & 2 & O08 \\
HSC J021844--044824 & 02:18:44.46 & $-$04:48:24.62 & $4.564$ & $19.8$ & $-26.4$ & 1 & 2 & W16 \\
HSC J021952--055957 & 02:19:52.67 & $-$05:59:57.17 & $3.863$ & $19.1$ & $-26.9$ & 1 & 2 & P15 \\
HSC J022156--055148 & 02:21:56.57 & $-$05:51:48.73 & $3.847$ & $21.1$ & $-24.9$ & 1 & 2 & P15 \\
HSC J022307--030840 & 02:23:07.95 & $-$03:08:40.16 & $3.675$ & $19.7$ & $-26.2$ & 1 & 2 & P15 \\
HSC J022320--031824 & 02:23:20.70 & $-$03:18:24.18 & $3.865$ & $19.4$ & $-26.6$ & 1 & 2 & P15 \\
HSC J022413--052724 & 02:24:13.40 & $-$05:27:24.79 & $3.779$ & $20.5$ & $-25.4$ & 1 & 2 & P15 \\
HSC J022527--042631 & 02:25:27.23 & $-$04:26:31.22 & $3.853$ & $21.4$ & $-24.6$ & 1 & 2 & L13 \\
HSC J022550--042142 & 02:25:50.66 & $-$04:21:42.21 & $3.860$ & $22.1$ & $-23.9$ & 1 & 2 & L13 \\
HSC J022739--041216 & 02:27:39.95 & $-$04:12:16.48 & $4.520$ & $22.9$ & $-23.4$ & 1 & 2 & L13 \\
HSC J022754--044535 & 02:27:54.62 & $-$04:45:35.35 & $3.741$ & $20.4$ & $-25.5$ & 1 & 2 & P15 \\
HSC J023002--043119 & 02:30:02.46 & $-$04:31:19.70 & $3.658$ & $20.9$ & $-25.0$ & 1 & 2 & P15 \\
HSC J023058--041357 & 02:30:58.66 & $-$04:13:57.91 & $4.014$ & $21.7$ & $-24.3$ & 1 & 2 & P15 \\
HSC J023519--042855 & 02:35:19.65 & $-$04:28:55.59 & $4.154$ & $21.4$ & $-24.6$ & 1 & 2 & P15 \\
HSC J084455+001848 & 08:44:55.08 & 00:18:48.58 & $3.692$ & $20.1$ & $-25.8$ & 1 & 2 & P15 \\
HSC J085051--002437 & 08:50:51.89 & $-$00:24:37.92 & $3.718$ & $20.3$ & $-25.6$ & 1 & 2 & P15 \\
HSC J085135+011300 & 08:51:35.61 & 01:13:00.59 & $3.607$ & $19.6$ & $-26.3$ & 1 & 2 & P15 \\
HSC J085151+020756 & 08:51:51.27 & 02:07:56.11 & $4.278$ & $19.1$ & $-27.0$ & 1 & 2 & P15 \\
HSC J085828+021214 & 08:58:28.62 & 02:12:14.85 & $3.527$ & $20.8$ & $-25.1$ & 1 & 2 & P15 \\
HSC J085907+002255 & 08:59:07.19 & 00:22:55.92 & $6.390$ & $22.8$ & $-24.0$ & 3 & 2 & M16 \\
HSC J090042+002415 & 09:00:42.11 & 00:24:15.88 & $3.637$ & $21.3$ & $-24.6$ & 1 & 2 & P15 \\
HSC J090242+014525 & 09:02:42.95 & 01:45:25.11 & $3.685$ & $19.9$ & $-26.0$ & 1 & 2 & P15 \\
HSC J090254+015510 & 09:02:54.87 & 01:55:10.85 & $6.010$ & $24.1$ & $-22.6$ & 3 & 2 & M17 \\
HSC J090314+021128 & 09:03:14.68 & 02:11:28.27 & $5.920$ & $23.6$ & $-23.1$ & 3 & 2 & M17 \\
HSC J090701+003745 & 09:07:01.95 & 00:37:45.15 & $3.675$ & $20.9$ & $-24.9$ & 1 & 2 & P15 \\
HSC J090833+014805 & 09:08:33.49 & 01:48:05.14 & $3.738$ & $21.2$ & $-24.7$ & 1 & 2 & P15 \\
HSC J095856+021047 & 09:58:56.69 & 02:10:47.78 & $4.200$ & $23.8$ & $-22.3$ & 1 & 2 & Mas12 \\
HSC J095901+024418 & 09:59:01.30 & 02:44:18.76 & $3.520$ & $23.1$ & $-22.7$ & 1 & 2 & Mas12 \\
HSC J095908+022707 & 09:59:08.11 & 02:27:07.52 & $5.070$ & $23.8$ & $-22.6$ & 2 & 2 & Mas12 \\
HSC J095928+015258 & 09:59:28.99 & 01:52:58.00 & $4.606$ & $24.0$ & $-22.3$ & 2 & 2 & M12 \\
HSC J095931+021332 & 09:59:31.01 & 02:13:32.89 & $3.650$ & $22.8$ & $-23.1$ & 1 & 2 & Mas12 \\
HSC J100025+014533 & 10:00:25.77 & 01:45:33.29 & $4.140$ & $22.6$ & $-23.5$ & 1 & 2 & Mas12 \\
HSC J100027+015750 & 10:00:27.95 & 01:57:50.15 & $3.410$ & $23.8$ & $-21.9$ & 1 & 2 & Mas12 \\
HSC J100051+023457 & 10:00:51.61 & 02:34:57.50 & $5.300$ & $23.6$ & $-22.9$ & 2 & 2 & Mas12 \\
HSC J100112+015107 & 10:01:12.62 & 01:51:07.57 & $3.840$ & $24.8$ & $-21.2$ & 1 & 2 & Mas12 \\
HSC J100144+013857 & 10:01:44.89 & 01:38:57.44 & $3.890$ & $23.3$ & $-22.7$ & 1 & 2 & Mas12 \\
HSC J100156+015218 & 10:01:56.55 & 01:52:18.94 & $4.450$ & $22.2$ & $-24.0$ & 1 & 2 & Mas12 \\
HSC J100233+022328 & 10:02:33.23 & 02:23:28.81 & $3.650$ & $22.6$ & $-23.3$ & 1 & 2 & Mas12 \\
HSC J100320+022930 & 10:03:20.90 & 02:29:30.03 & $4.412$ & $20.6$ & $-25.6$ & 1 & 2 & P15 \\
HSC J100338+015641 & 10:03:38.71 & 01:56:41.44 & $3.680$ & $20.7$ & $-25.2$ & 1 & 2 & L13 \\
HSC J100346+011911 & 10:03:46.33 & 01:19:11.15 & $3.557$ & $21.2$ & $-24.6$ & 1 & 2 & L13 \\
HSC J100426+022444 & 10:04:26.84 & 02:24:44.86 & $4.461$ & $22.2$ & $-24.0$ & 1 & 2 & L13 \\
HSC J114608--001745 & 11:46:08.95 & $-$00:17:45.60 & $3.921$ & $21.0$ & $-25.0$ & 1 & 2 & P15 \\
HSC J114706--010958 & 11:47:06.42 & $-$01:09:58.20 & $5.310$ & $19.4$ & $-27.1$ & 2 & 2 & Y17 \\
HSC J115122--000152 & 11:51:22.87 & $-$00:01:52.87 & $3.877$ & $20.8$ & $-25.2$ & 1 & 2 & P15 \\
HSC J115221+005536 & 11:52:21.27 & 00:55:36.56 & $6.370$ & $21.8$ & $-25.0$ & 3 & 2 & M16 \\
HSC J120138+010336 & 12:01:38.57 & 01:03:36.37 & $3.859$ & $19.9$ & $-26.0$ & 1 & 2 & P15 \\
HSC J120210--005425 & 12:02:10.09 & $-$00:54:25.52 & $3.596$ & $19.1$ & $-26.8$ & 1 & 2 & P15 \\
HSC J120246--005701 & 12:02:46.37 & $-$00:57:01.63 & $5.930$ & $23.7$ & $-22.9$ & 3 & 2 & M16 \\
HSC J120312--001118 & 12:03:12.65 & $-$00:11:18.77 & $4.592$ & $19.3$ & $-26.9$ & 1 & 2 & W16 \\
HSC J120621+002141 & 12:06:21.74 & 00:21:41.23 & $3.665$ & $19.9$ & $-26.0$ & 1 & 2 & P15 \\
HSC J120637--010424 & 12:06:37.95 & $-$01:04:24.71 & $3.728$ & $20.7$ & $-25.2$ & 1 & 2 & P15 \\
HSC J120754--000553 & 12:07:54.14 & $-$00:05:53.18 & $6.010$ & $23.9$ & $-22.8$ & 3 & 2 & M16 \\
HSC J120823+001027 & 12:08:23.84 & 00:10:27.68 & $5.273$ & $20.5$ & $-26.0$ & 2 & 2 & W16 \\
HSC J142046--011054 & 14:20:46.84 & $-$01:10:54.77 & $3.992$ & $20.2$ & $-25.9$ & 1 & 2 & P15 \\
HSC J142205--012812 & 14:22:05.63 & $-$01:28:12.36 & $3.909$ & $20.8$ & $-25.2$ & 1 & 2 & P15 \\
HSC J142329+004138 & 14:23:29.99 & 00:41:38.57 & $3.770$ & $19.4$ & $-26.5$ & 1 & 2 & P15 \\
HSC J142517--001540 & 14:25:17.72 & $-$00:15:40.88 & $6.180$ & $22.9$ & $-23.9$ & 3 & 2 & M17 \\
HSC J142548--002538 & 14:25:48.07 & $-$00:25:38.07 & $3.741$ & $20.9$ & $-25.0$ & 1 & 2 & P15 \\
HSC J142647+002740 & 14:26:47.82 & 00:27:40.07 & $3.692$ & $19.3$ & $-26.6$ & 1 & 2 & P15 \\
HSC J142920--000207 & 14:29:20.22 & $-$00:02:07.44 & $6.040$ & $23.0$ & $-23.7$ & 3 & 2 & M17 \\
HSC J143619--004855 & 14:36:19.27 & $-$00:48:55.34 & $4.001$ & $20.9$ & $-25.1$ & 1 & 2 & P15 \\
HSC J143634+005111 & 14:36:34.50 & 00:51:11.92 & $3.686$ & $21.3$ & $-24.6$ & 1 & 2 & P15 \\
HSC J144001--010702 & 14:40:01.30 & $-$01:07:02.17 & $6.130$ & $23.7$ & $-23.0$ & 3 & 2 & M17 \\
HSC J144137--001324 & 14:41:37.20 & $-$00:13:24.89 & $3.622$ & $21.4$ & $-24.5$ & 1 & 2 & P15 \\
HSC J144407--010152 & 14:44:07.64 & $-$01:01:52.65 & $4.540$ & $19.3$ & $-27.0$ & 1 & 2 & W16 \\
HSC J161143+553157 & 16:11:43.23 & 55:31:57.31 & $3.583$ & $20.1$ & $-25.8$ & 1 & 2 & P15 \\
HSC J162445+440410 & 16:24:45.39 & 44:04:10.04 & $3.639$ & $20.0$ & $-25.9$ & 1 & 2 & P15 \\
HSC J221644--001650 & 22:16:44.47 & $-$00:16:50.05 & $6.100$ & $22.8$ & $-23.9$ & 3 & 2 & M16 \\
HSC J221705--001307 & 22:17:05.71 & $-$00:13:07.67 & $4.668$ & $20.2$ & $-26.0$ & 1 & 2 & L13 \\
HSC J221917+010249 & 22:19:17.22 & 01:02:49.00 & $6.156$ & $23.5$ & $-23.2$ & 3 & 2 & K15 \\
HSC J222032+002537 & 22:20:32.50 & 00:25:37.64 & $4.193$ & $20.0$ & $-26.1$ & 1 & 2 & L13 \\
HSC J222221+011017 & 22:22:21.13 & 01:10:17.52 & $3.658$ & $21.0$ & $-24.9$ & 1 & 2 & P15 \\
HSC J222306+003118 & 22:23:06.94 & 00:31:18.65 & $3.780$ & $19.8$ & $-26.1$ & 1 & 2 & P15 \\
HSC J232522--002438 & 23:25:22.84 & $-$00:24:38.89 & $3.659$ & $20.6$ & $-25.3$ & 1 & 2 & P15 \\
HSC J232808--002757 & 23:28:08.99 & $-$00:27:57.28 & $4.131$ & $21.6$ & $-24.5$ & 1 & 2 & P15 \\
HSC J232850+004059 & 23:28:50.03 & 00:40:59.20 & $3.637$ & $21.5$ & $-24.4$ & 1 & 2 & P15 \\
HSC J233101--010604 & 23:31:01.64 & $-$01:06:04.15 & $3.498$ & $20.6$ & $-25.2$ & 1 & 2 & P15 \\
\end{longtable}
}
%%%%%%%%%%%%%%%%%%%%%%%%%%%%%%%%%%%%%%%

%%%%%%%%%%%%%%%%%%%%%%%%%%%%%%%%%%%%%%%%%%%%%%%%%%%%%%%%%%%%%%%%%
%%%%%%%%%%%%%%%%%%%%%%%%%%%%%%%%%%%%%%%%%%%%%%%%%%%%%%%%%%%%%%%%%
\section{Sample Selection} \label{sec:sample_selection}
%%%%%%%%%%%%%%%%%%%%%%%%%%%%%%%%%%%%%%%%%%%%%%%%%%%%%%%%%%%%%%%%%
%%%%%%%%%%%%%%%%%%%%%%%%%%%%%%%%%%%%%%%%%%%%%%%%%%%%%%%%%%%%%%%%%

%%%%%%%%%%%%%%%%%%%%%%%%%%%%%%%%%%%%%%%%%%%%%%%%%%%%%%%%%%%%%%%%%
\subsection{Source Selection} \label{sec:source_selection}
%%%%%%%%%%%%%%%%%%%%%%%%%%%%%%%%%%%%%%%%%%%%%%%%%%%%%%%%%%%%%%%%%

From the source catalogs created in Section \ref{sec:imaging_data}, 
we construct $z\sim 4-7$ dropout candidate catalogs 
based on the Lyman break color selection technique 
(e.g., \cite{1996ApJ...462L..17S}; \cite{2002ARA&A..40..579G}), 
i.e., by selecting sources which show clear 
Lyman break and blue UV continuum 
in their optical $grizy$ broadband spectral energy distributions (SEDs). 
As demonstrated in Figure \ref{fig:filters_SEDs}, 
$z \sim 4$, $z \sim 5$, $z \sim 6$, and $z \sim 7$ galaxy candidates can be selected 
based on their $gri$, $riz$, $izy$, and $zy$ colors, respectively.

First, 
we select sources with signal-to-noise ratio (S/N) $> 5$ within $1 \farcs 5$ diameter apertures 
in $i$ for $g$-dropouts, 
in $z$ for $r$-dropouts and $i$-dropouts, 
and 
in $y$ for $z$-dropouts. 
In addition, 
we require a $4.0 \sigma$ detection in $y$ for $i$-dropouts.  
We then select dropout galaxy candidates 
by using their broadband SED colors. 
Following the previous work that have used 
a similar filter set \citep{2009A&A...498..725H},   
we adopt 
\begin{eqnarray}
g-r &>& 1.0, \\
r-i &<& 1.0, \\
g-r &>& 1.5 (r-i) + 0.8,  
\end{eqnarray}
for $g$-dropouts, 
and 
\begin{eqnarray}
r-i &>& 1.2, \\
i-z &<& 0.7, \\
r-i &>& 1.5 (i-z) + 1.0,  
\end{eqnarray}
for $r$-dropouts. 
For $i$-dropouts, 
we apply the following criteria, 
\begin{eqnarray}
i-z &>& 1.5,  \\
z-y &<& 0.5, \\
i-z &>& 2.0(z-y) + 1.1. 
\end{eqnarray}
For $z$-dropouts, 
we use 
\begin{eqnarray}
z-y &>& 1.6. 
\end{eqnarray}
To remove low-$z$ source contaminations, 
we also require that 
sources be undetected ($<2 \sigma$) 
within $1 \farcs 5$ diameter apertures 
in $g$-band data for $r$-dropouts, 
in $g$- and $r$-band data for $i$-dropouts, 
and 
in $g$, $r$, and $i$-band data for $z$-dropouts. 
Since 
our $z$-dropout candidates are detected only in $y$-band images, 
we carefully check the single epoch observation images of the selected candidates 
to remove spurious sources and moving objects. 
Since this single epoch screening makes it difficult to 
find relatively faint $z$-dropouts in the UD layer, 
we focus on the D- and W-layer data in our $z$-dropout search. 
A detailed analysis for $z$-dropouts in the UD layer 
by using the latest available multiwavelength data sets, 
which is beyond the scope of this paper, 
will be presented in a forthcoming publication (Y. Harikane et al. in preparation).

Using the selection criteria described above, we select 
540,011 $g$-dropouts, 
38,944 $r$-dropouts, 
$537$ $i$-dropouts, 
and 
73 $z$-dropouts.  
Table \ref{tab:dropout_samples} summarizes our dropout galaxy candidate samples. 
The differences in the numbers of the selected candidates 
mainly come from the differences in the survey areas and depths.

%%%%%%%%%%%%%%%%%%%%%%%%%%%%%%%%%%%%%%
\begin{figure*}
 \begin{center}
  \includegraphics[width=15cm]{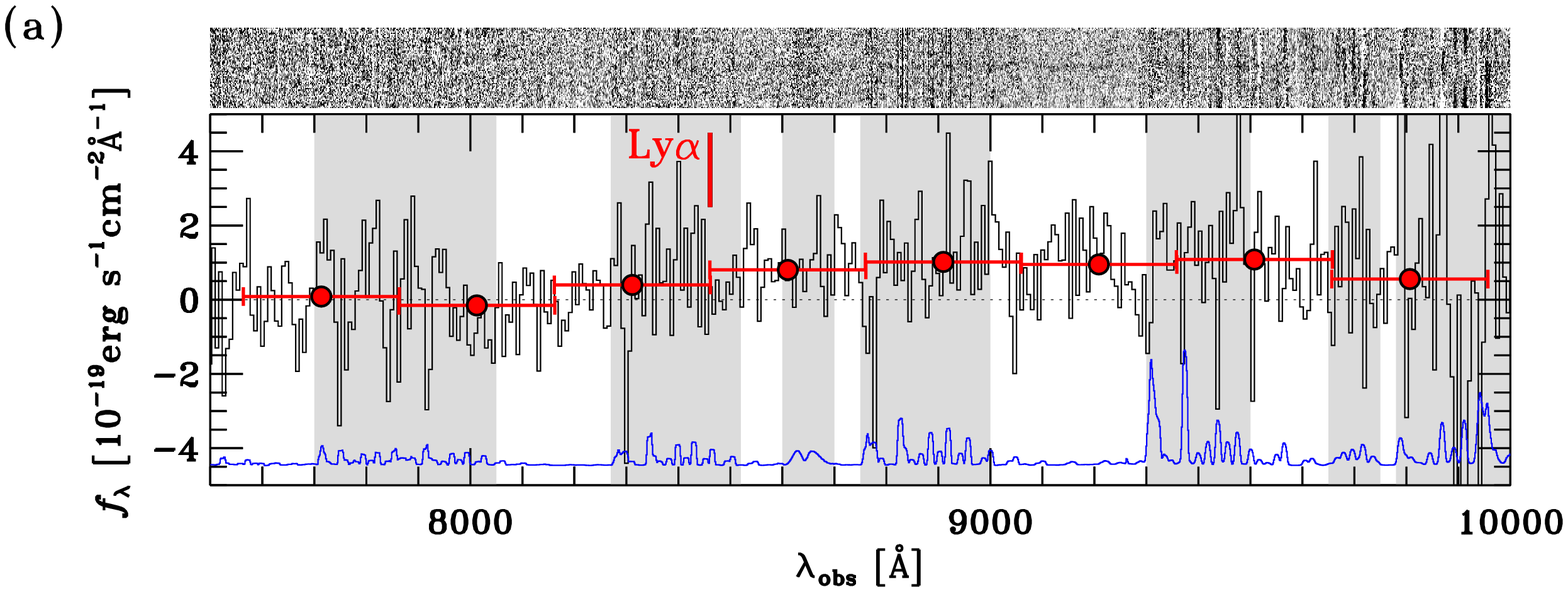} 
  \includegraphics[width=15cm]{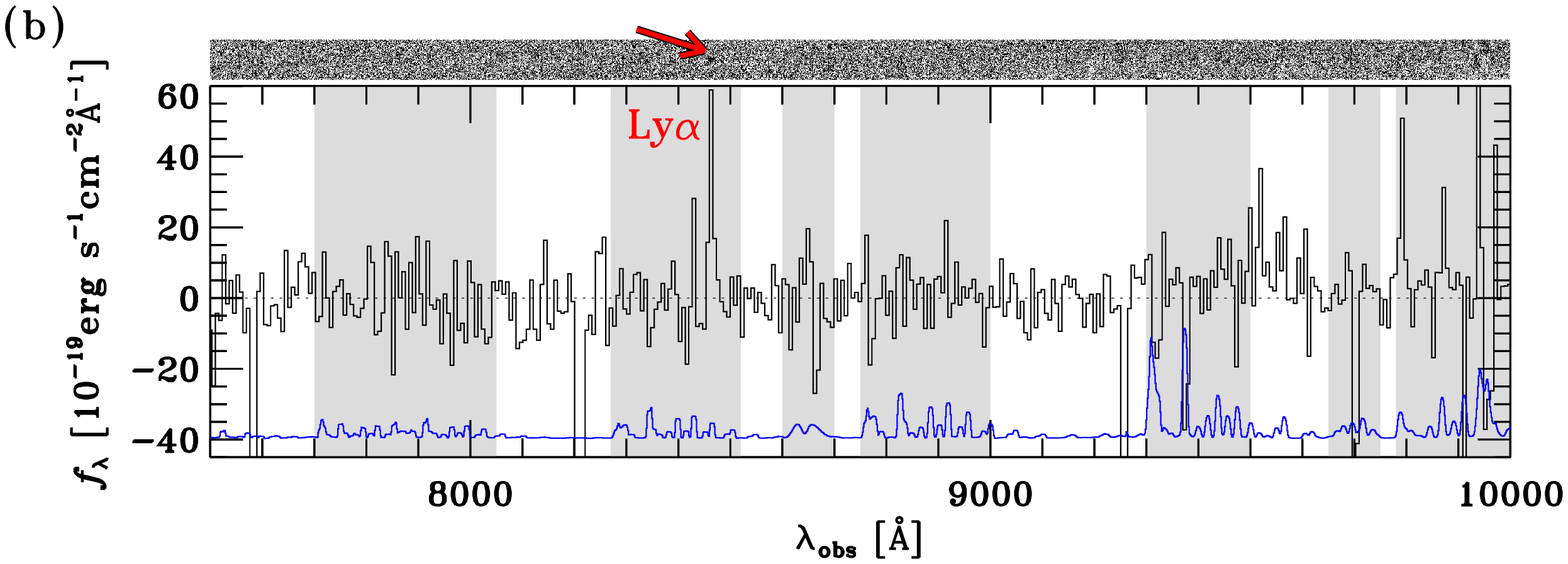} 
  \includegraphics[width=15cm]{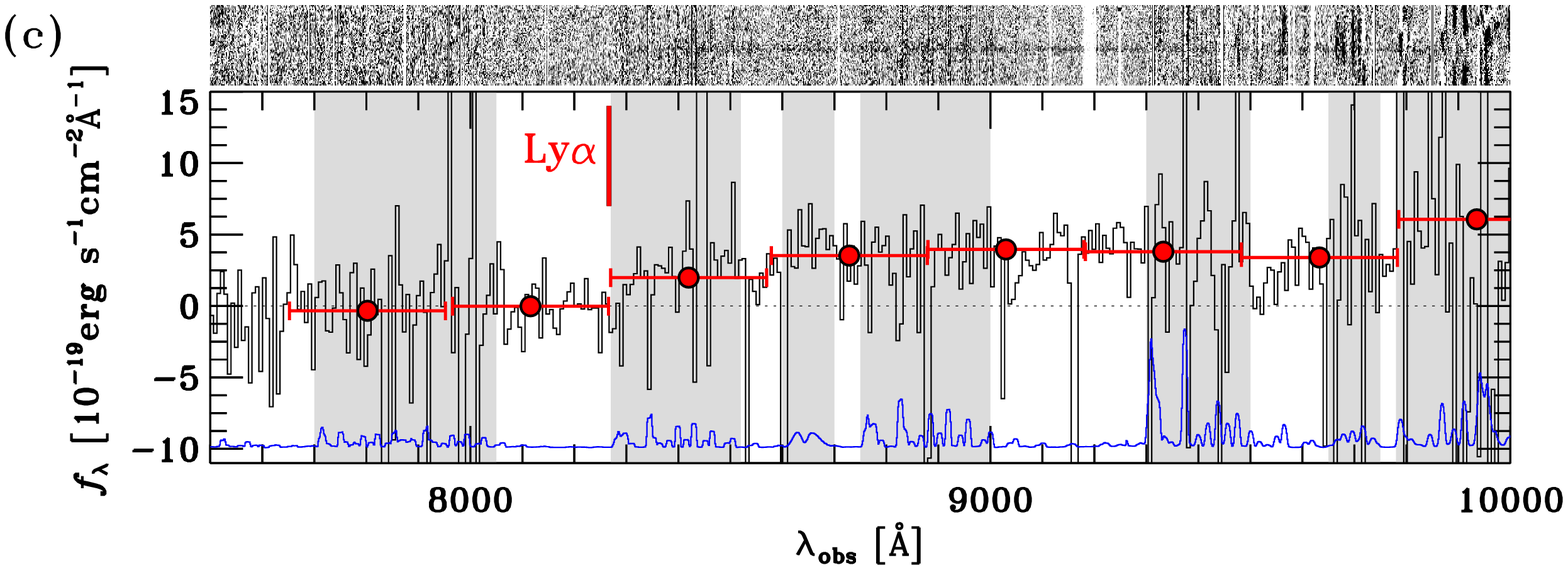} 
  \includegraphics[width=15cm]{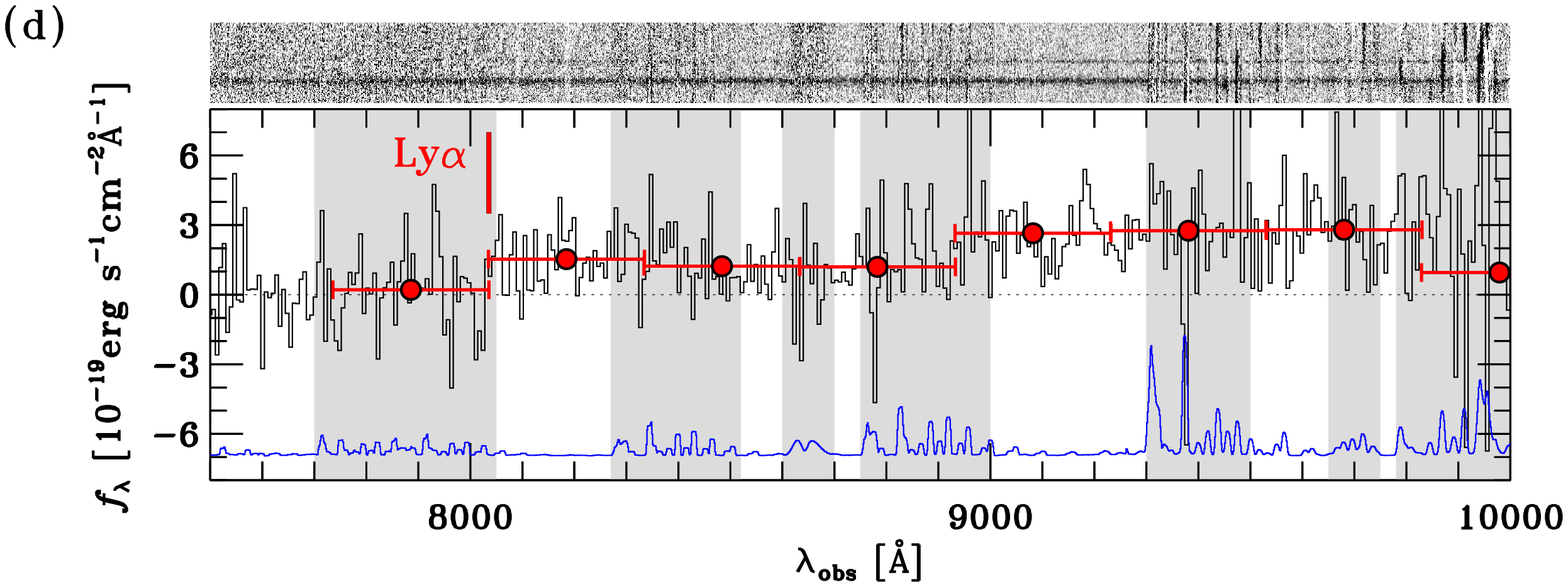} 
 \end{center}
\caption{
Optical spectra of $z \sim 6$ $i$-dropout galaxies: 
(a) HSC J090704+002624, 
(b) HSC J100332+024552, 
(c) HSC J084818+004509, 
and 
(d) HSC J084021+010311.  
In each figure, 
the top panel shows the two-dimensional spectrum (black is positive) 
and the bottom panel shows the one-dimensional spectrum. 
In the top panel, 
our dropout galaxy is located at the center in the spatial direction. 
The size along the spatial axis is 
$12\farcs7$ for (a)--(c) and $16\farcs8$ for (d).  
In the spectrum of (b), 
the Ly$\alpha$ emission line is marked with a red arrow. 
In the bottom panel, 
the object spectrum is shown with black histogram. 
All the spectra are smoothed by $8$--$9$ pixels ($11$--$12$ {\AA}). 
For the sources without Ly$\alpha$ in emission, 
we also plot the averaged spectra over about $300$ {\AA} bins with red filled circles 
and  
mark the wavelength of the Ly$\alpha$ transition with red vertical solid lines. 
The horizontal dotted line corresponds to zero flux density. 
The sky spectrum with an arbitrary normalization 
is plotted in blue (offset from zero). 
}\label{fig:spectra_idrop}
\end{figure*}
%%%%%%%%%%%%%%%%%%%%%%%%%%%%%%%%%%%%%%%

%%%%%%%%%%%%%%%%%%%%%%%%%%%%%%%%%%%%%%
\begin{figure*}
 \begin{center}
  \includegraphics[width=15cm]{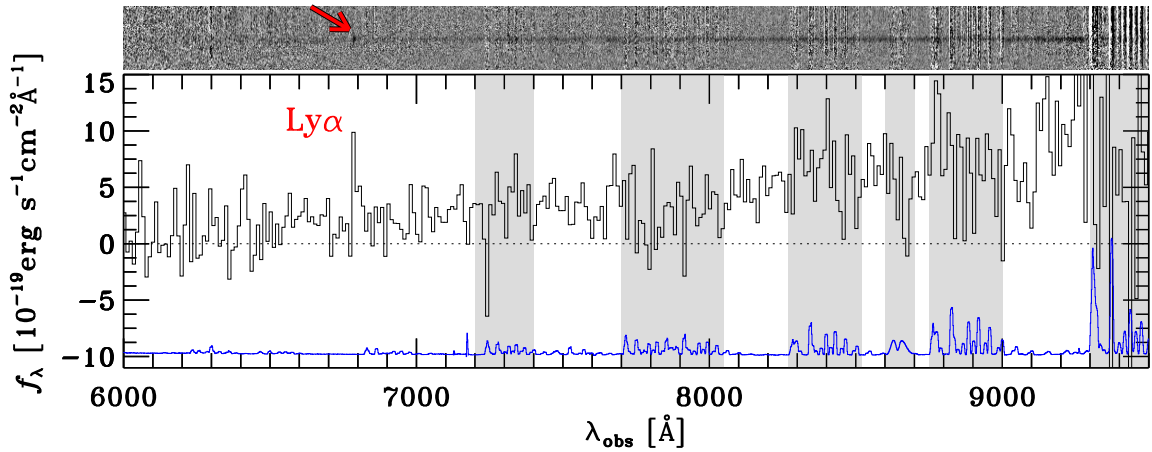} 
 \end{center}
\caption{
Same as Figure \ref{fig:spectra_idrop}, 
but for a $z \sim 5$ $r$-dropout galaxy, HSC J021930--050915.  
In the top panel, 
the size along the spatial axis is $12\farcs7$.  
}\label{fig:spectra_rdrop}
\end{figure*}
%%%%%%%%%%%%%%%%%%%%%%%%%%%%%%%%%%%%%%%

%%%%%%%%%%%%%%%%%%%%%%%%%%%%%%%%%%%%%%
\begin{figure*}
 \begin{center}
  \includegraphics[width=5.6cm]{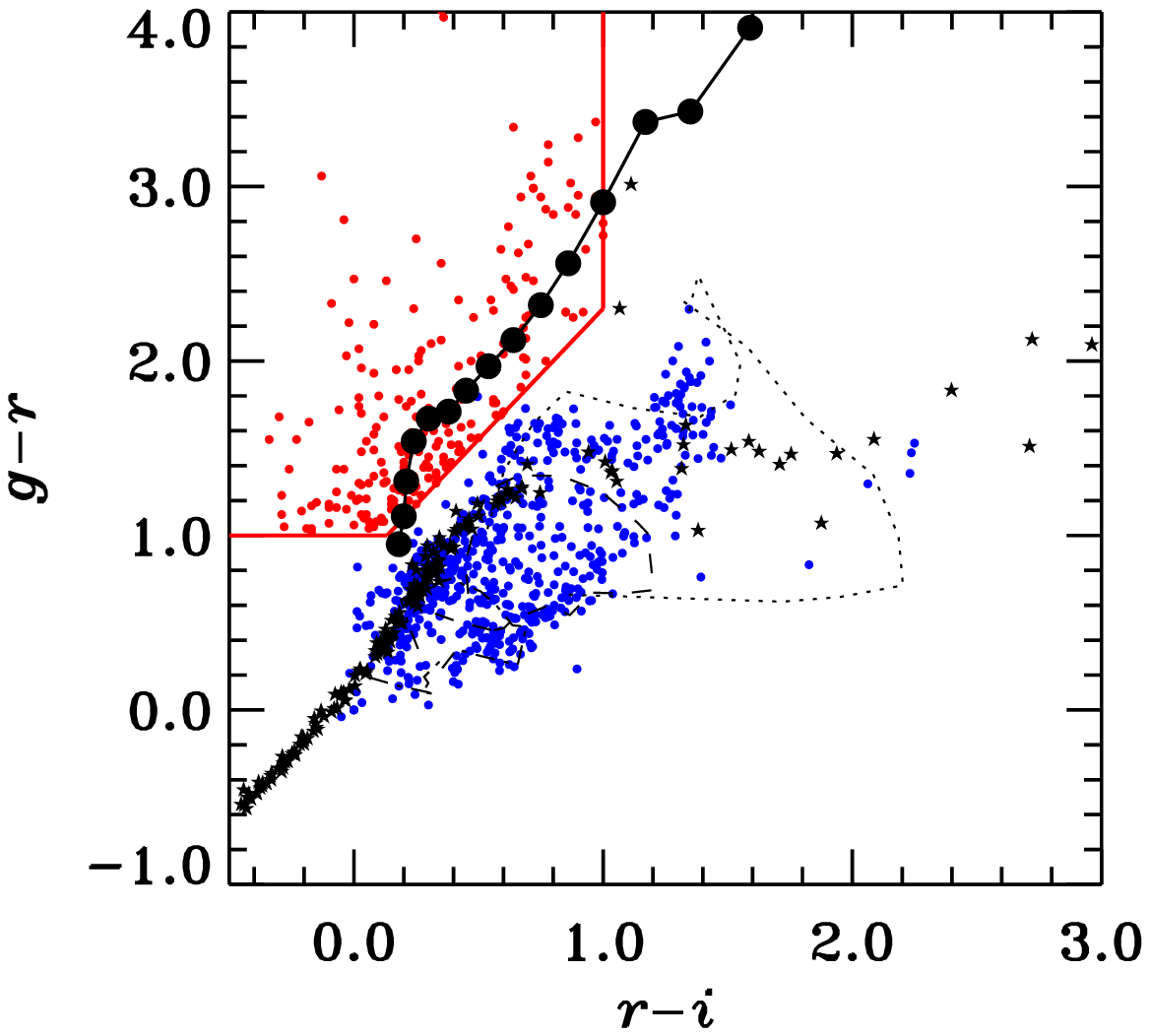} 
  \includegraphics[width=5.6cm]{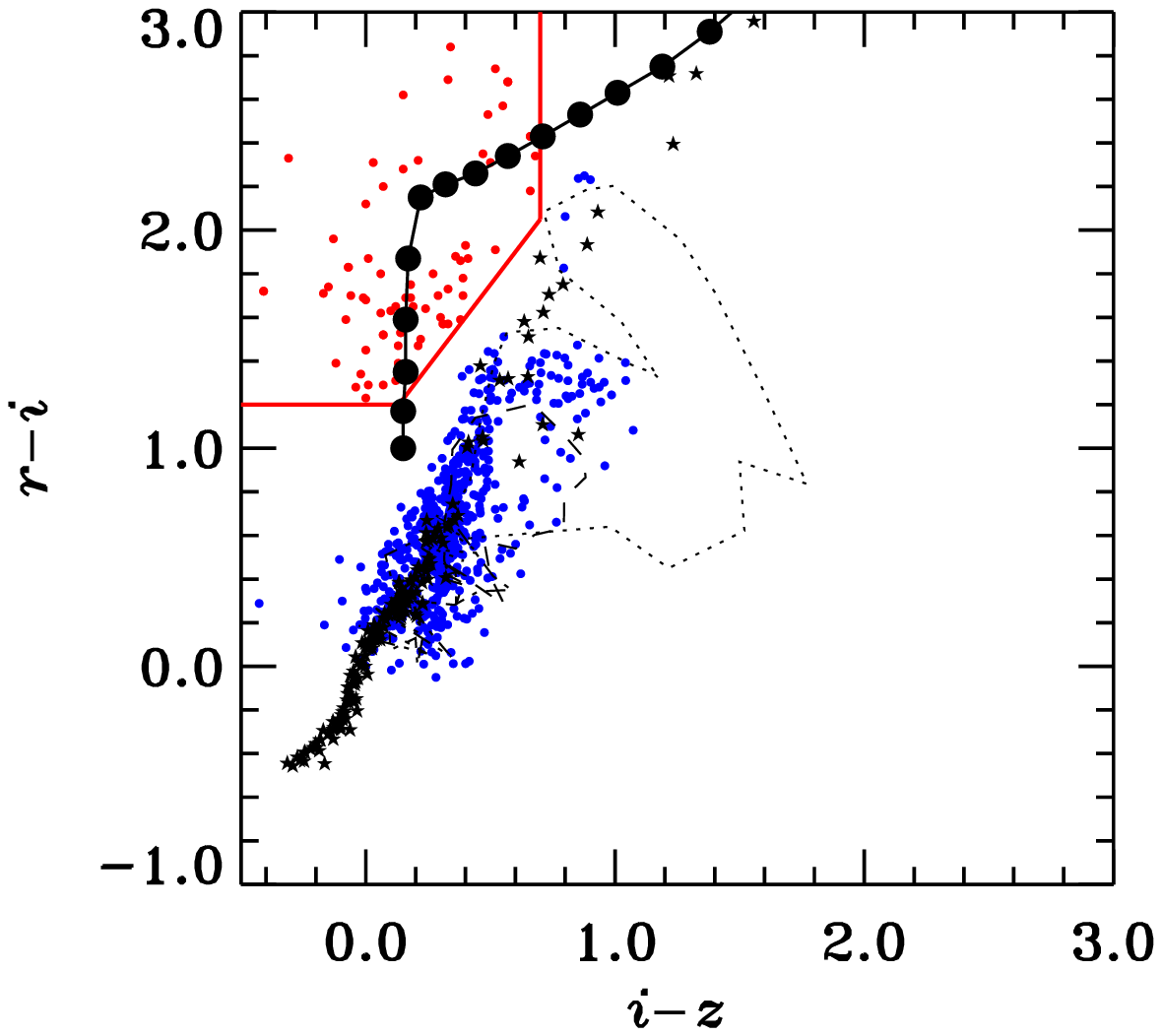} 
  \includegraphics[width=5.6cm]{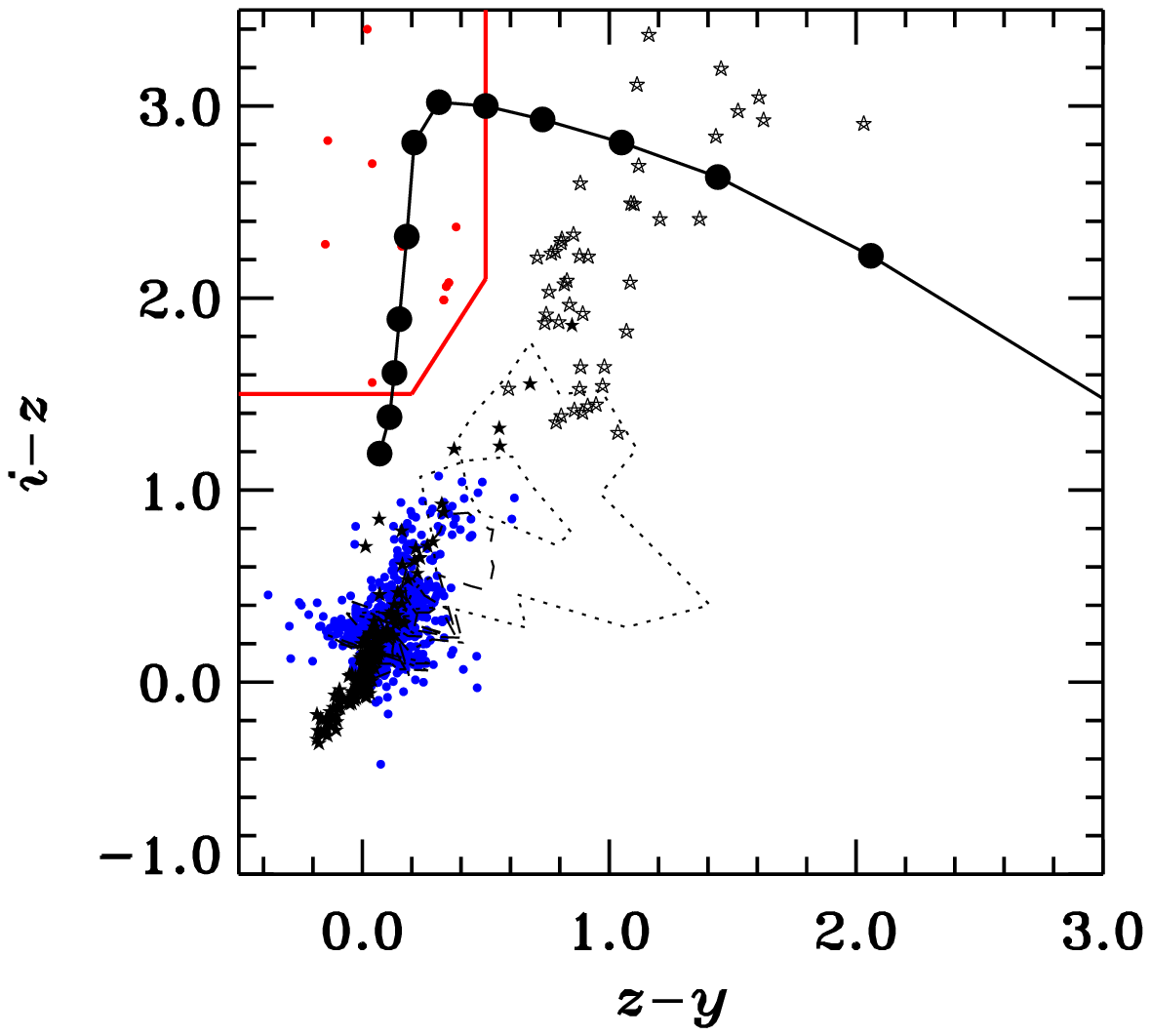} 
 \end{center}
\caption{
\textit{Left}: $g - r$ vs. $r - i$ two color diagram. 
The red circles are the spectroscopically identified galaxies in our $g$-dropout sample, 
and the blue circles are sources in the UD-COSMOS field 
with spectroscopic redshifts of $z = 0-3$ measured by the VVDS survey.
The black solid line indicates the track of young star-forming galaxy spectra 
produced with the \citet{2003MNRAS.344.1000B} model. 
The black filled circles on the black solid line show their redshift 
from $z=3.1$ to $z=4.5$ 
with an interval of $\Delta z = 0.1$. 
The red solid lines show the color selection criteria for our $g$-dropouts. 
The dotted, dashed, and dot-dashed lines are 
typical spectra of elliptical, Sbc, and irregular galaxies 
\citep{1980ApJS...43..393C} redshifted from $z=0$ to $z=2$. 
The filled and open stars indicate 
Galactic stars taken from \citet{1983ApJS...52..121G}
and L /T dwarfs from \citet{2004AJ....127.3553K}. 
\textit{Middle}: $r - i$ vs. $i - z$ two color diagram. 
The red circles are the spectroscopically identified galaxies in our $r$-dropout sample. 
The redshift range of the black filled circles are from $z=4.2$ to $z=5.5$. 
The redshift ranges of the dotted, dashed, and dot-dashed lines are  from $z=0$ to $z=3$.
The other symbols are the same as in the left panel. 
\textit{Right}: $i - z$ vs. $z - y$ two color diagram. 
The red circles are the spectroscopically identified galaxies in our $i$-dropout sample. 
The redshift range of the black filled circles are from $z=5.4$ to $z=6.5$.  
The other symbols are the same as in the middle panel. 
}\label{fig:two_color_diagram}
\end{figure*}
%%%%%%%%%%%%%%%%%%%%%%%%%%%%%%%%%%%%%%%

In our samples, 
five sources are identified 
through our spectroscopic follow-up observations with FOCAS 
(Section \ref{sec:spectroscopic_data}).   
We find the five LBG candidates, 
HSC J090704+002624, 
HSC J100332+024552, 
HSC J084818+004509, 
HSC J084021+010311, 
and HSC J021930--050915, 
are real high-$z$ galaxies at 
$z \simeq 5.96$, 
$z = 5.957$, 
$z \simeq 5.80$, 
$z \simeq 5.61$, 
and $z = 4.580$. 
The first four galaxies are included in our $i$-dropout sample, 
and the last one is in our $r$-dropout sample. 
The first and the last three galaxies were selected for our follow-up targets 
because they are relatively bright among sources in our samples 
that could be targeted during 
our observing runs and had not been spectroscopically observed. 
The second galaxy was a mask filler source 
that was randomly chosen 
from our $i$-dropout candidates within the field-of-view of FOCAS 
centered on a primary target, a bright LAE. 
Figures \ref{fig:spectra_idrop} and \ref {fig:spectra_rdrop} 
show the one-dimensional and two-dimensional spectra 
of the five identified galaxies.
For HSC J100332+024552 and HSC J021930--050915, 
we detect an emission line that shows 
an asymmetric profile with a steeply rising edge at the shorter wavelength of the peak  
and a slowly decaying red tail, which are characteristic features of Ly$\alpha$ 
at high redshift \citep{2006ApJ...648....7K,2006PASJ...58..313S}.\footnote{We confirm that 
no other emission lines are detected in their spectra, 
which excludes the possibilities that 
the detected line is 
a strong emission line at lower $z$, 
i.e., H$\beta$, or [{\sc Oiii}] for HSC J100332+024552, 
and 
H$\beta$, [{\sc Oii}], or [{\sc Oiii}] for HSC J021930--050915.  
In other words, 
the single line detections in our spectra  
cannot completely rule out the possibilities that 
the detected lines are 
H$\alpha$ or [{\sc Oii}] for HSC J100332+024552 
and 
H$\alpha$ for HSC J021930--050915. 
However, 
their asymmetric line profiles suggest that 
the detected line is likely to be redshifted Ly$\alpha$, 
not H$\alpha$ or [{\sc Oii}] (e.g., \cite{2006ApJ...648....7K,2006PASJ...58..313S}). 
}
Their redshifts are determined based on the Ly$\alpha$ emission line. 
For the other three sources,  
their Ly$\alpha$ break feature 
and low-S/N absorption line features in their continua 
are used for their redshift determinations,
although their uncertainties are relatively large.  
Since we have taken only two exposures for HSC J084818+004509 
due to a technical problem in our observations, 
the reduced spectrum is severely affected by cosmic rays. 
This source has also been observed with LDSS3.  
However, the number of exposures with LDSS3 is also only two 
and it is difficult to remove cosmic rays in its reduced spectrum, 
although the Ly$\alpha$ break feature in its continuum is confirmed. 
Note that HSC J084818+004509 has been reported as a $z=5.78$ galaxy 
by the Subaru high-$z$ exploration of low-luminosity quasars (SHELLQs) survey 
\citep{2016ApJ...828...26M}, 
whose redshift determination result is broadly consistent with our result. 
Although these five sources are likely to be high-$z$ galaxies because of these observational results, 
it should be noted that 
it is difficult to completely rule out the possibilities that 
they are foreground sources such as Galactic brown dwarfs based on these low-S/N spectra. 
The nature of these sources will be checked by future follow-up observations.

%%%%%%%%%%%%%%%%%%%%%%%%%%%%%%%%%%%%%%
\begin{figure}
 \begin{center}
  \includegraphics[width=7cm]{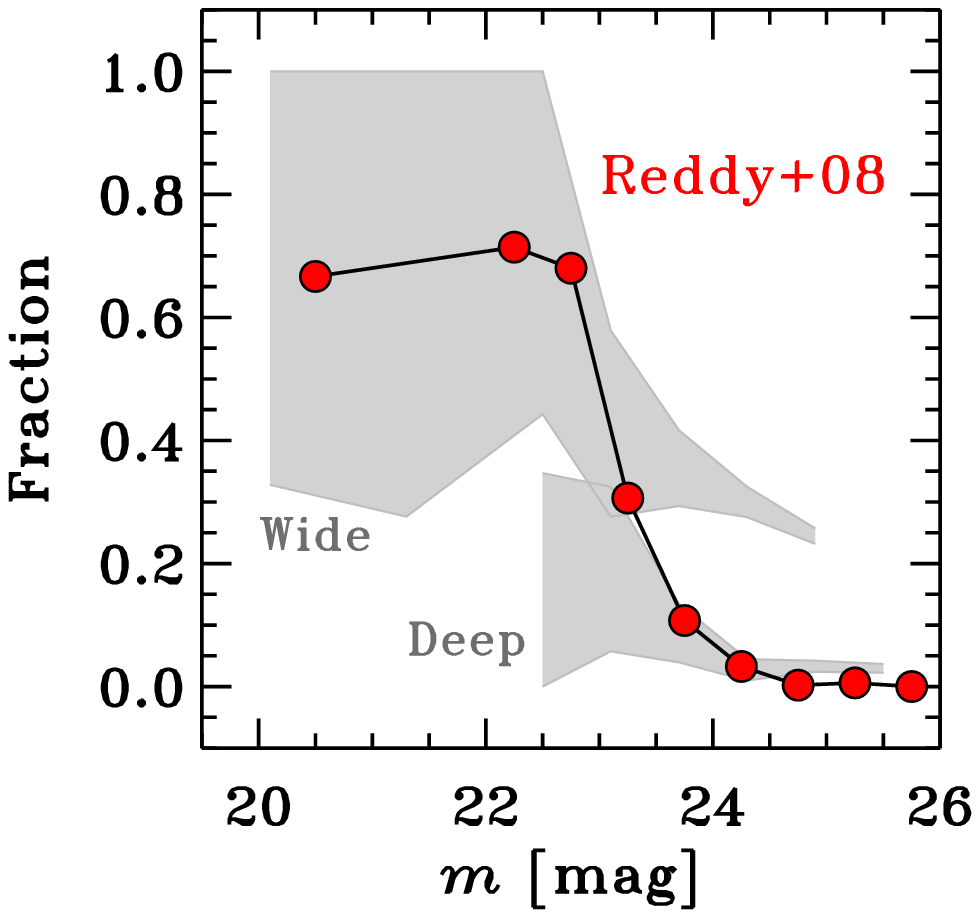} 
  \includegraphics[width=7cm]{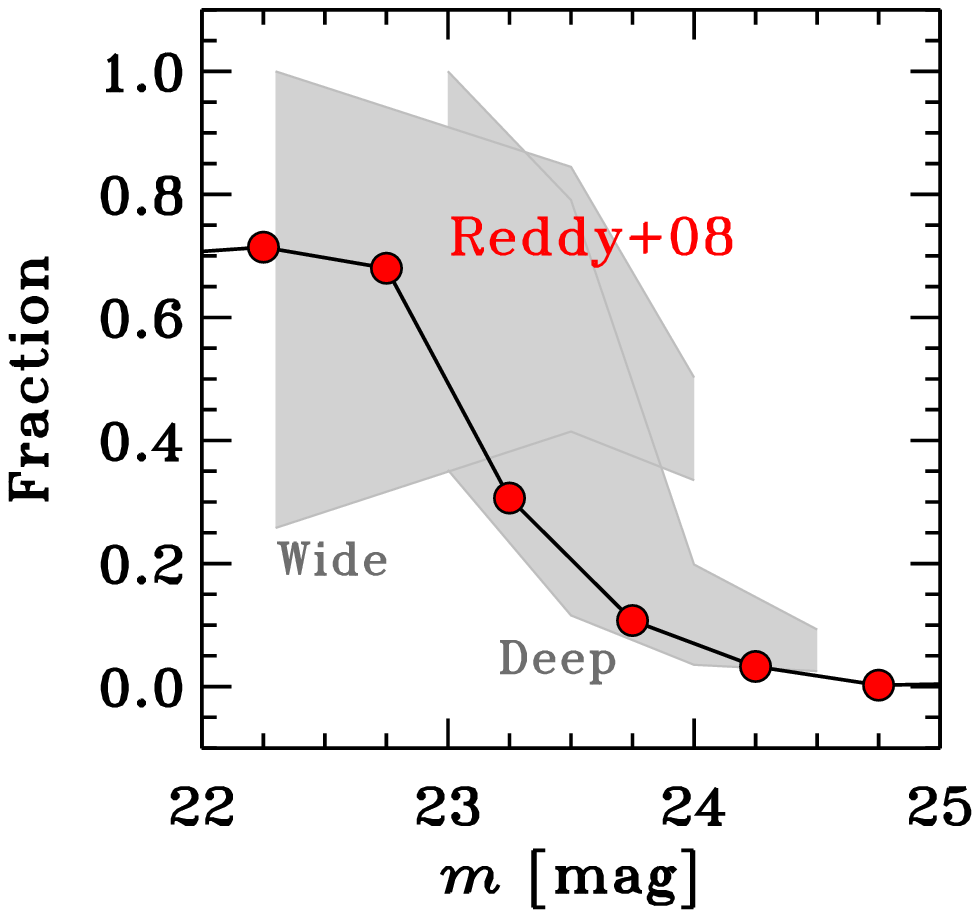} 
 \end{center}
\caption{
Estimated contamination fractions 
as a function of apparent magnitude 
for our $g$-dropout (top) and $r$-dropout (bottom) samples. 
The gray shaded regions represent  
the $1\sigma$ uncertainties of the estimated contamination fractions 
for the Wide and Deep layers. 
For the apparent magnitude, 
$i$- and $z$-band magnitudes are used 
for $g$-dropouts and $r$-dropouts, respectively. 
The red filled circles are the results of \citet{2008ApJS..175...48R} 
for their LBG sample. 
}\label{fig:contami_fraction}
\end{figure}
%%%%%%%%%%%%%%%%%%%%%%%%%%%%%%%%%%%%%%%

%%%%%%%%%%%%%%%%%%%%%%%%%%%%%%%%%%%%%%
\begin{table}
  \tbl{
  Estimated contamination fractions 
  for the $z \sim 4$ and $z \sim 5$ samples 
  selected from the W- and D-layer data. 
  }{%
  \begin{tabular}{cccc}
      \hline
  magnitude & fraction  & magnitude & fraction \\
      \hline
  \multicolumn{2}{c}{$z \sim 4$ in D} & \multicolumn{2}{c}{$z \sim 5$ in D} \\
22.5 & $0.08^{+0.27}_{-0.08}$ & 23.0 & $0.94^{+0.06}_{-0.59}$\\
23.1 & $0.16^{+0.17}_{-0.10}$ & 23.5 & $0.34^{+0.46}_{-0.22}$\\
23.7 & $0.08^{+0.05}_{-0.04}$ & 24.0 & $0.10^{+0.10}_{-0.06}$\\
24.3 & $0.03^{+0.02}_{-0.02}$ & 24.5 & $0.06^{+0.04}_{-0.03}$\\
24.9 & $0.03^{+0.01}_{-0.01}$ & --- & --- \\
25.5 & $0.03^{+0.01}_{-0.01}$ & --- & --- \\
  \hline
  \multicolumn{2}{c}{$z \sim 4$ in W} & \multicolumn{2}{c}{$z \sim 5$ in W} \\
20.1 & $0.91^{+0.09}_{-0.59}$ & 22.3 & $0.73^{+0.27}_{-0.47}$\\
21.3 & $0.79^{+0.21}_{-0.51}$ & 23.5 & $0.59^{+0.25}_{-0.18}$\\ 
22.5 & $0.68^{+0.32}_{-0.23}$ & 24.0 & $0.41^{+0.09}_{-0.08}$\\
23.1 & $0.41^{+0.17}_{-0.13}$ & --- & --- \\
23.7 & $0.35^{+0.07}_{-0.06}$ & --- & --- \\
24.3 & $0.30^{+0.03}_{-0.02}$ & --- & --- \\
24.9 & $0.24^{+0.01}_{-0.01}$ & --- & --- \\
 \hline
    \end{tabular}}\label{tab:contaminant_fraction}
%\begin{tabnote}
%hogehoge
%\end{tabnote}
\end{table}
%%%%%%%%%%%%%%%%%%%%%%%%%%%%%%%%%%%%%%

In addition, 
we incorporate the results of our spectroscopic observations for high-$z$ galaxies 
with Magellan/IMACS (Section \ref{sec:spectroscopic_data}). 
We also check the spectroscopic catalogs shown in other studies 
(\cite{2008ApJ...675.1076S}; \cite{2008ApJS..176..301O}; \cite{2010AJ....140..546W}; 
\cite{2012MNRAS.422.1425C}; \cite{2012ApJ...755..169M}; \cite{2012ApJ...760..128M}; 
\cite{2013AJ....145....4W}; \cite{2013A&A...559A..14L}; \cite{2015ApJ...798...28K}; 
\cite{2015ApJS..218...15K}\footnote{
We use the MOSDEF Spectroscopic redshift catalog 
that was released on 2016 August 16.}; 
\cite{2016ApJ...819...24W}; \cite{2016ApJ...826..114T}; 
\cite{2016ApJS..225...27M};
\cite{2016ApJ...828...26M}; \cite{2017A&A...597A..79P}; \cite{2017A&A...600A.110T}; 
\cite{2017AJ....153..184Y}; 
\cite{2017ApJ...841..111M}; 
\cite{2017arXiv170405854M}; \cite{2017arXiv170500733S}; 
R. Higuchi et al. in preparation; 
see also \cite{2016ApJS..227...11B}). 
We adopt their classifications between galaxies and AGNs in their catalogs.  
For the catalogs of 
the VIMOS VLT Deep Survey (VVDS; \cite{2013A&A...559A..14L}) 
and 
the VIMOS Ultra Deep Survey (VUDS; \cite{2017A&A...600A.110T}), 
we take into account sources 
whose redshifts are $>70-75${\%} correct, 
i.e., sources with redshift reliability flags of 2, 3, 4, 9, 12, 13, 14, and 19.  
Here we focus on sources with spectroscopic redshifts 
$z_{\rm spec} > 3$ in these catalogs. 
Our contamination estimates with sources at $z_{\rm spec} < 3$ 
are presented in the next section.

In total, 
358 dropouts in our sample have been spectroscopically identified 
by our observations and the other studies. 
Among these identified sources, 
270 sources are found to be galaxies at $z_{\rm spec}>3$, 
and the other 
88 sources are AGNs. 
These sources are listed in Table \ref{tab:spectroscopy_identifications}.

Figure \ref{fig:two_color_diagram} shows the distributions of 
the spectroscopically identified galaxies at $z_{\rm spec} > 3$ in our dropout samples 
in the two-color diagrams. 
We also plot sources in the UD-COSMOS field 
with spectroscopic redshifts of $z_{\rm spec} < 3$ 
that are measured by the VVDS. 
In addition, 
the tracks of model spectra of young star-forming galaxies 
that are produced with the stellar population synthesis code {\sc galaxev} 
\citep{2003MNRAS.344.1000B} are shown. 
As model parameters 
a Salpeter initial mass function \citep{1955ApJ...121..161S}, 
an age of $70$ Myr after the initial star formation, 
and metallicity of $Z / Z_\odot = 0.2$ are adopted. 
We use the \citet{2000ApJ...533..682C} dust extinction formula 
with reddening of $E(B-V) = 0.16$. 
The IGM absorption is considered 
following the prescription of \citet{1995ApJ...441...18M}. 
The colors of the spectroscopically identified galaxies are 
broadly consistent with those expected from the model spectra.

%%%%%%%%%%%%%%%%%%%%%%%%%%%%%%%%%%%%%%
\begin{figure*}
 \begin{center}
  \includegraphics[width=11cm]{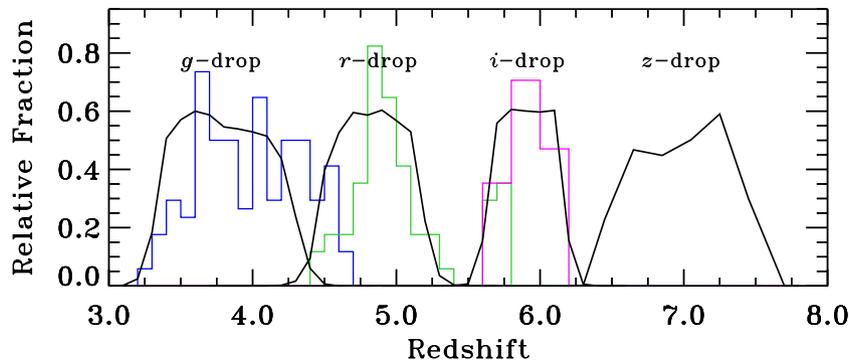} 
 \end{center}
\caption{Selection completeness estimates for our 
$z \sim 4$, $z \sim 5$, $ z \sim 6$, and $z \sim 7$ samples 
The black curves correspond to 
the results of our Monte Carlo calculations 
averaged over the W, D, and  UD layers 
described in Section \ref{sec:selection_completeness}. 
The average redshifts of these samples are roughly $3.8$, $4.9$, $5.9$, and $6.9$. 
The blue, green, and magenta histograms are 
the redshift distributions of the spectroscopically identified galaxies  
in $g$-, $r$-, and $i$-dropout samples, respectively. 
}\label{fig:completeness}
\end{figure*}
%%%%%%%%%%%%%%%%%%%%%%%%%%%%%%%%%%%%%%%

%%%%%%%%%%%%%%%%%%%%%%%%%%%%%%%%%%%%%%%%%%%%%%%%%%%%%%%%%%%%%%%%%
\subsection{Contamination} \label{sec:contaminations}
%%%%%%%%%%%%%%%%%%%%%%%%%%%%%%%%%%%%%%%%%%%%%%%%%%%%%%%%%%%%%%%%%

Some foreground objects such as red galaxies at intermediate redshifts  
can satisfy our color criteria by photometric errors, 
although intrinsically they do not enter the color selection window. 
To estimate the numbers of such contaminants in our dropout samples, 
we use shallower HSC data of COSMOS that are created 
with a subset of the real HSC data for the UD-COSMOS field. 
We use two shallower data sets 
whose depths are comparable with those in the W layer and D layer. 
We assume that 
the UD-COSMOS data are sufficiently deep 
and the contamination rates in our dropout selections for the UD-COSMOS are small. 
First, we select objects which 
do not satisfy our selection criteria from the UD-COSMOS catalog. 
We then regard them as foreground interlopers in the W-layer-depth and D-layer-depth COSMOS samples 
if they satisfy our selection criteria for the W-layer and D-layer dropouts, respectively,  
and calculate their number counts.  
Based on comparisons between the surface number densities of interlopers 
and those of the selected dropouts, 
we estimate 
the fractions of foreground interlopers, 
which are shown in Figure \ref{fig:contami_fraction} 
and Table \ref{tab:contaminant_fraction}. 
The fractions of foreground interlopers at magnitude fainter than $24.0$ mag 
are estimated to be 
less than about $10${\%} for the D-layer samples 
and 
less than about $30-40${\%} for the W-layer samples. 
At the brighter magnitude bins, 
our dropout samples in the wide and deep layers 
are more contaminated by the foreground interlopers. 
Note that similar results have been obtained by \citet{2008ApJS..175...48R}.
We subtract the number counts of foreground interlopers  
from the number counts of our dropouts 
and consider both of the uncertainties 
in Section \ref{subsec:luminosity_function}. 
For a sanity check, 
we derive the interloper fraction in the W-layer samples 
by using the spec-$z$ catalog of VVDS, 
which covers a small portion of our W-layer fields. 
Although the number of objects which are included 
in both our samples and the VVDS catalog is small, 
the interloper fraction for the $z \sim 4$ W-layer sample is estimated to be about 40{\%}, 
which is consistent with the results estimated from the shallower HSC data.

For the $z \sim 6-7$ dropout samples, 
we cannot estimate the surface number densities of interlopers 
by adopting this method, 
since the number densities of such sources 
in the shallower depth COSMOS field data 
are too low. 
Instead, 
we make use of the spectroscopic observation results 
taken by our study as well as in the literature.
Based on the spectroscopic redshift catalog 
created in Section \ref{sec:source_selection}, 
31 sources in our $z \sim 6$ dropout sample 
are spectroscopically identified 
in our follow-up observations and in the other studies 
and all the sources are at $z > 5.5$. 
Although 
it is unclear whether the other candidates are real high-$z$ sources or foreground interlopers, 
we assume that the contamination fraction of interlopers is negligibly small 
based on the limited spectroscopy results. 
For $z\sim 7$ dropout sample, 
none of our candidates 
have been followed up with spectroscopy.
We will carry out follow-up spectroscopy for our $z$-dropout candidates 
in the near future.

It should be noted that 
our sample is contaminated 
not only by low-$z$ interlopers but also by high-$z$ AGNs. 
We take into account the AGN contamination 
in our samples in Section \ref{sec:results_and_discussion}.

%%%%%%%%%%%%%%%%%%%%%%%%%%%%%%%%%%%%%%%%%%%%%%%%%%%%%%%%%%%%%%%%%
\subsection{Selection Completeness} \label{sec:selection_completeness}
%%%%%%%%%%%%%%%%%%%%%%%%%%%%%%%%%%%%%%%%%%%%%%%%%%%%%%%%%%%%%%%%%

We estimate the selection completeness of our dropout galaxies 
by running a suite of Monte Carlo simulations 
with an input mock catalog of high-$z$ galaxies. 
In the mock catalog, 
the size distribution of galaxies follows 
recent results of galaxy 
log-normal size distributions and 
size-luminosity relations as a function of redshift 
based on \textit{Hubble} legacy data sets 
(\cite{2015ApJS..219...15S}; 
see also \cite{2010ApJ...709L..21O}; \cite{2012ApJ...756L..12M}; 
\cite{2013ApJ...777..155O}; \cite{2015ApJ...804..103K}; 
\cite{2016MNRAS.457..440C}; \cite{2016A&A...593A..22R}). 
The Sersic index $n$ is fixed at $n = 1.5$, 
which is also suggested from the results of   
\citet{2015ApJS..219...15S}. 
A uniform distribution of the intrinsic ellipticities 
in the range of $0.0$--$0.8$ is assumed, 
since the observational results of $z \sim 3-5$ dropout galaxies 
have roughly uniform distributions \citep{2006ApJ...652..963R}. 
Position angles are randomly chosen.  
To produce galaxy SEDs, 
we use the stellar population synthesis model of 
{\sc galaxev} \citep{2003MNRAS.344.1000B}. 
We adopt the Salpeter initial mass function \citep{1955ApJ...121..161S}
with lower and upper mass cutoffs of $0.1 M_\odot$ and $100 M_\odot$, 
a constant rate of star formation, 
age of $25$ Myr,  
metallicity of $Z/Z_\odot = 0.2$, 
and 
\citet{2000ApJ...533..682C} dust extinction ranging from 
$E(B-V) = 0.0$--$0.4$ 
so that we can cover from very blue continua with $\beta \simeq -3.0$ 
to moderately red ones with $\beta \simeq -1.0$. 
The IGM absorption is taken into account 
by using the prescription of \citet{1995ApJ...441...18M}.

Different simulations are carried out for the W, D, and UD layers 
by using the SynPipe software 
\citep{2017arXiv170501599H,murata2017}, 
which utilizes GalSim v1.4 \citep{2015A&C....10..121R} and the HSC pipeline.   
We insert large numbers of artificial sources 
into HSC images of individual CCDs at the single exposure level.  
Next we stack the single exposure images 
and create source catalogs 
in the same manner as the real ones. 
We then select high-$z$ galaxy candidates with the same selection criteria  
and calculate the selection completeness as a function of magnitude and redshift, $C(m,z)$, 
averaged over UV slope $\beta$ weighted with 
the $\beta$ distribution of \citet{2014ApJ...793..115B}. 
For the $\beta$ distribution of very bright sources at $M_{\rm UV} \lesssim -22$ mag 
where \citet{2014ApJ...793..115B} do not probe,  
we extrapolate their results for fainter magnitudes.

Figure \ref{fig:completeness} shows 
the results of our selection completeness estimates as a function of redshift.  
The average redshift values are roughly 
$z \sim 3.8$ for $g$-dropouts, 
$z \sim 4.9$ for $r$-dropouts, 
$z \sim 5.9$  for $i$-dropouts, 
and  
$z \sim 6.9$  for $z$-dropouts. 
In Figure \ref{fig:completeness}, 
we also show the redshift distributions of the spectroscopically identified galaxies 
in our samples (Section \ref{sec:source_selection}). 
The redshift distributions of the spectroscopically identified galaxies  
are broadly consistent with 
the results of our selection completeness simulations, 
although the distributions of the spectroscopically identified galaxies 
in the $g$- and $r$-dropout samples 
appear to be shifted toward slightly higher redshift. 
This is probably because 
the spectroscopically identified galaxies are biased to 
ones with strong Ly$\alpha$ emission. 
In particular, 
the redshift distribution of the spectroscopically identified $r$-dropouts 
has a secondary peak at around $z=5.7$, 
which is caused by $z=5.7$ Ly$\alpha$ emitters 
found by Subaru Suprime-Cam and HSC narrowband surveys in the literature.

%%%%%%%%%%%%%%%%%%%%%%%%%%%%%%%%%%%%%%%%%%%%%%%%%%%%%%%%%%%%%%%%%
%%%%%%%%%%%%%%%%%%%%%%%%%%%%%%%%%%%%%%%%%%%%%%%%%%%%%%%%%%%%%%%%%
\section{Results and Discussion} \label{sec:results_and_discussion}
%%%%%%%%%%%%%%%%%%%%%%%%%%%%%%%%%%%%%%%%%%%%%%%%%%%%%%%%%%%%%%%%%
%%%%%%%%%%%%%%%%%%%%%%%%%%%%%%%%%%%%%%%%%%%%%%%%%%%%%%%%%%%%%%%%%

%%%%%%%%%%%%%%%%%%%%%%%%%%%%%%%%%%%%%%%%%%%%%%%%%%%%%%%%%%%%%%%%%
\subsection{The UV Luminosity Functions} \label{subsec:luminosity_function}
%%%%%%%%%%%%%%%%%%%%%%%%%%%%%%%%%%%%%%%%%%%%%%%%%%%%%%%%%%%%%%%%%

%%%%%%%%%%%%%%%%%%%%%%%%%%%%%%%%%%%%%%%
\begin{table}
  \tbl{
  Estimated galaxy UV LFs at $z \sim 4$, $z \sim 5$, $z \sim 6$, and $z \sim 7$ based on the HSC SSP data. 
  }{%
  \begin{tabular}{cccc}
      \hline
  $M_{\rm UV}$ & $\Phi$  & $M_{\rm UV}$ & $\Phi$ \\
  (mag) & ($10^{-4}$ mag$^{-1}$ Mpc$^{-3}$) & (mag) & ($10^{-4}$ mag$^{-1}$ Mpc$^{-3}$)  \\
      \hline
  \multicolumn{2}{c}{$z \sim 4$} & \multicolumn{2}{c}{$z \sim 6$} \\
$-23.837$ & $0.00055^{+0.00156}_{-0.00055}$ & $-24.244$ & $0.00001^{+0.00019}_{-0.00001}$ \\
$-23.637$ & $0.00113^{+0.00205}_{-0.00103}$ & $-23.944$ & $0.00002^{+0.00019}_{-0.00002}$ \\
$-23.437$ & $0.00325^{+0.00380}_{-0.00197}$ & $-23.644$ & $0.00007^{+0.00026}_{-0.00007}$ \\
$-23.237$ & $0.00630^{+0.00474}_{-0.00286}$ & $-23.344$ & $0.00016^{+0.00025}_{-0.00011}$ \\
$-23.037$ & $0.00885^{+0.00490}_{-0.00333}$ & $-23.044$ & $0.00147^{+0.00107}_{-0.00066}$ \\
$-22.837$ & $0.02171^{+0.00827}_{-0.00702}$ & $-22.744$ & $0.00621^{+0.00292}_{-0.00231}$ \\
$-22.637$ & $0.04771^{+0.01277}_{-0.01343}$ & $-22.444$ & $0.02553^{+0.00360}_{-0.00243}$ \\
$-22.437$ & $0.10648^{+0.02046}_{-0.02616}$ & $-22.144$ & $0.05615^{+0.01242}_{-0.00777}$ \\
$-22.237$ & $0.19365^{+0.02885}_{-0.04072}$ & $-21.844$ & $0.21313^{+0.05322}_{-0.04056}$ \\
$-22.037$ & $0.37561^{+0.04174}_{-0.06703}$ & $-21.544$ & $0.49479^{+0.11521}_{-0.08826}$ \\
$-21.837$ & $0.72645^{+0.05566}_{-0.10819}$ & --- & --- \\
$-21.637$ & $1.30260^{+0.05961}_{-0.15856}$ & --- & --- \\
$-21.437$ & $2.27743^{+0.05760}_{-0.23933}$ & --- & --- \\
$-21.237$ & $3.71089^{+0.07428}_{-0.40147}$ & --- & --- \\
$-21.037$ & $5.60696^{+0.08491}_{-0.62452}$ & --- & --- \\
$-20.837$ & $7.96770^{+0.10421}_{-0.91524}$ & --- & --- \\
$-20.637$ & $10.05840^{+0.10550}_{-1.18857}$ & --- & --- \\
$-20.437$ & $12.74950^{+0.25563}_{-0.24866}$ & --- & --- \\
$-20.237$ & $16.72980^{+0.30374}_{-0.29621}$ & --- & --- \\
$-20.037$ & $23.61950^{+0.41061}_{-0.40085}$ & --- & --- \\
$-19.837$ & $29.82960^{+0.57881}_{-0.56363}$ & --- & --- \\  
\hline
  \multicolumn{2}{c}{$z \sim 5$} &   \multicolumn{2}{c}{$z \sim 7$} \\ 
$-24.241$ & $0.00003^{+0.00013}_{-0.00003}$ & $-24.165$ & $0.00001^{+0.00019}_{-0.00001}$ \\
$-23.491$ & $0.00012^{+0.00190}_{-0.00012}$ & $-23.665$ & $0.00010^{+0.00039}_{-0.00009}$ \\
$-22.991$ & $0.00397^{+0.00422}_{-0.00258}$ & $-23.165$ & $0.00091^{+0.00080}_{-0.00044}$ \\
$-22.741$ & $0.02156^{+0.00813}_{-0.00692}$ & --- & --- \\
$-22.491$ & $0.06576^{+0.01501}_{-0.01736}$ & --- & --- \\
$-22.241$ & $0.16906^{+0.02681}_{-0.03629}$ & --- & --- \\
$-21.991$ & $0.40066^{+0.04266}_{-0.06929}$ & --- & --- \\
$-21.741$ & $0.81996^{+0.05653}_{-0.11351}$ & --- & --- \\
$-21.491$ & $1.44029^{+0.09020}_{-0.16802}$ & --- & --- \\
$-21.241$ & $2.50578^{+0.12027}_{-0.28970}$ & --- & --- \\
$-20.991$ & $3.91303^{+0.15802}_{-0.46063}$ & --- & --- \\
  \hline
\end{tabular}}\label{tab:UVLF}
%\begin{tabnote}
%This is table note.
%\end{tabnote}
\end{table}
%%%%%%%%%%%%%%%%%%%%%%%%%%%%%%%%%%%%%%%

We derive the rest-frame UV luminosity functions of $z\sim4-7$ galaxies 
by applying the effective volume method \citep{1999ApJ...519....1S}. 
Based on the results of the selection completeness simulations, 
we estimate the effective survey volume per unit area 
as a function of apparent magnitude, 
\begin{equation}
V_{\rm eff} (m) 
= \int C(m,z) \frac{dV(z)}{dz} dz, 
\end{equation}
where $C(m,z)$ is the selection completeness estimated in Section \ref{sec:selection_completeness}, 
i.e., the probability that a galaxy 
with apparent magnitude $m$ at redshift $z$ is detected and 
satisfies the selection criteria, 
and 
$dV(z)/dz$ is the differential comoving volume as a function of redshift 
(e.g., \cite{1999astro.ph..5116H}).

The space number densities of dropouts that are corrected for incompleteness and contamination effects 
are obtained by calculating 
\begin{equation}
\psi(m) 
= \frac{n_{\rm raw}(m) - n_{\rm con}(m)}{V_{\rm eff}(m)}, 
\end{equation}
where $n_{\rm raw}(m)$ is the surface number density of selected dropouts in an apparent magnitude bin of $m$, 
and $n_{\rm con}(m)$ is the surface number density of interlopers in the magnitude bin 
estimated in Section \ref{sec:contaminations}. 
To calculate the surface number densities, 
we use the effective area values summarized in Table \ref{tab:HSCdata}. 
The $1\sigma$ uncertainties are calculated 
by taking account of Poisson confidence limits \citep{1986ApJ...303..336G} 
on the numbers of the sources. 
To calculate the $1\sigma$ uncertainties of the space number densities of dropouts, 
we consider the uncertainties of the surface number densities of selected dropouts 
and those of interlopers.  
We restrict our analysis 
for the $z \sim 4-5$ D- and W-layer samples 
to the magnitude ranges where the contamination rate estimates are available. 
Note that the $z\sim 4$ UD-layer sample 
includes several very bright candidates with magnitude brighter than $22.0$ mag. 
However, 
three of them have been spectroscopically observed 
and all of the three are at $z_{\rm spec} < 1$ \citep{2009ApJS..184..218L}, 
while many fainter sources have been identified at $z_{\rm spec} > 3$ 
as checked in Section \ref{sec:source_selection}. 
Although the number of observed very bright sources is small,   
we do not use dropout candidates with magnitude brighter than $22.0$ mag 
in the $z\sim 4$ UD-layer sample.

We convert the number densities of dropouts as a function of apparent magnitude, $\psi(m)$, 
into the UV LFs, $\Phi[M_{\rm UV}(m)]$, i.e., the number densities of dropouts  
as a function of rest-frame UV absolute magnitude.
We calculate the absolute UV magnitudes of dropouts from their apparent magnitudes 
using their average redshifts $\bar z$:   
\begin{equation}
M_{\rm UV} 
= m + 2.5 \log(1+\bar z) - 5 \log \left( \frac{d_{\rm L}(\bar z) }{ 10 \, {\rm pc}} \right) + (m_{\rm UV} - m),  
\end{equation}
where $d_{\rm L}$ is the luminosity distance in units of parsecs 
and 
$(m_{\rm UV} - m)$ 
is the $K$-correction term 
between the magnitude at rest-frame UV 
and the magnitude in the bandpass that we use. 
We set the $K$-correction term to be $0$ 
by assuming that dropout galaxies have flat UV continua, i.e., constant $f_\nu$ in the rest-frame UV 
(e.g., Figure 3 of \cite{2006ApJ...642..653S} and 
Figure 7 of \cite{2010A&A...523A..74V}). 
For the apparent magnitude $m$, 
we use $i$-band magnitudes for $g$-dropouts, 
$z$-band magnitudes for $r$- and $i$-dropouts, 
and 
$y$-band magnitudes for $z$-dropouts. 
The central wavelength of the $i$-band 
corresponds to $\sim 1600$ {\AA} in the rest-frame of $g$-dropouts, 
and that of the $z$-band is $\sim 1300-1500$ {\AA} 
in the rest-frame of $r$- and $i$-dropouts, on average.   
Note that the $y$-band probes slightly shorter wavelength 
in the rest-frame of $z$-dropouts, 
about $1230$ {\AA}.

The top panel of Figure \ref{fig:UVLF_wAGN} 
shows our derived LF for dropouts at $z \sim 4$ 
and those taken from the previous galaxy work of 
\citet{2015ApJ...803...34B} 
and \citet{2015ApJ...810...71F}, 
which are based on the \textit{Hubble} legacy survey data, 
and that of \citet{2010A&A...523A..74V}, 
which is based on the CFHT deep legacy survey data. 
The previous studies have derived their UV LF estimates 
in the UV magnitude range of $M_{\rm UV} > - 23$ mag. 
Our results are broadly consistent with the previous results 
in this magnitude range. 
However, at $M_{\rm UV} < -23$ mag, 
where no previous high-$z$ galaxy studies have probed, 
our results appear to have a hump 
and follow a shallower slope than 
the extrapolation of the exponential cutoff from the fainter bins. 
Figure \ref{fig:UVLF_wAGN} also shows  
our LF results for the $z \sim 5$ dropout sample 
and the results of the previous galaxy studies. 
We find that the situation is similar to that for the $z \sim 4$ dropout sample. 
In Figure \ref{fig:UVLF_wAGN2}, 
we present the results of our LF estimates for the $z \sim 6-7$ dropout samples. 
For $z \sim 6$, we also plot the previous results 
taken from \citet{2015ApJ...803...34B}, 
\citet{2015ApJ...810...71F}, 
and \citet{2015MNRAS.452.1817B}. 
For $z \sim 7$, the previous estimates by  
\citet{2013MNRAS.432.2696M}, 
\citet{2013ApJ...768..196S}, 
\citet{2015ApJ...803...34B}, 
\citet{2015ApJ...810...71F}, 
\citet{2017MNRAS.466.3612B}, and \citet{2017arXiv170204867I} 
are shown for comparison. 
These previous work has presented their estimates 
in the magnitude range of $M_{\rm UV} > - 22.5$ ($-23.0$) mag at $z \sim 6$ ($z \sim 7$).  
Our results are in good agreement with the previous results in these magnitude ranges. 
However, at the brighter magnitude ranges, 
our LF results seem to have a hump 
compared to the simple extrapolation of the exponentially declining shape. 
Note that the effect of the Eddington bias \citep{1913MNRAS..73..359E}, 
which can cause an apparent increase of the number of bright sources 
due to photometric scatter from sources in fainter bins, 
should be small at these bright-end hump features. 
This is because their magnitude ranges are much brighter than 
the limiting magnitudes of the samples.

To investigate the bright-end hump features, 
we plot the UV LFs of AGNs taken from the literature in Figure \ref{fig:UVLF_wAGN}. 
We find that the bright-end hump features in our LF results for dropouts 
are broadly consistent with the UV LFs of AGNs  
obtained by \citet{2011ApJ...728L..26G}. 
Our LF results are also consistent with 
those of \citet{2017arXiv170405996A} at the very bright end of $M_{\rm UV} \sim -25$ mag, 
but our results are larger at $M_{\rm UV} \sim -23$ mag 
than their results as well as those of \citet{2016ApJ...832..208N}. 
This is probably because 
they focus on $z\sim4$ quasars with stellar morphology 
while the selection of ours and \citet{2011ApJ...728L..26G} 
can also identify galaxies with faint AGNs whose morphology is extended 
(see also \citet{2017arXiv170405996A}). 
In Figure \ref{fig:UVLF_wAGN}, 
we also compare our bright-end LF results with those of AGNs at $z \sim 5$ 
obtained by \citet{2012ApJ...756..160I}, \citet{2013ApJ...768..105M}, and \citet{2016ApJ...832..208N}. 
Although the uncertainties of our estimates are large, 
our results are in agreement with these AGN results. 
In addition, Figure \ref{fig:UVLF_wAGN2} shows that 
our bright-end LF results for dropouts are broadly consistent with 
those of AGNs at $z \sim 6$ taken from 
\citet{2010AJ....139..906W}, \citet{2015ApJ...798...28K}, and \citet{2016ApJ...833..222J}.

In our dropout selection, 
we probe redshifted Ly$\alpha$ break features of high-$z$ galaxies. 
However, 
high-$z$ AGNs also have similar Ly$\alpha$ break features. 
It is thus expected that our dropout sample is contaminated by AGNs  
(e.g., for $i$-dropout selection, see Figure 1 of \cite{2016ApJ...828...26M}). 
Actually, as described in Section \ref{sec:source_selection}, 
our dropout samples include spectroscopically confirmed AGNs. 
Based on our spectroscopy results as well as those in the literature,    
we derive the galaxy fraction of 
spectroscopically confirmed dropouts, 
i.e., 
the number of spectroscopically confirmed high-$z$ galaxies 
divided by the sum of the numbers of 
spectroscopically confirmed high-$z$ galaxies and AGNs, 
in our $z \sim 4-6$ samples in each magnitude bin 
(Figures \ref{fig:UVLF_wAGN} and \ref{fig:UVLF_wAGN2}). 
As shown in Figure \ref{fig:UVLF_wAGN}, 
the $z \sim 4$ galaxy fraction is smaller than $20${\%} at $M_{\rm UV} < -23$ mag, 
but it increases with increasing magnitude 
and it reaches about $100${\%} at $M_{\rm UV} > -22$ mag. 
Similarly, 
in Figure \ref{fig:UVLF_wAGN2}, 
the galaxy fraction for the $z \sim 6$ sample is less (more) than 
$50${\%} at $M_{\rm UV} < -23$ mag ($M_{\rm UV} > -23$ mag). 
These results suggest that 
our bright-end LF estimates are significantly contaminated by AGNs. 
The very wide area of the HSC SSP allows us to 
bridge the UV LFs of high-$z$ galaxies and AGNs, 
both of which can be selected with redshifted Ly$\alpha$ break features.  
Note that we also show the results of the faint end of the AGN UV LFs 
(\cite{2015A&A...578A..83G}; 
\cite{2017arXiv170407750P})  
in the magnitude range of $M_{\rm UV} \gtrsim -22$ mag 
in Figures \ref{fig:UVLF_wAGN} and \ref{fig:UVLF_wAGN2}. 
We find that our results are much larger than their results, 
which also suggests that the AGN contamination is not significant 
in this faint magnitude range.

Because it is not easy to distinguish galaxies from AGNs in our dropout samples 
solely based on the ground-based optical imaging data, 
we first investigate the shape of the UV LFs of dropouts 
by focusing on the magnitude range where the galaxy fraction is large. 
Figure \ref{fig:UVLF_selected} shows the UV LFs of dropouts at $z \sim 4-7$ 
based on our Subaru HSC results, 
previous \textit{Hubble} results (\cite{2015ApJ...803...34B}; \cite{2017arXiv170204867I}), 
and other ground-based telescope results \citep{2017MNRAS.466.3612B}.  
The combination of our results with the previous work  
reveals the shapes of the UV LFs for high-$z$ dropout sources 
in a very wide magnitude range of $-26 \lesssim M_{\rm UV} \lesssim -14$ mag 
for the first time. 
Our wide area survey reveals that 
the UV LFs of dropouts have bright end humps 
that are related to the significant contribution of light from AGNs. 
To characterize the UV LFs of dropout galaxies, 
we focus on the LF estimates at $M_{\rm UV} > -23$ mag, 
where the galaxy fraction is significantly large. 
We fit a Schechter function \citep{1976ApJ...203..297S} to the data points, 
\begin{equation}
\phi(L) dL 
	= \phi^\ast \left( \frac{L}{L^\ast} \right)^\alpha 
		\exp \left( - \frac{L}{L^\ast} \right) d \left( \frac{L}{L^\ast} \right), 
\end{equation} 
where 
$\phi^\ast$ is the overall normalization, 
$L^\ast$ is the characteristic luminosity, 
and 
$\alpha$ is the faint-end slope. 
We define a Schechter function expressed in terms of absolute magnitude 
$\Phi(M_{\rm UV})$ as $\phi(L) dL = \Phi(M_{\rm UV}) dM_{\rm UV}$, i.e., 
\begin{eqnarray}
\Phi(M_{\rm UV}) 
	&=& \frac{\ln 10}{2.5} \phi^\ast 10^{-0.4 (M_{\rm UV} - M_{\rm UV}^\ast) (\alpha +1)} \nonumber \\
	&& \times \exp \left( - 10^{-0.4 (M_{\rm UV} - M_{\rm UV}^\ast)} \right), 
\label{eq:schechter}
\end{eqnarray}
where 
$M^\ast_{\rm UV}$ is the characteristic magnitude.  
We fit this function to the observed LFs derived from the results 
of our observations and the previous \textit{Hubble} results of 
\citet{2015ApJ...803...34B} and \citet{2017arXiv170204867I}. 
Varying the three parameters, 
we search for the best-fit set of ($\phi^\ast$, $M^\ast_{\rm UV}$, $\alpha$) 
that minimizes $\chi^2$. 
The best-fit parameters are 
summarized in Table \ref{tab:LF_best_fit_parameters_wAGN}  
and the best-fit Schechter function is plotted in Figure \ref{fig:UVLF_selected}.

Figure \ref{fig:UVLF_selected_all} summarizes 
the UV LF estimates at $z \sim 4-7$ and their best-fit Schechter functions.  
In Figure \ref{fig:Schechter_fit}, 
we show the $1\sigma$ and $2\sigma$ confidence intervals 
for the combinations of the Schechter parameters. 
We find that 
$M_{\rm UV}^\ast$ shows little evolution 
while 
the other two parameters decrease with increasing redshift 
as already pointed out in the previous work 
(e.g., \cite{2015ApJ...803...34B}; \cite{2015MNRAS.452.1817B}; 
\cite{2015ApJ...810...71F}).

Note that 
there are on-going projects in our HSC SSP collaboration 
to search for high-$z$ quasars by using selection techniques 
that are optimized for quasars. 
The exact shapes of the quasar UV LFs at $z \sim 4$, $z \sim 5$, and $z \sim 6-7$ 
are presented in \citet{2017arXiv170405996A}, M. Niida et al. in preparation, 
and Y. Matsuoka et al. in preparation, respectively.

%%%%%%%%%%%%%%%%%%%%%%%%%%%%%%%%%%%%%%%%%%%%%%%%%%%%%%%%%%%%%%%%%
\subsection{The Galaxy UV Luminosity Functions} 
%%%%%%%%%%%%%%%%%%%%%%%%%%%%%%%%%%%%%%%%%%%%%%%%%%%%%%%%%%%%%%%%%

\begin{figure*}
 \begin{center}
  \includegraphics[width=16cm]{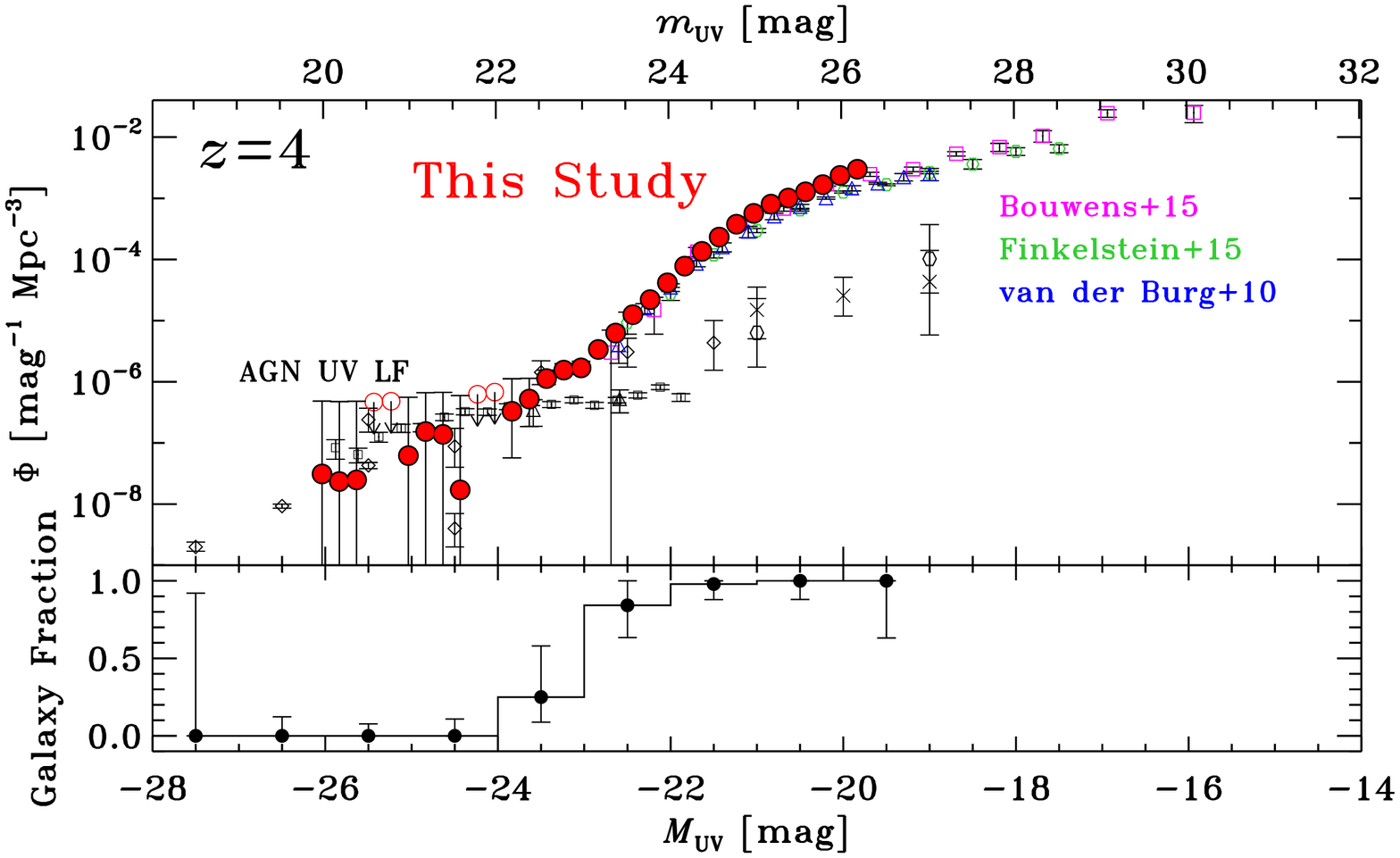} 
  \includegraphics[width=16cm]{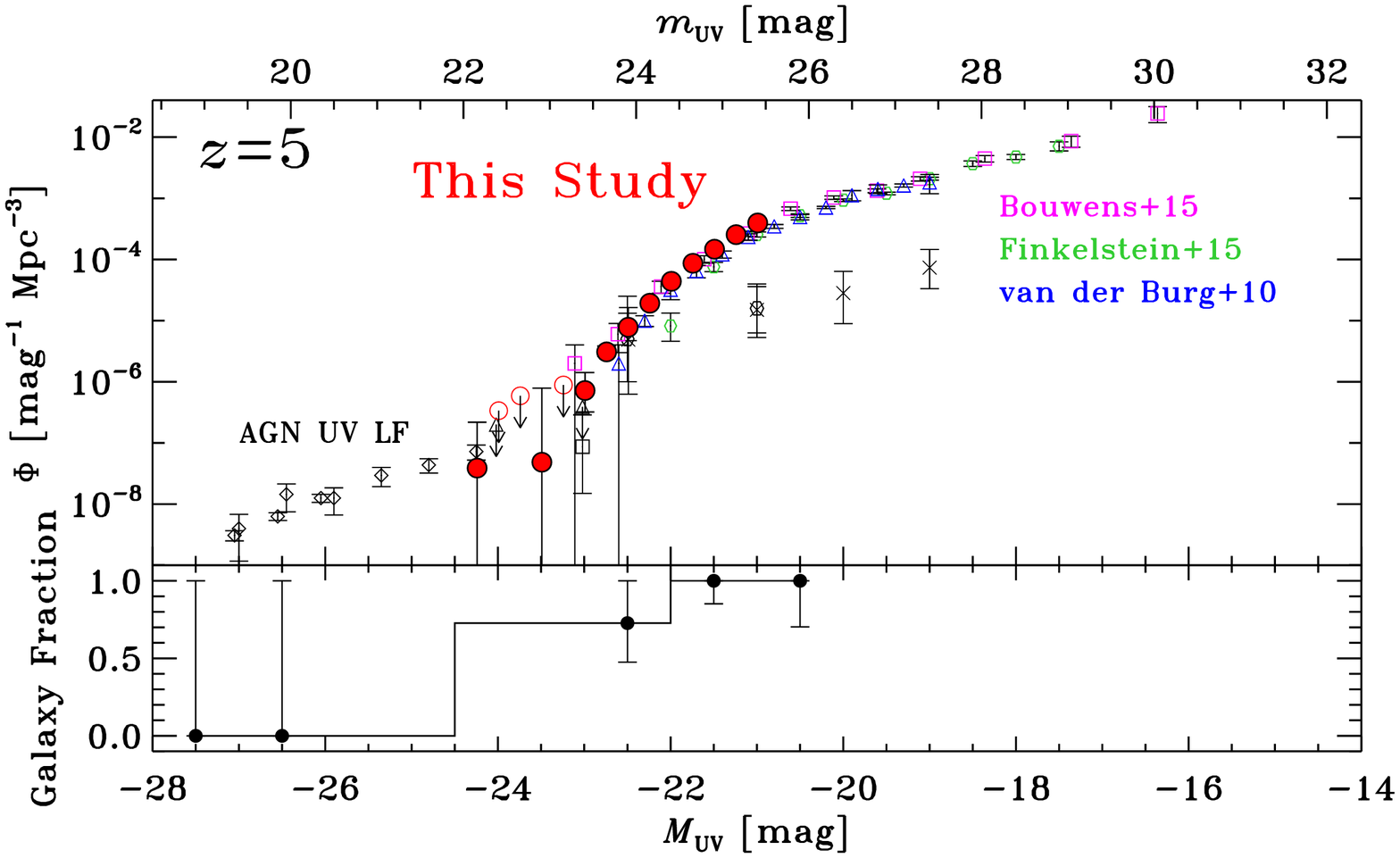} 
 \end{center}
\caption{
\textit{Top}: 
rest-frame UV luminosity functions of dropouts before quasar contamination correction 
at $z \sim 4$ and $z \sim 5$. 
The red circles show our results based on the HSC SSP survey data. 
The red open circles with a downward arrow 
denote the $1\sigma$ upper limits. 
For comparison, 
we also show previous results for galaxies taken from  
\authorcite{2015ApJ...803...34B} (\yearcite{2015ApJ...803...34B}; open magenta squares) at $z \sim 4-5$, 
\authorcite{2015ApJ...810...71F} (\yearcite{2015ApJ...810...71F}; open green hexagons) at $z \sim 4-5$, 
and 
\authorcite{2010A&A...523A..74V} (\yearcite{2010A&A...523A..74V}; open blue triangles) at $z \sim 4-5$. 
Our LF estimates are broadly in agreement with previous results 
in the magnitude range where previous results are available. 
In addition, 
we plot previous results for quasars taken from 
\authorcite{2011ApJ...728L..26G} (\yearcite{2011ApJ...728L..26G}; open black diamonds) at $z \sim 4$, 
\authorcite{2015A&A...578A..83G} (\yearcite{2015A&A...578A..83G}; black crosses) at $z \sim 4-5$, 
\authorcite{2017arXiv170407750P} (\yearcite{2017arXiv170407750P}; open black hexagons) at $z \sim 4-5$, 
\authorcite{2016ApJ...832..208N} (\yearcite{2016ApJ...832..208N}; open black triangle) at $z \sim 4-5$, 
\authorcite{2017arXiv170405996A} (\yearcite{2017arXiv170405996A}; open black squares) at $z=4$, 
\authorcite{2013ApJ...768..105M} (\yearcite{2013ApJ...768..105M}; open black diamonds) at $z \sim 5$,  
and 
\authorcite{2012ApJ...756..160I} (\yearcite{2012ApJ...756..160I}; open black squares) at $z \sim 5$. 
\textit{Bottom}: 
fraction of galaxies in our $z \sim 4-5$ dropout samples 
based on spectroscopy results. 
For the denominator of the fraction, 
the sum of the numbers of galaxies and quasars is used. 
}\label{fig:UVLF_wAGN}
\end{figure*}

\begin{figure*}
 \begin{center}
  \includegraphics[width=16cm]{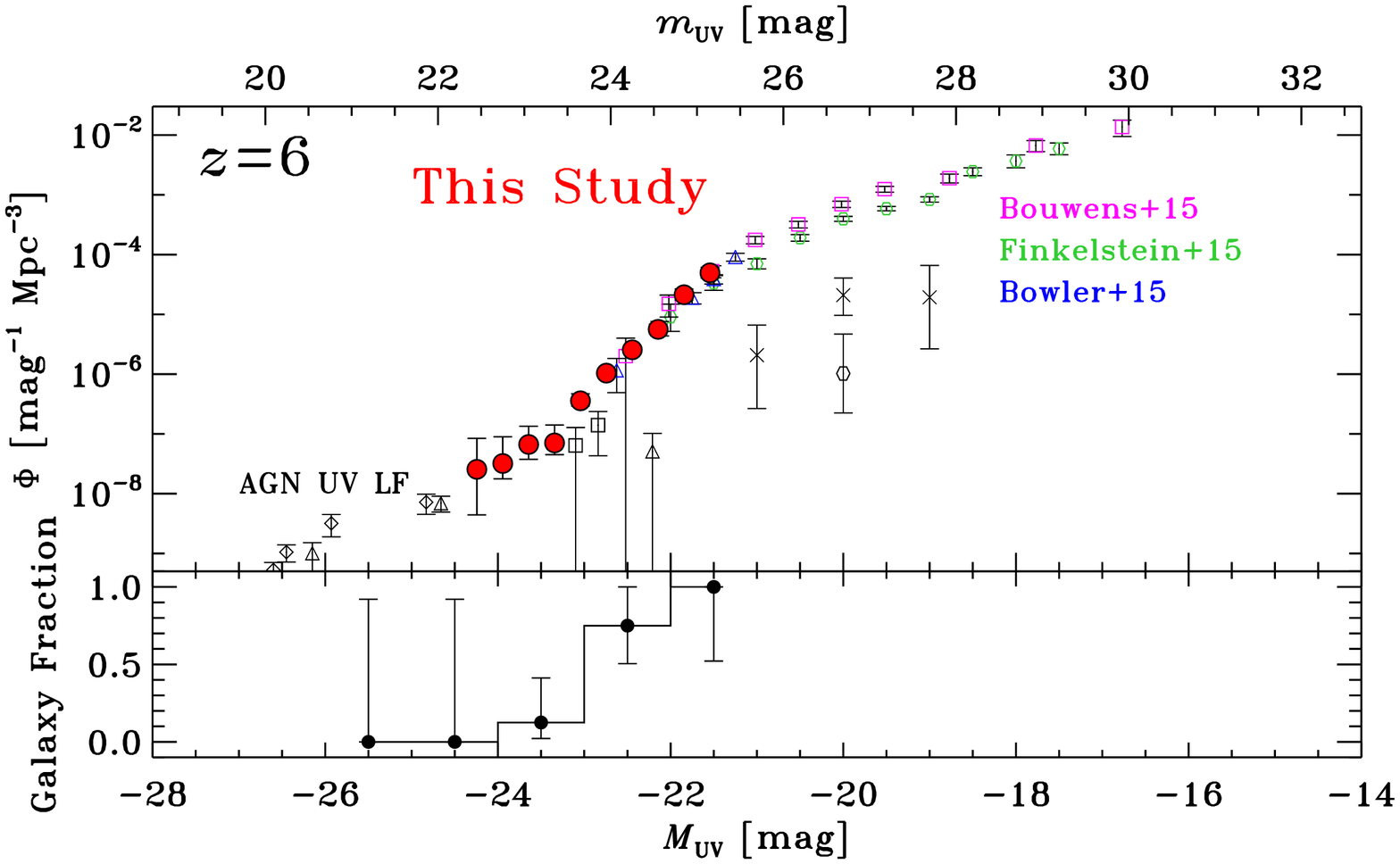} 
  \includegraphics[width=16cm]{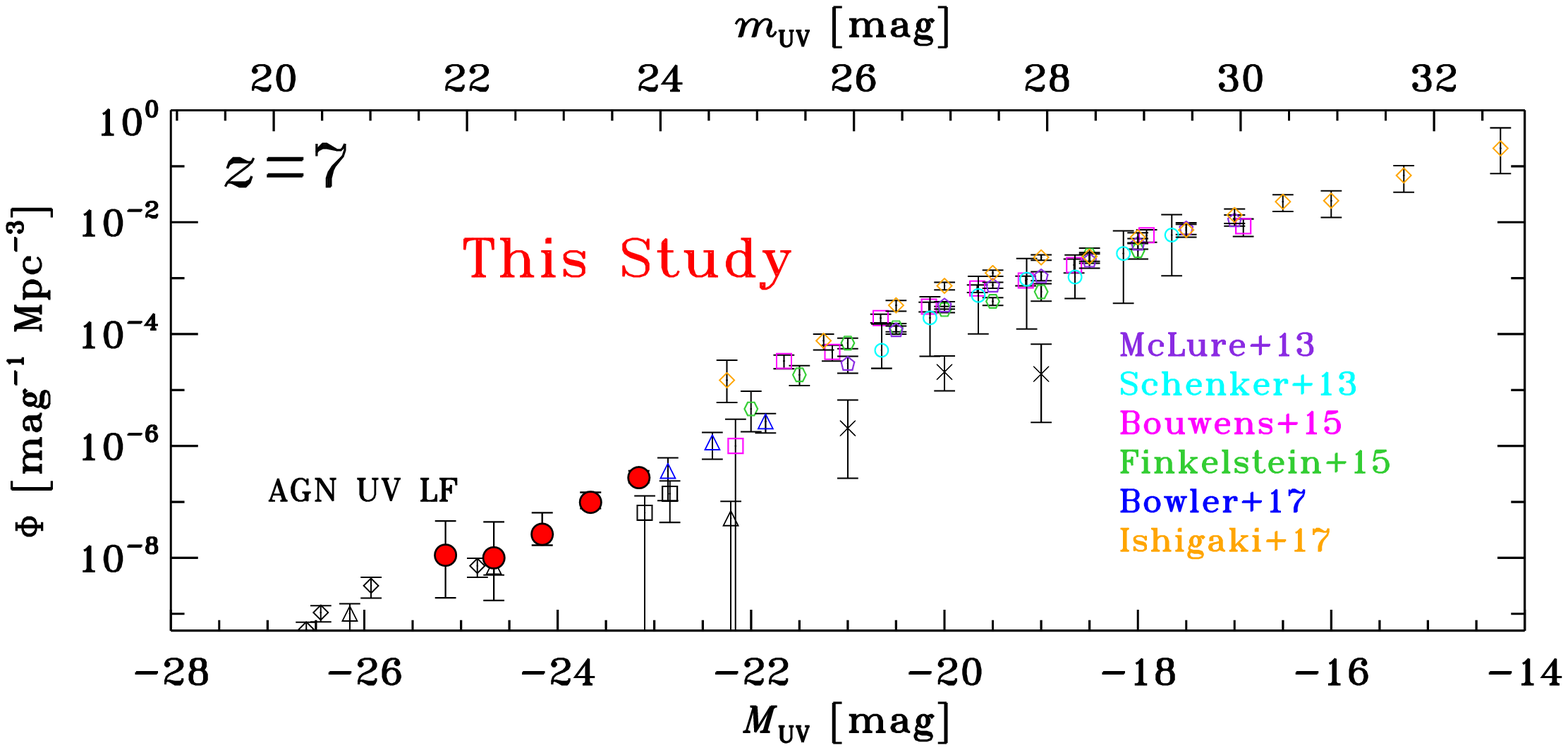} 
 \end{center}
\caption{
Same as Figure \ref{fig:UVLF_wAGN}, but for $z \sim 6$ and $z \sim 7$. 
For comparison, 
we also show previous results for galaxies taken from  
\authorcite{2013MNRAS.432.2696M} (\yearcite{2013MNRAS.432.2696M}; open purple pendagons) at $z \sim 7$, 
\authorcite{2013ApJ...768..196S} (\yearcite{2013ApJ...768..196S}; open cyan circles) at $z \sim 7$, 
\authorcite{2015ApJ...803...34B} (\yearcite{2015ApJ...803...34B}; open magenta squares) at $z \sim 6-7$, 
\authorcite{2015ApJ...810...71F} (\yearcite{2015ApJ...810...71F}; open green hexagons) at $z \sim 6-7$, 
\authorcite{2015MNRAS.452.1817B} (\yearcite{2015MNRAS.452.1817B}; open blue triangles) at $z \sim 6$, 
\authorcite{2017MNRAS.466.3612B} (\yearcite{2017MNRAS.466.3612B}; open blue triangles) at $z \sim 7$, 
and 
\authorcite{2017arXiv170204867I} (\yearcite{2017arXiv170204867I}; open orange diamonds) at $z \sim 7$. 
In addition, 
we plot previous results for quasars taken from 
\authorcite{2015A&A...578A..83G} (\yearcite{2015A&A...578A..83G}; black crosses) at $z \sim 6$, 
\authorcite{2017arXiv170407750P} (\yearcite{2017arXiv170407750P}; open black hexagons) at $z \sim 6$, 
\authorcite{2010AJ....139..906W} (\yearcite{2010AJ....139..906W}; open black triangles) at $z \sim 6$,  
\authorcite{2015ApJ...798...28K} (\yearcite{2015ApJ...798...28K}; open black squares) at $z \sim 6$, 
and 
\authorcite{2016ApJ...833..222J} (\yearcite{2016ApJ...833..222J}; open black diamonds) at $z \sim 6$. 
}\label{fig:UVLF_wAGN2}
\end{figure*}

\begin{figure*}
 \begin{center}
  \includegraphics[width=14cm]{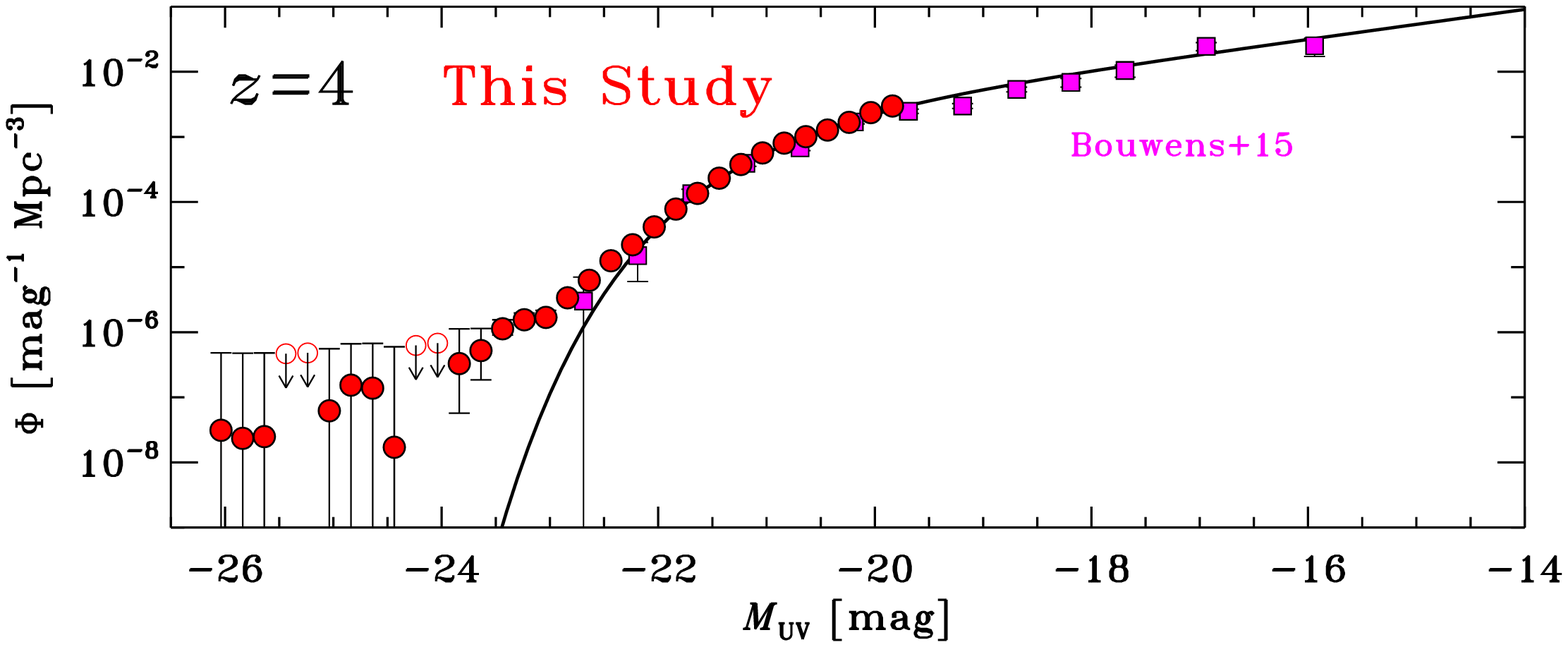} 
  \includegraphics[width=14cm]{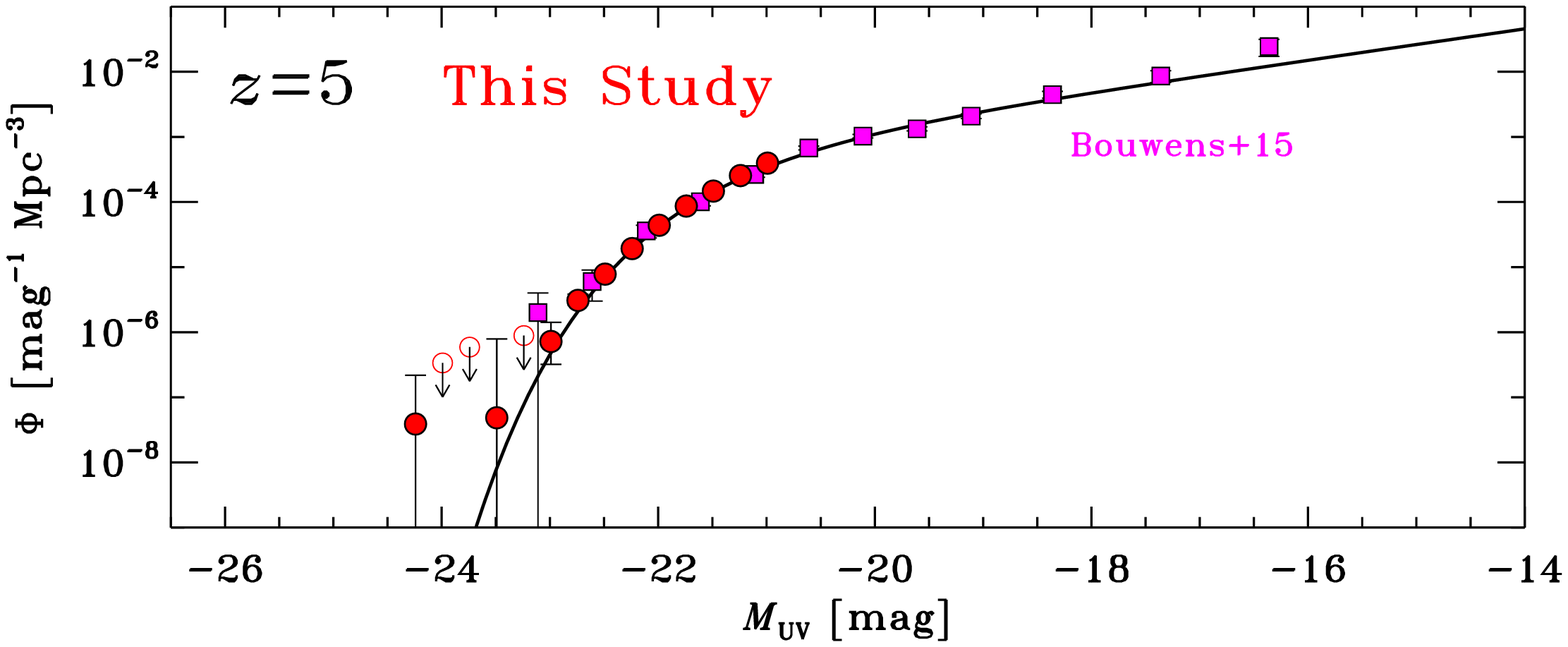} 
  \includegraphics[width=14cm]{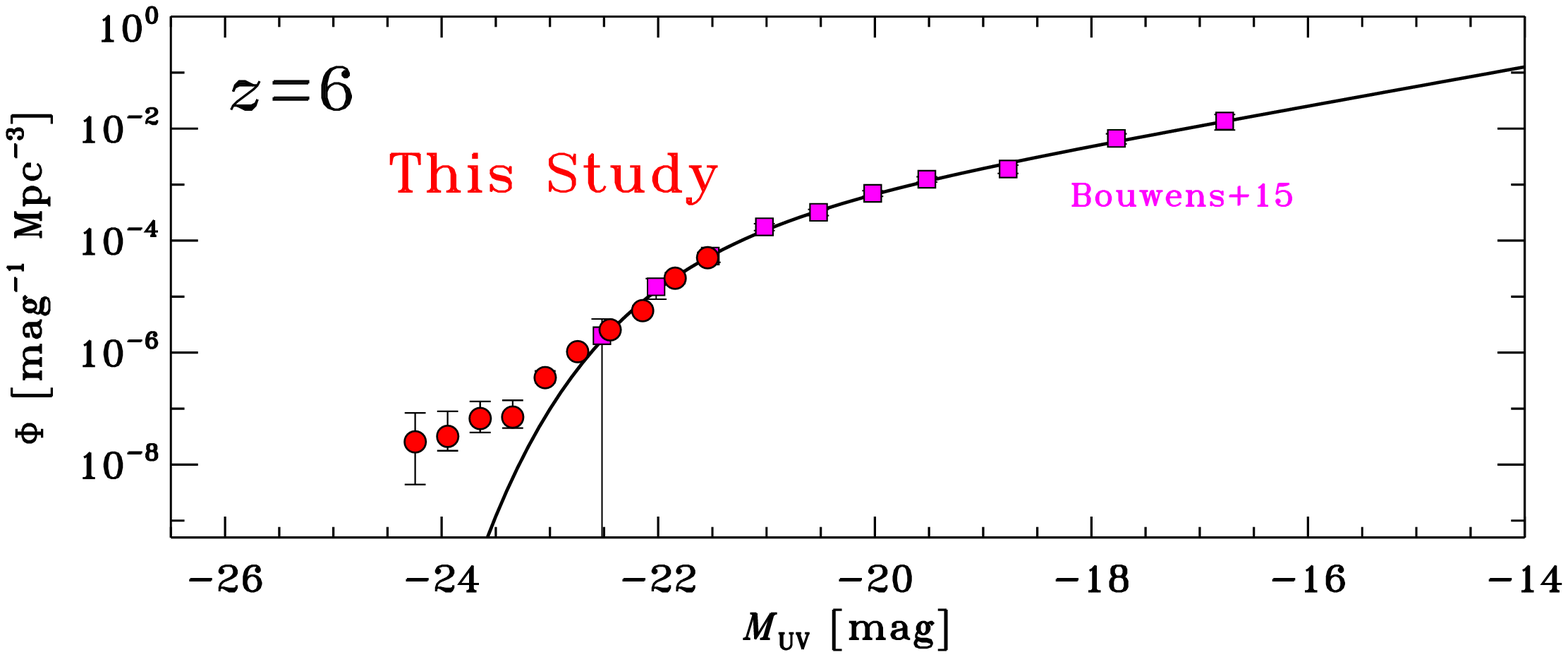} 
  \includegraphics[width=14cm]{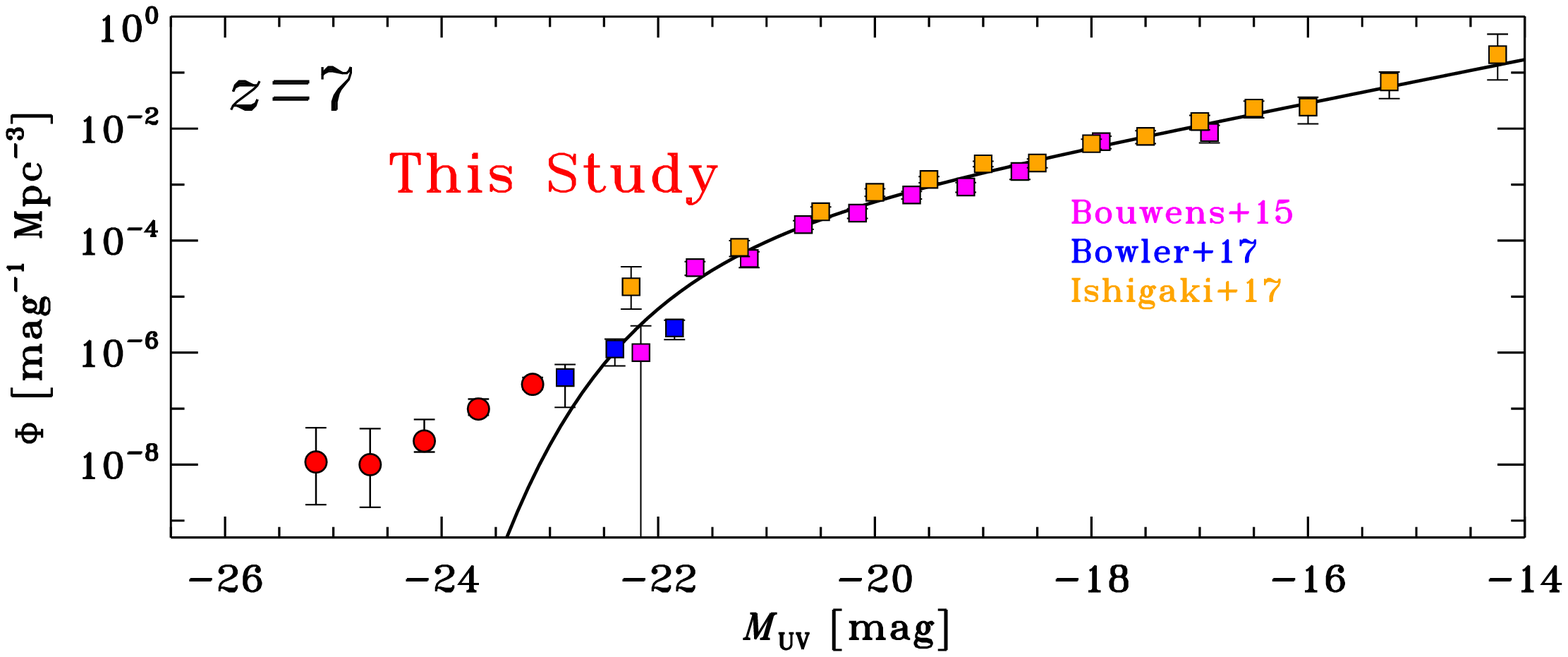} 
 \end{center}
\caption{
Rest-frame UV luminosity functions of dropouts 
at $z \sim 4$, $z \sim 5$, $z \sim 6$, and $z \sim 7$ from top to bottom. 
The red circles show our results based on the HSC SSP survey data. 
The red open circles with a downward arrow are the $1\sigma$ upper limits. 
For comparison, 
we also show previous results for galaxies taken from  
\authorcite{2015ApJ...803...34B} (\yearcite{2015ApJ...803...34B}; filled magenta squares) at $z \sim 4-7$, 
\authorcite{2017MNRAS.466.3612B} (\yearcite{2017MNRAS.466.3612B}; filled blue squares) at $z \sim 7$, 
and 
\authorcite{2017arXiv170204867I} (\yearcite{2017arXiv170204867I}; filled orange squares) at $z \sim 7$. 
The solid lines are the best-fit Schechter functions. 
}\label{fig:UVLF_selected}
\end{figure*}

\begin{figure*}
 \begin{center}
  \includegraphics[width=15cm]{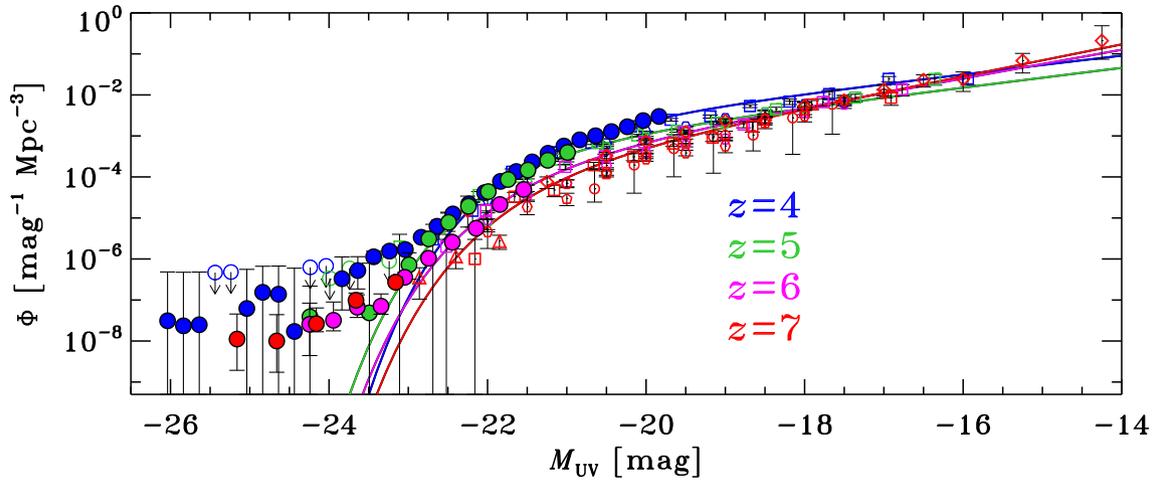} 
 \end{center}
\caption{
Rest-frame UV luminosity functions of dropouts and 
the best-fit Schechter functions 
at $z \sim 4$ (blue), $z \sim 5$ (green), $z \sim 6$ (magenta), and $z \sim 7$ (red). 
The filled circles and open circles with an arrow
correspond to our results, 
and 
the pentagons, the small open circles, 
the squares, the hexagons, the triangles, and the diamonds are 
the results of 
\citet{2013MNRAS.432.2696M}, 
\citet{2013ApJ...768..196S}, 
\citet{2015ApJ...803...34B}, 
\citet{2015ApJ...810...71F}, 
\citet{2017MNRAS.466.3612B}, 
and 
\citet{2017arXiv170204867I}, respectively. 
}\label{fig:UVLF_selected_all}
\end{figure*}

\begin{figure*}
 \begin{center}
  \includegraphics[width=5.6cm]{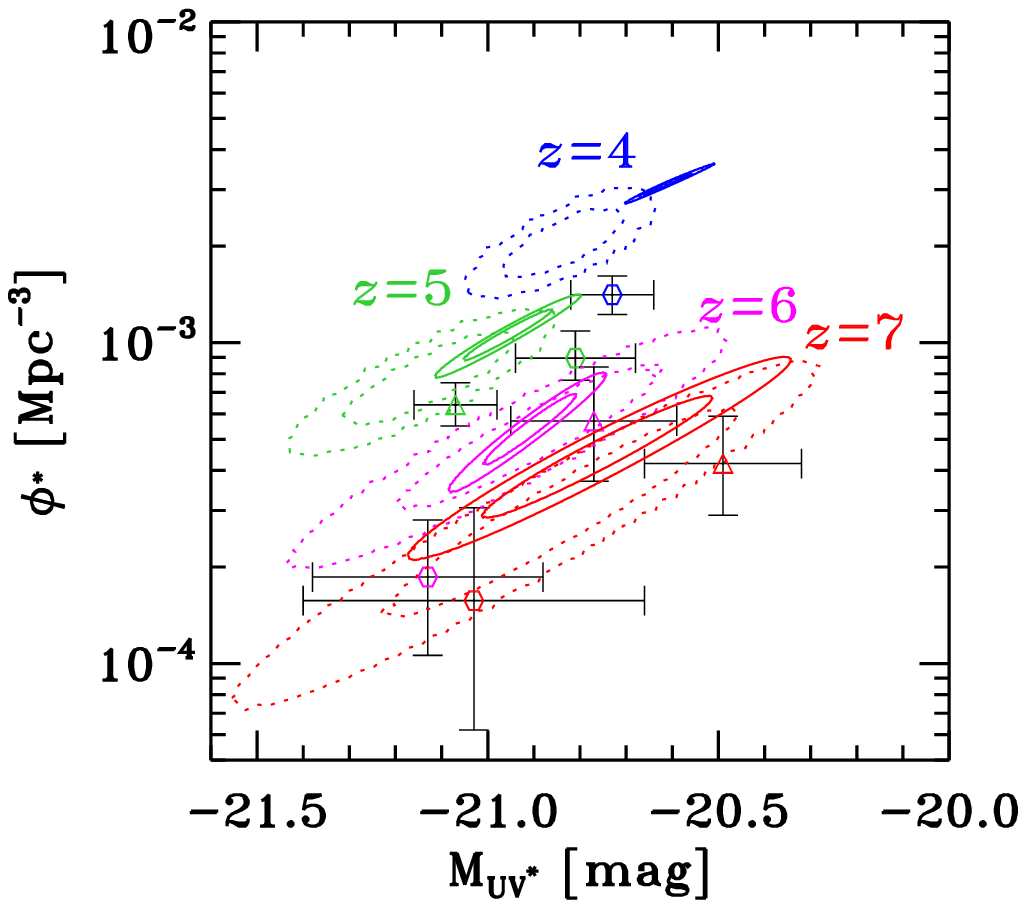} 
  \includegraphics[width=5.6cm]{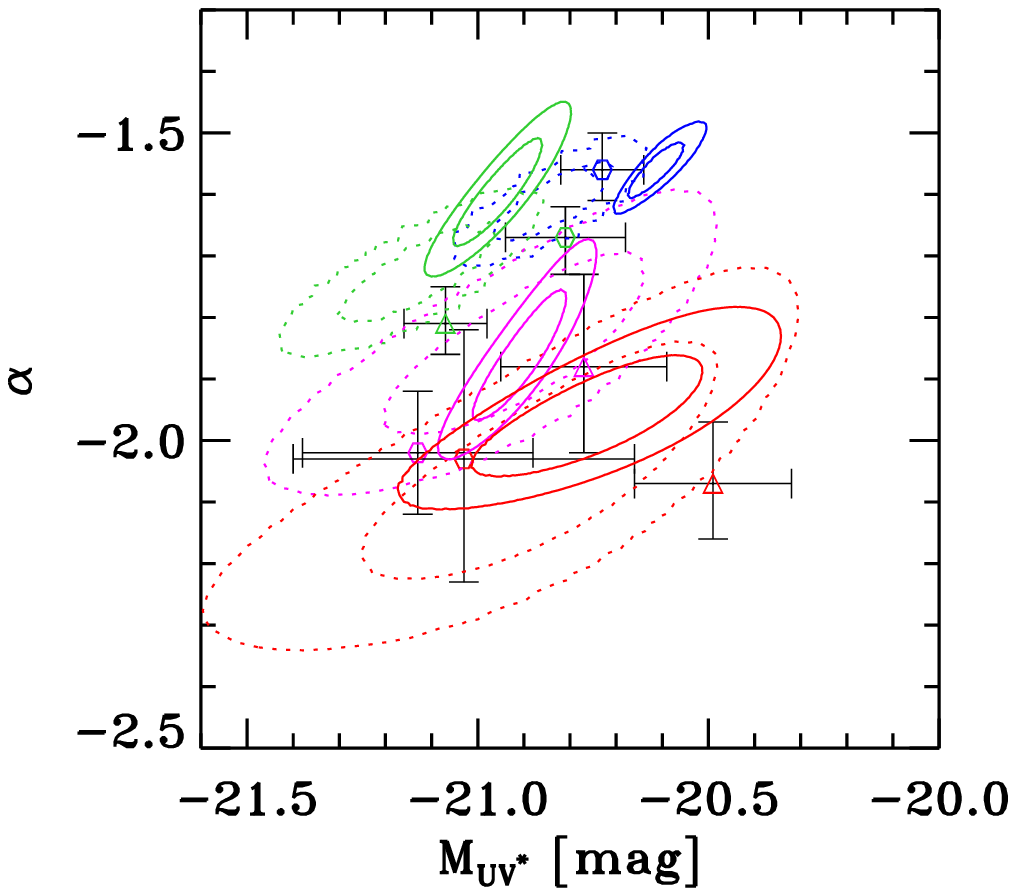} 
  \includegraphics[width=5.6cm]{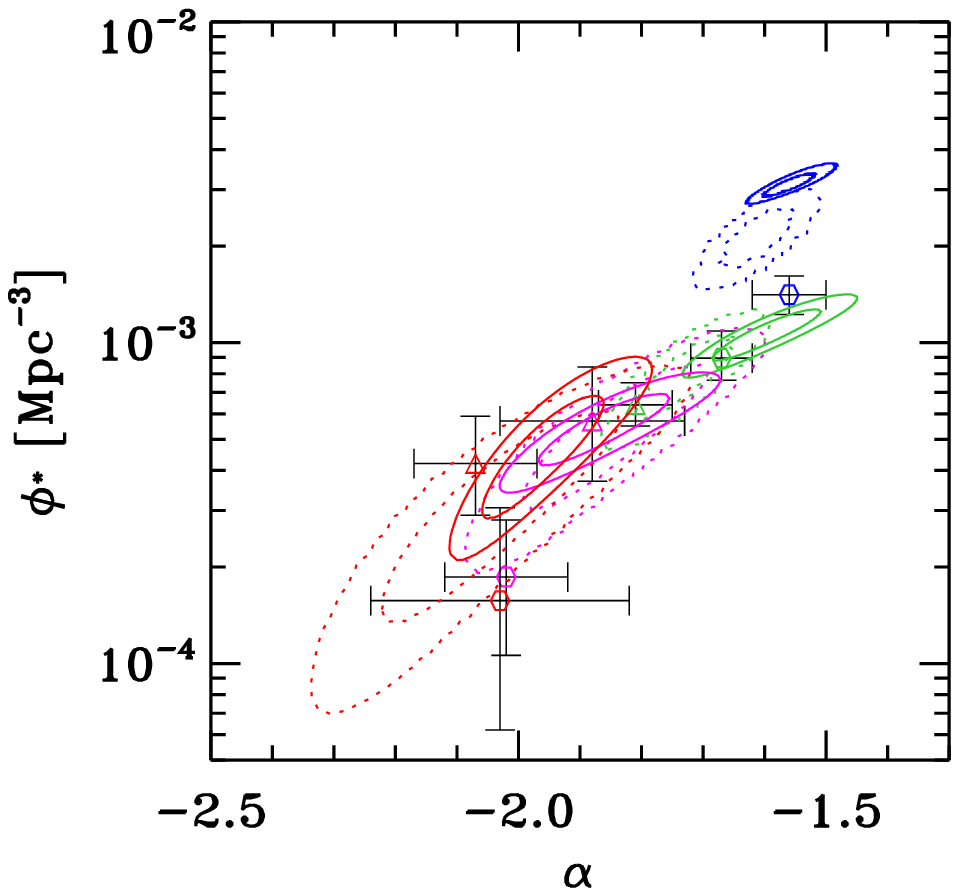} 
 \end{center}
\caption{
$1\sigma$ and $2 \sigma$ confidence intervals on the Schechter parameters, 
$M_{\rm UV}^\ast$, $\phi^\ast$, and $\alpha$. 
The blue, green, magenta, and red solid contours correspond to 
our results for the galaxy UV LFs at $z \sim 4$, $z \sim 5$, $z \sim 6$, and $z \sim 7$, respectively. 
The dotted contours correspond to the results of \citet{2015ApJ...803...34B}. 
The hexagons show the results of \citet{2015ApJ...810...71F} 
and the triangles are those of \citet{2015MNRAS.452.1817B} and \citet{2017MNRAS.466.3612B}. 
}\label{fig:Schechter_fit}
\end{figure*}

In what follows, we estimate the galaxy UV LFs in as wide a magnitude range as possible 
by taking into account the contributions of AGNs in our LF estimates, 
although the associated uncertainties are not small.  
To subtract the AGN contributions, 
we take advantage of the galaxy fraction estimates based on the spectroscopy results 
shown in Figures \ref{fig:UVLF_wAGN} and \ref{fig:UVLF_wAGN2}; 
we multiply the UV LFs by the spectroscopic galaxy fraction, 
both of which are derived in Section \ref{subsec:luminosity_function}. 
Since the number of spectroscopically confirmed sources 
in our $z \sim 5$ and $z \sim 7$ samples are not large, 
we apply the same galaxy fraction values for $z \sim 5$ ($z \sim 7$)  
as those for the $z \sim 4$ ($z \sim 6$) sample,  
assuming that the galaxy fraction has little evolution.

\begin{figure*}
 \begin{center}
  \includegraphics[width=14cm]{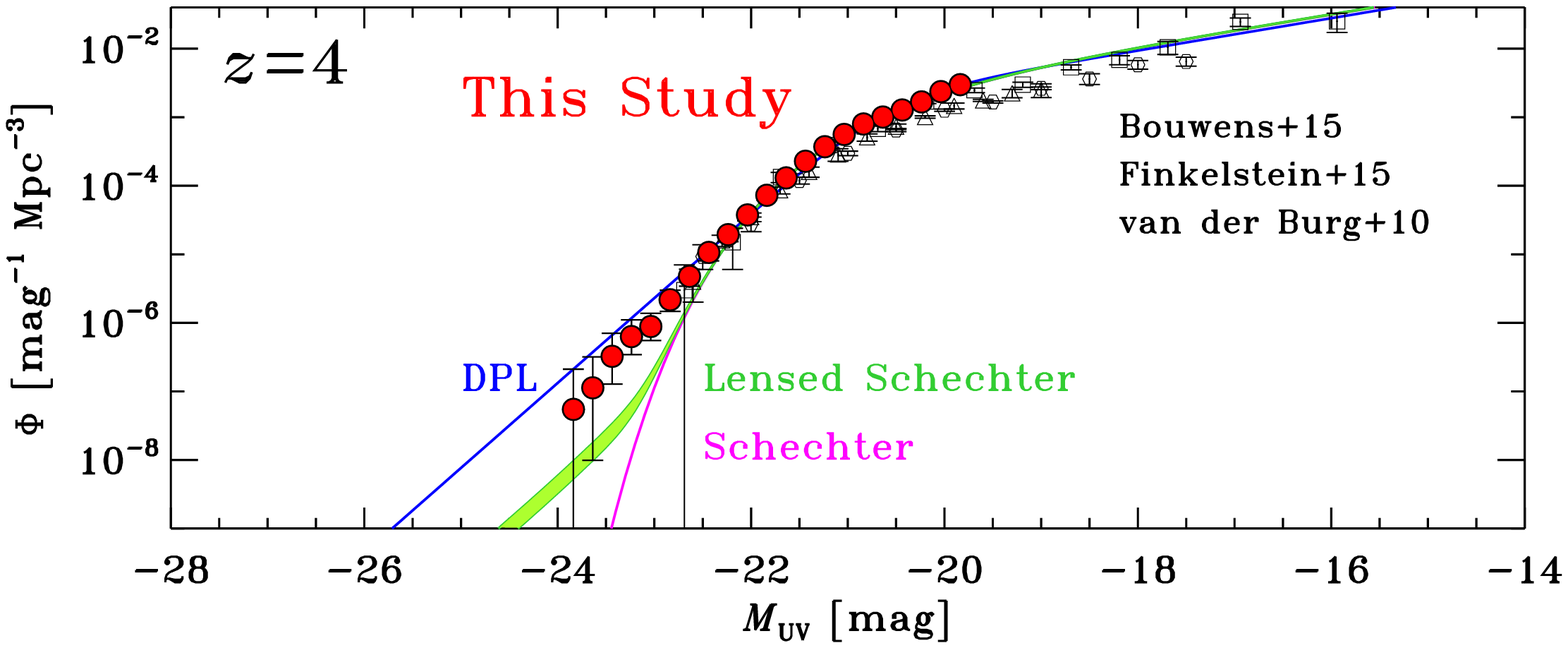} 
  \includegraphics[width=14cm]{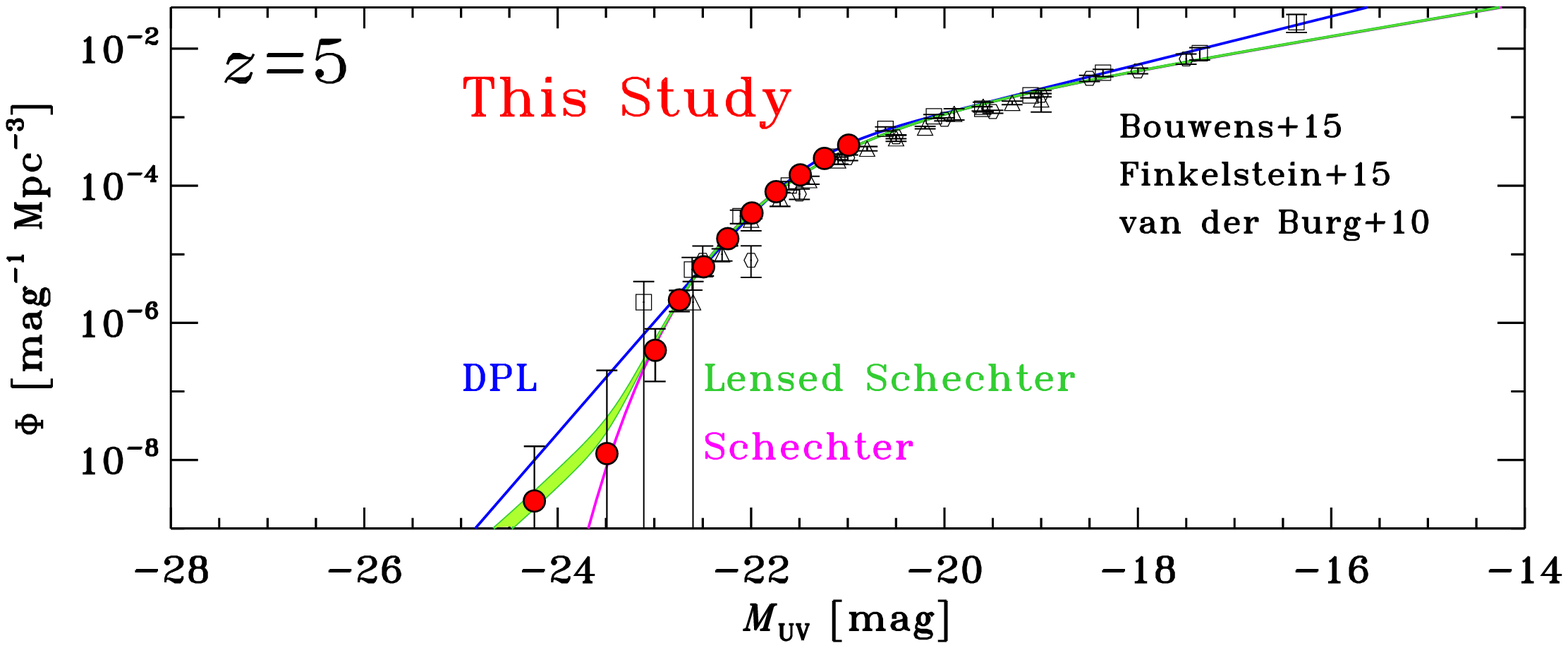} 
  \includegraphics[width=14cm]{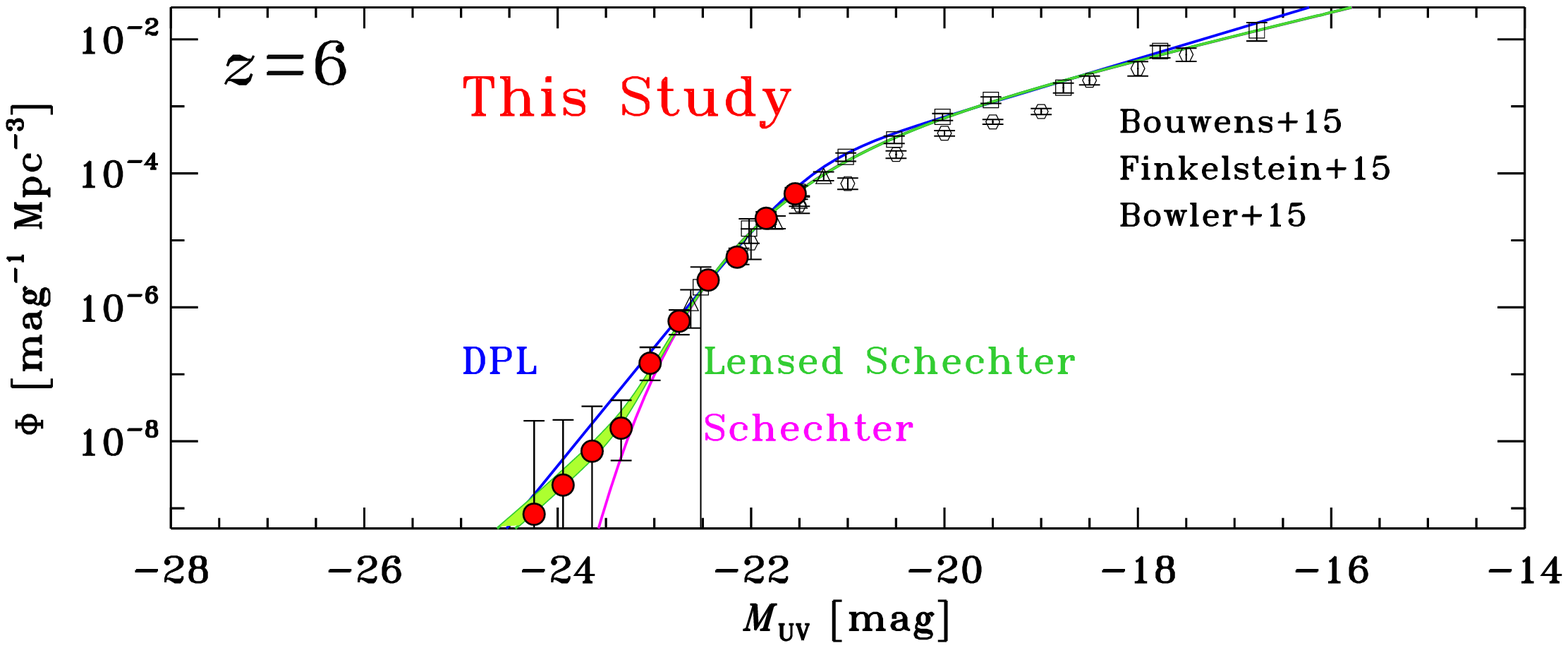} 
  \includegraphics[width=14cm]{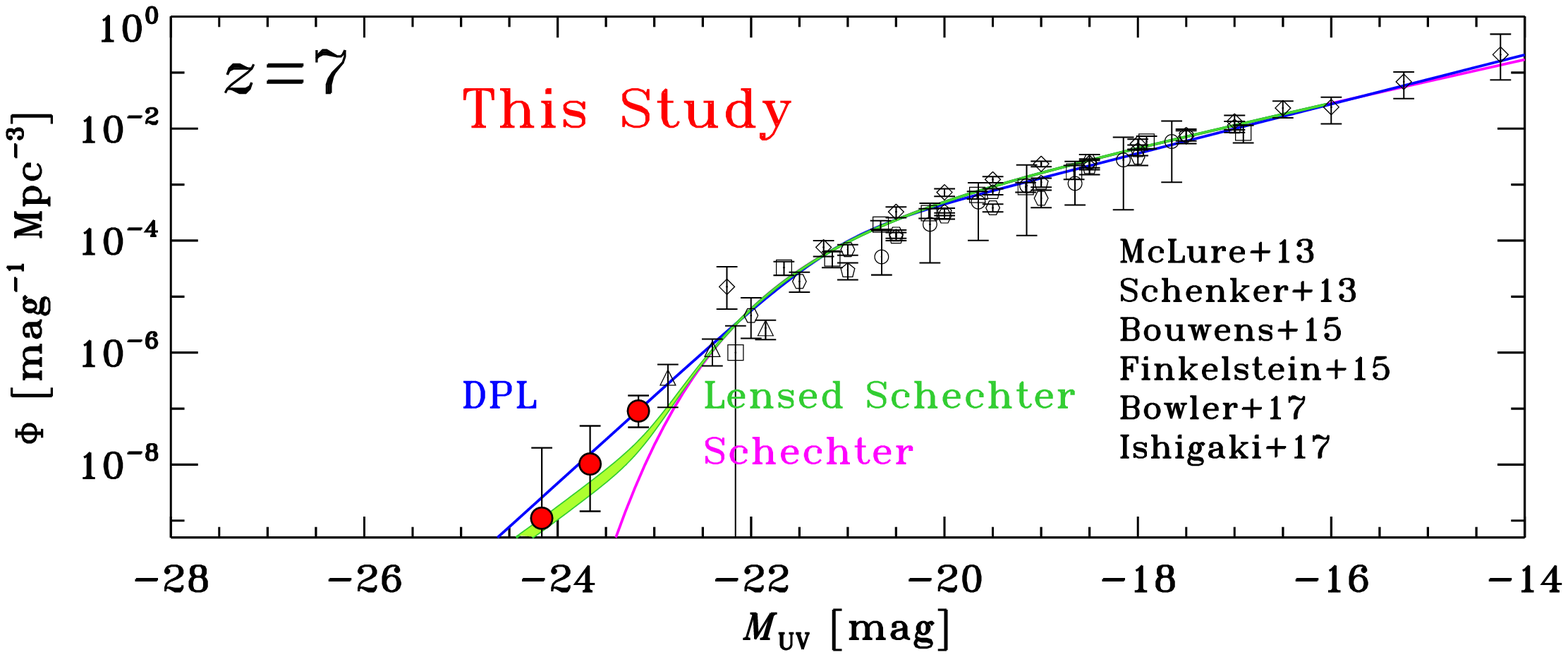} 
 \end{center}
\caption{
Rest-frame UV luminosity functions of galaxies that take into account quasar contamination correction 
at $z \sim 4$, $z \sim 5$, $z \sim 6$, and $z \sim 7$ from top to bottom. 
The green shaded region corresponds to 
the best-fit Schechter functions that take into account the effect of gravitational lensing 
with the two cases of the optical depth estimates \citep{2011ApJ...742...15T,2015MNRAS.450.1224B}. 
The magenta curve is the best-fit Schechter function without considering the lensing effect 
and  
the blue curve is the best-fit DPL function. 
For comparison, 
we also show previous results for galaxies taken from  
\authorcite{2013MNRAS.432.2696M} (\yearcite{2013MNRAS.432.2696M}; open pendagons) at $z \sim 7$, 
\authorcite{2013ApJ...768..196S} (\yearcite{2013ApJ...768..196S}; open circles) at $z \sim 7$, 
\authorcite{2015ApJ...803...34B} (\yearcite{2015ApJ...803...34B}; open squares) at $z \sim 4-7$, 
\authorcite{2015ApJ...810...71F} (\yearcite{2015ApJ...810...71F}; open hexagons) at $z \sim 4-7$, 
\authorcite{2017MNRAS.466.3612B} (\yearcite{2017MNRAS.466.3612B}; open triangles) at $z \sim 7$, 
and 
\authorcite{2017arXiv170204867I} (\yearcite{2017arXiv170204867I}; open diamonds) at $z \sim 7$. 
}\label{fig:UVLF}
\end{figure*}

Figure \ref{fig:UVLF} and Table \ref{tab:UVLF} 
show our estimates of the galaxy UV LFs from $z \sim 4$ to $z \sim 7$. 
We confirm that 
our results are consistent with the previous results 
in the UV magnitude range fainter than $-23$ mag, 
as is also the case with our results before considering the contribution of AGNs. 
This is because 
the number densities of AGNs are negligibly small compared with galaxies in this magnitude range. 
In the brighter magnitude range of $M_{\rm UV} < -23$ mag, 
we find that 
our LF estimates for $z \sim 4$, $6$, and $7$ still appear to have a hump, 
although the uncertainties are large. 
To characterize the derived galaxy UV LFs, 
we compare the following three functions.

One form is a Schechter function (Equation \ref{eq:schechter}). 
We adopt the best-fit Schechter functions 
that are obtained for the magnitude range where the galaxy fraction is large 
(Section \ref{subsec:luminosity_function}).  
Table \ref{tab:LF_best_fit_parameters} 
summarizes the adopted parameter values and the reduced $\chi^2$.

Another functional form is a double power-law (DPL) function 
(e.g., \cite{2012MNRAS.426.2772B}), 
\begin{equation}
\phi(L) dL 
= \phi^\ast \left[ \left( \frac{L}{L^\ast} \right)^{-\alpha} + \left( \frac{L}{L^\ast} \right)^{-\beta}   \right]^{-1} \frac{dL}{L^\ast}, 
\end{equation}
where 
the definitions of 
$\phi^\ast$, $M^\ast_{\rm UV}$, and $\alpha$ are 
the same as those in Equation (\ref{eq:schechter}), 
and $\beta$ is the bright-end power-law slope. 
We define a DPL function 
as a function of absolute magnitude 
$\Phi (M_{\rm UV})$ as 
$\phi(L)dL = \Phi (M_{\rm UV}) dM_{\rm UV}$, 
\begin{eqnarray}
\Phi(M_{\rm UV}) 
	&=& \frac{\ln 10}{2.5} \phi^\ast \nonumber \\
	&& \hspace{-2em} \times \left[10^{0.4(\alpha+1)(M_{\rm UV} - M_{\rm UV}^\ast)} + 10^{0.4(\beta+1)(M_{\rm UV} - M_{\rm UV}^\ast)} \right]^{-1}. 
\label{eq:dpl}
\end{eqnarray} 
We derive the best-fit parameters of Equation (\ref{eq:dpl}) 
by a $\chi^2$ minimization fit to the observed galaxy UV LFs 
obtained in this study and  and the previous \textit{Hubble} studies by 
\citet{2015ApJ...803...34B} and \citet{2017arXiv170204867I}. 
Table \ref{tab:LF_best_fit_parameters} shows 
the best-fit set of the parameters.

The other form is 
a modified Schechter function that considers 
the effect of gravitational lens magnification by foreground sources 
(e.g., \cite{2011Natur.469..181W}; \cite{2011ApJ...742...15T}; \cite{2015ApJ...805...79M}; \cite{2015MNRAS.450.1224B}). 
To take into account the magnification effect on the observed shape of the galaxy UV LFs, 
we basically follow the method presented by \citet{2011Natur.469..181W}. 
A gravitationally lensed Schechter function can be estimated with 
the convolution between the intrinsic Schechter function  
and the magnification distribution of a Singular Isothermal Sphere (SIS), $dP/d\mu$, 
weighted by the strong lensing optical depth $\tau_{\rm m}$, 
which is the fraction of strongly lensed random lines of sight.  
The overall magnification distribution 
can be modeled by using the probability distribution for magnification 
of multiply imaged sources over a fraction $\tau_{\rm m}$ of the sky. 
To conserve total flux on the cosmic sphere centered on an observer, 
we need to consider the de-magnification of unlensed sources:  
\begin{equation}
\mu_{\rm demag}
	= \frac{1 - \left\langle \mu_{\rm mult} \right\rangle \tau_{\rm m}}{1 - \tau_{\rm m}}, 
\end{equation} 
where $\left\langle \mu_{\rm mult} \right\rangle = 4$  
is the mean magnification of multiply imaged sources. 
For a given LF $\phi(L)$, 
a gravitationally lensed LF $\phi_{\rm lensed}(L)$ can then be obtained by 
\begin{eqnarray}
\phi_{\rm lensed} (L)
	&=& (1 - \tau_{\rm m}) \frac{1}{\mu_{\rm demag}} \phi \left( \frac{L}{\mu_{\rm demag}} \right) \nonumber \\
	&\hspace{1em}& + \tau_{\rm m} \int^\infty_0 d \mu \frac{1}{\mu} 
			\left( \frac{dP_{\rm m,1}}{d\mu} + \frac{dP_{\rm m,2}}{d\mu} \right) \phi\left( \frac{L}{\mu} \right), 
\end{eqnarray}
where 
\begin{equation}
\frac{dP_{\rm m,1}}{d\mu}
= 
\left\{ 
	\begin{array}{ll}
		\frac{2}{(\mu - 1)^3} & ({\rm for} \,\,\, \mu > 2) \\
		0 & ({\rm for} \,\,\, 0< \mu < 2)
	\end{array}
\right. 
\end{equation}
is the magnification distribution 
as a function of magnification factor $\mu$ 
for the brighter image 
in a strongly lensed system given for an SIS 
and 
\begin{equation}
\frac{dP_{\rm m,2}}{d\mu}
	= \frac{2}{(\mu + 1)^3} \,\,\, ({\rm for} \,\,\, \mu > 0) 
\end{equation}
is the magnification probability distribution of the second image. 
We consider two cases of optical depth estimate results 
to cover a possible range of systematic uncertainties. 
One is based on the high-resolution ray-tracing simulations of \citet{2011ApJ...742...15T}. 
From their results of the probability distribution function of lensing magnification,  
the optical depth values are estimated to be 
$\tau_{\rm m} = (0.00231, \, 0.00315, \, 0.00380, \, 0.00446)$ 
at $z=(4, \, 5, \, 6, \, 7)$. 
The other is based on a calibrated Faber-Jackson relation \citep{1976ApJ...204..668F}  
obtained by \citet{2015MNRAS.450.1224B}: 
$\tau_{\rm m} = (0.0041, \, 0.0054, \, 0.0065, \, 0.0072)$ 
at $z=(4, \, 5, \, 6, \, 7)$. 
Note that these optical depth estimates would correspond to upper limits, 
because 
some fraction of lensed dropouts might be too close to foreground lensing galaxies 
to be selected as dropouts in our samples. 
For the Schechter function parameters, 
we adopt the best-fit values obtained in Section \ref{subsec:luminosity_function}.  
The adopted parameters and the reduced $\chi^2$ values are 
summarized in Table \ref{tab:LF_best_fit_parameters}.

In Figure \ref{fig:UVLF}, 
we show the best-fit functions of these three functional forms 
with the derived galaxy UV LF results. 
We find that 
the bright-end shapes of the observed galaxy UV LFs cannot be explained 
by the Schechter functions, 
although the excess at $z\sim5$ is not significant.  
The significance values of the excesses from the Schechter functions 
are 
$5.2 \sigma$, 
$0.4\sigma$, 
$2.3\sigma$, 
and 
$2.5\sigma$
at $z \sim 4$, $5$, $6$, and $7$, respectively. 
Because the AGN UV LFs are constrained relatively well at $z \sim 4$, 
we check whether the bright-end shape of the galaxy UV LF 
has an excess 
if we use the best-fit AGN UV LF for subtraction of 
the AGN contribution. 
We confirm that similar results are obtained if we use the best-fit AGN UV LFs 
taken from \citet{2017arXiv170405996A} and \citet{2011ApJ...728L..26G}. 
In Figure \ref{fig:UVLF}, 
it seems that 
the DPL and the lensed Schechter functional forms provide better fits to the observed galaxy UV LFs 
than the original Schechter functional form.  
If this is the case, 
the results would suggest that 
bright-end galaxies are significantly affected by gravitational lensing, 
a high fraction of apparently bright galaxies are blended merging galaxies, 
and/or 
negative feedback for star formation in massive galaxies might be inefficient. 
Note that 
the observed galaxy UV LF data points at $z \sim 4$ are better described with the DPL 
and the significance of the hump feature at $M_{\rm UV} < -22.5$ mag 
from the lensed Schechter function is about $4.7 \sigma$. 
At higher redshifts, 
the significance values of the excess from the lensed Schechter function 
are $< 1\sigma$ at $z \sim 5-6$ and about $1.6 \sigma$ at $z \sim 7$. 
The bright-end LFs at $z \sim 5-7$ could be explained solely by the gravitational lensing effect, 
unless a significant number of lensed dropouts are missed 
due to their foreground galaxies that are too close to them on the sky.
To investigate whether our bright-end dropout galaxies 
are strongly affected by gravitational lensing, 
we will check their environments and 
identify foreground sources around them which can act as lenses 
(e.g., \cite{2015MNRAS.450.1224B}) 
in future analyses.  
To examine the possibility that 
a fraction of our bright-end galaxies are blended merging galaxies, 
higher resolution imaging data taken with \textit{Hubble} 
are needed 
(e.g., \cite{2017MNRAS.466.3612B}). 
The \textit{Hubble} data will also be useful 
for determining the quasar  contamination rate, 
because quasars should show up as point sources with \textit{Hubble}.

It should be noted, however, that 
there remain not only statistical uncertainties but also systematic ones 
in our LF estimates particularly at the bright end. 
For example, 
in our selection completeness estimates for bright-end sources, 
we have extrapolated the UV slope $\beta$ distribution in the literature 
and have not taken into account the effect of Ly$\alpha$ emission
because of lack of appropriate references. 
However, our effective volume estimates would not 
be 
correct 
if the real $\beta$ or Ly$\alpha$ equivalent width (EW) distribution is 
significantly different from the used ones, 
as may already be implied in Figure \ref{fig:completeness}. 
To check these possibilities directly, 
we will derive the $\beta$ distribution of $z \sim 4-5$ 
bright dropout galaxies by using our multi-band HSC data 
and will derive the Ly$\alpha$ EW distributions based on spectroscopy results. 
Here we investigate the robustness of our results against possible uncertainties 
in the selection completeness estimates  
by simply assuming that the uncertainty is $15${\%}.
\footnote{
\citet{2015ApJ...803...34B} have considered 
$\sim 10${\%} systematic errors in their
selection volume estimates. 
Here we adopt a slightly more pessimistic value of $15${\%} than theirs.}
We repeat
the Schechter and DPL function fittings 
for the $z \sim 4$ galaxy UV LF with the larger uncertainties. 
The best-fit Schechter parameters are found to be 
($M_{\rm UV}^\ast$ [mag], $\phi^\ast$ [$10^{-3}$ Mpc$^{-3}$], $\alpha$) 
$=$ 
($-20.83^{+0.05}_{-0.03}$, $1.96^{+0.22}_{-0.18}$, $-1.62^{+0.04}_{-0.04}$) 
and the best-fit DPL function parameters are 
($M_{\rm UV}^\ast$ [mag], $\phi^\ast$ [$10^{-3}$ Mpc$^{-3}$], $\alpha$, $\beta$) 
$=$ 
($-21.16^{+0.08}_{-0.08}$, $0.88^{+0.12}_{-0.11}$, $-1.80^{+0.04}_{-0.03}$, $-4.74^{+0.14}_{-0.16}$), 
both of which are slightly different from those listed 
in Tables  \ref{tab:LF_best_fit_parameters_wAGN} and \ref{tab:LF_best_fit_parameters}. 
However, even in this case, the bright-end excess feature is confirmed. 
The significance value of the excess from the best-fit Schechter function is 
$3.8 \sigma$, 
and that from the lensed Schechter function is 
$3.4 \sigma$. 
There are also other possible sources of systematic uncertainties.  
The galaxy fraction estimates based on the spectroscopy results 
still have large uncertainties, particularly for $z \sim 5-7$, 
because the number of sources with spectroscopic redshifts is limited. 
In addition, 
although we carefully construct our dropout samples 
by checking their detections in the multi-band stacked images for the $z \sim 4-6$ samples 
and in the single epoch observation images for the $z \sim 7$ sample, 
they may still include some transient objects such as supernovae. 
This is because, 
if transient objects are bright in our observations with long wavelength bands 
but faint in the observations with short ones, 
they can mimic Ly$\alpha$ break features. 
Improved constraints on the form of the bright end 
based on follow-up spectroscopic observations 
and wider area imaging from the on-going HSC SSP 
will reduce the remaining uncertainties 
on the UV LF estimates in the near future.

%%%%%%%%%%%%%%%%%%%%%%%%%%%%%%%%%%%%%%
%%%%%%%%%%%%%%%%%%%%%%%%%%%%%%%%%%%%%%%

%%%%%%%%%%%%%%%%%%%%%%%%%%%%%%%%%%%%%%
%%%%%%%%%%%%%%%%%%%%%%%%%%%%%%%%%%%%%%%

%%%%%%%%%%%%%%%%%%%%%%%%%%%%%%%%%%%%%%
%%%%%%%%%%%%%%%%%%%%%%%%%%%%%%%%%%%%%%%

%%%%%%%%%%%%%%%%%%%%%%%%%%%%%%%%%%%%%%
%%%%%%%%%%%%%%%%%%%%%%%%%%%%%%%%%%%%%%%

%%%%%%%%%%%%%%%%%%%%%%%%%%%%%%%%%%%%%%
%%%%%%%%%%%%%%%%%%%%%%%%%%%%%%%%%%%%%%%

%%%%%%%%%%%%%%%%%%%%%%%%%%%%%%%%%%%%%%
%%%%%%%%%%%%%%%%%%%%%%%%%%%%%%%%%%%%%%%

%%%%%%%%%%%%%%%%%%%%%%%%%%%%%%%%%%%%%%%
\begin{table*}
  \tbl{
  Best-fit parameters of the Schechter functions 
  for the rest-frame UV LFs at $z \sim 4-7$. 
  (1) Average redshift. (2) Characteristic magnitude. (3) Normalization. (4) Faint end slope. 
  (5) Reduced $\chi^2$. 
  }{%
  \begin{tabular}{cccccc}
      \hline
      Dropout Sample & $\left\langle z \right\rangle$ & $M_{\rm UV}^\ast$  &  $\phi^\ast$  &  $\alpha$  &  $\chi^2_\nu$\\ 
      & & (mag) & ($10^{-3}$ Mpc$^{-3}$) & & \\
      & (1) & (2) & (3) & (4) & (5) \\
      \hline
      $g$ & $4$ & $-20.63^{+0.05}_{-0.02}$ & $3.04^{+0.23}_{-0.10}$ & $-1.57^{+0.03}_{-0.02}$  & $7.0$ \\
      $r$ & $5$ & $-20.96^{+0.06}_{-0.05}$ & $1.06^{+0.13}_{-0.11}$ & $-1.60^{+0.06}_{-0.05}$  & $1.3$ \\
      $i$ & $6$ & $-20.91^{+0.07}_{-0.06}$ & $0.54^{+0.10}_{-0.08}$ & $-1.87^{+0.07}_{-0.06}$  & $0.8$ \\
      $z$ & $7$ & $-20.77^{+0.16}_{-0.15}$ & $0.438^{+0.150}_{-0.107}$ & $-1.97^{+0.07}_{-0.05}$  & $2.0$ \\
      \hline
    \end{tabular}}\label{tab:LF_best_fit_parameters_wAGN}
%\begin{tabnote}
%This is table note.
%\end{tabnote}
\end{table*}
%%%%%%%%%%%%%%%%%%%%%%%%%%%%%%%%%%%%%%%

%%%%%%%%%%%%%%%%%%%%%%%%%%%%%%%%%%%%%%%
\begin{table*}
  \tbl{
  $\chi^2$ values of the best-fit Schechter and DPL functions 
  for the rest-frame UV LFs at $z \sim 4-7$. 
  (1) Average redshift. (2) Characteristic magnitude. (3) Normalization. (4) Faint end slope. 
  (5) Bright end power-law slope for the DPL function. (6) Reduced $\chi^2$. 
  }{%
  \begin{tabular}{ccccccc}
      \hline
      Dropout Sample & $\left\langle z \right\rangle$ & $M_{\rm UV}^\ast$  &  $\phi^\ast$  &  $\alpha$  &  $\beta$  &  $\chi^2_\nu$\\ 
      & & (mag) & ($10^{-3}$ Mpc$^{-3}$) & & & \\
      & (1) & (2) & (3) & (4) & (5) & (6) \\
      \hline
      \multicolumn{6}{c}{Schechter function} \\
      $g$ & $4$ & $-20.63$ & $3.04$ & $-1.57$ & --- & $6.4$ \\
      $r$ & $5$ & $-20.96$ & $1.06$ & $-1.60$ & --- & $1.1$ \\
      $i$ & $6$ & $-20.91$ & $0.54$ & $-1.87$ & --- & $0.9$ \\
      $z$ & $7$ & $-20.77$ & $0.438$ & $-1.97$ & --- & $2.5$ \\
      \hline
      \multicolumn{6}{c}{DPL function} \\
      $g$ & $4$ & $-20.58^{+0.06}_{-0.07}$ & $2.31^{+0.22}_{-0.23}$ & $-1.59^{+0.04}_{-0.05}$ & $-4.10^{+0.06}_{-0.07}$ & $5.4$ \\
      $r$ & $5$ & $-21.44^{+0.07}_{-0.07}$ & $0.36^{+0.05}_{-0.05}$ & $-1.88^{+0.05}_{-0.04}$ & $-5.07^{+0.17}_{-0.18}$ & $1.0$ \\
      $i$ & $6$ & $-21.40^{+0.09}_{-0.10}$ & $0.175^{+0.039}_{-0.036}$ & $-2.08^{+0.05}_{-0.06}$ & $-5.43^{+0.20}_{-0.23}$ & $0.8$ \\
      $z$ & $7$ & $-21.10^{+0.25}_{-0.25}$ & $0.156^{+0.086}_{-0.059}$ & $-2.10^{+0.11}_{-0.10}$ & $-4.90^{+0.38}_{-0.42}$ & $1.1$ \\
      \hline
      \multicolumn{6}{c}{Lensed Schechter function with the optical depth estimates of \citet{2011ApJ...742...15T}} \\
      $g$ & $4$ & $-20.63$ & $3.04$ & $-1.57$ & --- & $6.3$ \\
      $r$ & $5$ & $-20.96$ & $1.06$ & $-1.60$ & --- & $1.1$ \\
      $i$ & $6$ & $-20.91$ & $0.54$ & $-1.87$ & --- & $0.6$ \\
      $z$ & $7$ & $-20.77$ & $0.438$ & $-1.97$ & --- & $2.2$ \\
      \multicolumn{6}{c}{Lensed Schechter function with the optical depth estimates of \citet{2015MNRAS.450.1224B}} \\
      $g$ & $4$ & $-20.63$ & $3.04$ & $-1.57$ & --- & $6.2$ \\
      $r$ & $5$ & $-20.96$ & $1.06$ & $-1.60$ & --- & $1.1$ \\
      $i$ & $6$ & $-20.91$ & $0.54$ & $-1.87$ & --- & $0.6$ \\
      $z$ & $7$ & $-20.77$ & $0.438$ & $-1.97$ & --- & $2.1$ \\
      \hline
    \end{tabular}}\label{tab:LF_best_fit_parameters}
%\begin{tabnote}
%This is table note.
%\end{tabnote}
\end{table*}
%%%%%%%%%%%%%%%%%%%%%%%%%%%%%%%%%%%%%%%

%%%%%%%%%%%%%%%%%%%%%%%%%%%%%%%%%%%%%%%%%%%%%%%%%%%%%%%%%%%%%%%%%
%%%%%%%%%%%%%%%%%%%%%%%%%%%%%%%%%%%%%%%%%%%%%%%%%%%%%%%%%%%%%%%%%
\section{Summary} \label{sec:summary}
%%%%%%%%%%%%%%%%%%%%%%%%%%%%%%%%%%%%%%%%%%%%%%%%%%%%%%%%%%%%%%%%%
%%%%%%%%%%%%%%%%%%%%%%%%%%%%%%%%%%%%%%%%%%%%%%%%%%%%%%%%%%%%%%%%%

In this paper,
we have identified 
579,565 dropout candidates at $z \sim 4-7$ by the standard color selection technique 
from the $100$ deg$^2$ deep optical imaging data of the HSC SSP survey.  
Among these dropout candidates, 
358 dropouts have spectroscopic redshifts 
obtained by our follow-up observations and in the literature. 
Combining our bright-end UV LF estimates 
with those from the complementary ultra-deep \textit{Hubble} legacy surveys, 
we have derived the UV LFs of dropouts from $z \sim7$ to $z \sim 4$ 
in a very wide UV magnitude range of $-26 < M_{\rm UV} < -14$ mag, 
which corresponds to the luminosity range of $\sim 0.002$ -- $100 \, L_{\rm UV}^\ast$. 
We have derived the best-fit Schechter parameters of $M_{\rm UV}^\ast$, $\alpha$, and $\phi^\ast$, 
by fitting Schechter functions to the UV LFs 
in the magnitude range of $M_{\rm UV} > -23$ mag, 
where the contribution of high-$z$ galaxies is dominant according to the spectroscopic results. 
We have found that 
there is little evolution in $M_{\rm UV}^\ast$ 
and the other Schechter function parameters, $\alpha$ and $\phi^\ast$, 
decrease with increasing redshift, as the previous work has already pointed out.  
Since our HSC SSP data bridge the LFs of galaxies and AGNs with great statistical accuracies, 
we have carefully subtracted the contribution of high-$z$ AGNs 
to investigate the bright end of the galaxy UV LFs 
by making use of the galaxy fraction 
as a function of UV magnitude that is derived from the spectroscopic results.  
To characterize the shapes of the derived galaxy UV LFs, 
we have compared the three functional forms:  
a Schechter function, a DPL function, 
and a modified Schechter function that takes into account 
the effect of gravitational lens magnification by foreground sources. 
We have found that the Schechter function cannot explain 
the shapes of the bright-end galaxy UV LFs at $> 2 \sigma$ significance.  
Instead, the galaxy UV LFs are better described with 
either the DPL or the lensed Schechter function.  
If this is true, 
the results would indicate that 
bright-end galaxies are significantly affected by gravitational lensing magnification, 
a significant number of bright-end galaxies are 
merger systems that are apparently blended at ground-based resolution, 
and/or 
AGN feedback for star formation suppression at high redshift is inefficient.

%%%%%%%%%%%%%%%%%%%%%%%%%%%%%%%%%%%%%%%%%%%%%%%%%%%%%%%%%%%%%%%%%
%%%%%%%%%%%%%%%%%%%%%%%%%%%%%%%%%%%%%%%%%%%%%%%%%%%%%%%%%%%%%%%%%
\begin{ack}

We thank the anonymous referee for valuable comments and suggestions 
that improved the manuscript. 
We greatly appreciate the support of the HSC pipeline team, 
particularly Jim Bosch, Hisanori Furusawa, Michitaro Koike, 
Robert Lupton, Paul Price, Tadafumi Takata, Yoshihiko Yamada, 
and Naoki Yasuda. 
We acknowledge Song Huang, Naoki Yasuda, Ryoma Murata, and Hiroko Niikura 
for their helpful advices to make use of SynPipe. 
We thank 
Rychard Bouwens and Masafumi Ishigaki 
for providing us with the machine-readable tables of their results, 
and 
Alex Hagen  
for checking the acronym of our program name. 
The HSC collaboration 
includes the astronomical communities of Japan and Taiwan, and Princeton University. 
The HSC instrumentation and software were developed by 
the National Astronomical Observatory of Japan (NAOJ), 
the Kavli Institute for the Physics and Mathematics of the Universe (Kavli IPMU), 
the University of Tokyo, the High Energy Accelerator Research Organization (KEK), 
the Academia Sinica Institute for Astronomy and Astrophysics in Taiwan (ASIAA), 
and Princeton University.  
Funding was contributed by the FIRST program from Japanese Cabinet Office, 
the Ministry of Education, Culture, Sports, Science and Technology (MEXT), 
the Japan Society for the Promotion of Science (JSPS),  
Japan Science and Technology Agency  (JST),  the Toray Science Foundation,
NAOJ, Kavli IPMU, KEK, ASIAA,  and Princeton University.
This paper makes use of software developed for the Large Synoptic Survey Telescope. 
We thank the LSST Project for making their code available as free software at 
http://dm.lsst.org.
This work was partially performed using the computer facilities of
the Institute for Cosmic Ray Research, The University of Tokyo. 
This work was supported by JSPS KAKENHI Grant Number JP15K17602.

\end{ack}

%%%%%%%%%%%%%%%%%%%%%%%%%%%%%%%%%%%%%%%%%%%%%%%%%%%%%%%%%%%%%%%%%
%%%%%%%%%%%%%%%%%%%%%%%%%%%%%%%%%%%%%%%%%%%%%%%%%%%%%%%%%%%%%%%%%

%%%%%%%%%%%%%%%%%%%%%%%%%%%%%%%%%%%%%%%%%%%%%%%%%%%%%%%%%%%%%%%%%
%%%%%%%%%%%%%%%%%%%%%%%%%%%%%%%%%%%%%%%%%%%%%%%%%%%%%%%%%%%%%%%%%

%\bibliographystyle{apj}
%\bibliography{apj-jour,ref}

\begin{thebibliography}{}
\expandafter\ifx\csname natexlab\endcsname\relax\def\natexlab#1{#1}\fi

\bibitem[{{Abazajian} {et~al.}(2004){Abazajian}, {Adelman-McCarthy},
  {Ag{\"u}eros}, {Allam}, {Anderson}, {Anderson}, {Annis}, {Bahcall}, {Baldry},
  {Bastian}, {Berlind}, {Bernardi}, {Blanton}, {Bochanski}, {Boroski},
  {Briggs}, {Brinkmann}, {Brunner}, {Budav{\'a}ri}, {Carey}, {Carliles},
  {Castander}, {Connolly}, {Csabai}, {Doi}, {Dong}, {Eisenstein}, {Evans},
  {Fan}, {Finkbeiner}, {Friedman}, {Frieman}, {Fukugita}, {Gal}, {Gillespie},
  {Glazebrook}, {Gray}, {Grebel}, {Gunn}, {Gurbani}, {Hall}, {Hamabe},
  {Harris}, {Harris}, {Harvanek}, {Heckman}, {Hendry}, {Hennessy}, {Hindsley},
  {Hogan}, {Hogg}, {Holmgren}, {Ichikawa}, {Ichikawa}, {Ivezi{\'c}}, {Jester},
  {Johnston}, {Jorgensen}, {Kent}, {Kleinman}, {Knapp}, {Kniazev}, {Kron},
  {Krzesinski}, {Kunszt}, {Kuropatkin}, {Lamb}, {Lampeitl}, {Lee}, {Leger},
  {Li}, {Lin}, {Loh}, {Long}, {Loveday}, {Lupton}, {Malik}, {Margon},
  {Matsubara}, {McGehee}, {McKay}, {Meiksin}, {Munn}, {Nakajima}, {Nash},
  {Neilsen}, {Newberg}, {Newman}, {Nichol}, {Nicinski}, {Nieto-Santisteban},
  {Nitta}, {Okamura}, {O'Mullane}, {Ostriker}, {Owen}, {Padmanabhan},
  {Peoples}, {Pier}, {Pope}, {Quinn}, {Richards}, {Richmond}, {Rix}, {Rockosi},
  {Schlegel}, {Schneider}, {Scranton}, {Sekiguchi}, {Seljak}, {Sergey},
  {Sesar}, {Sheldon}, {Shimasaku}, {Siegmund}, {Silvestri}, {Smith}, {Smol{\v
  c}i{\'c}}, {Snedden}, {Stebbins}, {Stoughton}, {Strauss}, {SubbaRao},
  {Szalay}, {Szapudi}, {Szkody}, {Szokoly}, {Tegmark}, {Teodoro}, {Thakar},
  {Tremonti}, {Tucker}, {Uomoto}, {Vanden Berk}, {Vandenberg}, {Vogeley},
  {Voges}, {Vogt}, {Walkowicz}, {Wang}, {Weinberg}, {West}, {White}, {Wilhite},
  {Xu}, {Yanny}, {Yasuda}, {Yip}, {Yocum}, {York}, {Zehavi}, {Zibetti}, \&
  {Zucker}}]{2004AJ....128..502A}
{Abazajian}, K., {Adelman-McCarthy}, J.~K., {Ag{\"u}eros}, M.~A., {et~al.}
  2004, \aj, 128, 502

\bibitem[{{Adelberger} \& {Steidel}(2000)}]{2000ApJ...544..218A}
{Adelberger}, K.~L., \& {Steidel}, C.~C. 2000, \apj, 544, 218

\bibitem[{{Aihara} {et~al.}(2017{\natexlab{a}}){Aihara}, {Armstrong},
  {Bickerton}, {Bosch}, {Coupon}, {Furusawa}, {Hayashi}, {Ikeda}, {Kamata},
  {Karoji}, {Kawanomoto}, {Koike}, {Komiyama}, {Lupton}, {Mineo}, {Miyatake},
  {Miyazaki}, {Morokuma}, {Obuchi}, {Oishi}, {Okura}, {Price}, {Takata},
  {Tanaka}, {Tanaka}, {Tanaka}, {Uchida}, {Uraguchi}, {Utsumi}, {Wang},
  {Yamada}, {Yamanoi}, {Yasuda}, {Arimoto}, {Chiba}, {Finet}, {Fujimori},
  {Fujimoto}, {Furusawa}, {Goto}, {Goulding}, {Gunn}, {Harikane}, {Hattori},
  {Hayashi}, {Helminiak}, {Higuchi}, {Hikage}, {Ho}, {Hsieh}, {Huang}, {Huang},
  {Imanishi}, {Iwata}, {Jaelani}, {Jian}, {Kashikawa}, {Katayama}, {Kojima},
  {Konno}, {Koshida}, {Leauthaud}, {Lee}, {Lin}, {Lin}, {Mandelbaum},
  {Matsuoka}, {Medezinski}, {Miyama}, {Momose}, {More}, {More}, {Mukae},
  {Murata}, {Murayama}, {Nagao}, {Nakata}, {Niikura}, {Nishizawa}, {Oguri},
  {Okabe}, {Ono}, {Onodera}, {Onoue}, {Ouchi}, {Pyo}, {Shibuya}, {Shimasaku},
  {Simet}, {Speagle}, {Spergel}, {Strauss}, {Sugahara}, {Sugiyama}, {Suto},
  {Suzuki}, {Tait}, {Takada}, {Terai}, {Toba}, {Turner}, {Uchiyama}, {Umetsu},
  {Urata}, {Usuda}, {Yeh}, \& {Yuma}}]{2017arXiv170208449A}
{Aihara}, H., {Armstrong}, R., {Bickerton}, S., {et~al.} 2017{\natexlab{a}},
  ArXiv e-prints, arXiv:1702.08449

\bibitem[{{Aihara} {et~al.}(2017{\natexlab{b}}){Aihara}, {Arimoto},
  {Armstrong}, {Arnouts}, {Bahcall}, {Bickerton}, {Bosch}, {Bundy}, {Capak},
  {Chan}, {Chiba}, {Coupon}, {Egami}, {Enoki}, {Finet}, {Fujimori}, {Fujimoto},
  {Furusawa}, {Furusawa}, {Goto}, {Goulding}, {Greco}, {Greene}, {Gunn},
  {Hamana}, {Harikane}, {Hashimoto}, {Hattori}, {Hayashi}, {Hayashi},
  {He{\l}miniak}, {Higuchi}, {Hikage}, {Ho}, {Hsieh}, {Huang}, {Huang},
  {Ikeda}, {Imanishi}, {Inoue}, {Iwasawa}, {Iwata}, {Jaelani}, {Jian},
  {Kamata}, {Karoji}, {Kashikawa}, {Katayama}, {Kawanomoto}, {Kayo}, {Koda},
  {Koike}, {Kojima}, {Komiyama}, {Konno}, {Koshida}, {Koyama}, {Kusakabe},
  {Leauthaud}, {Lee}, {Lin}, {Lin}, {Lupton}, {Mandelbaum}, {Matsuoka},
  {Medezinski}, {Mineo}, {Miyama}, {Miyatake}, {Miyazaki}, {Momose}, {More},
  {More}, {Moritani}, {Moriya}, {Morokuma}, {Mukae}, {Murata}, {Murayama},
  {Nagao}, {Nakata}, {Niida}, {Niikura}, {Nishizawa}, {Obuchi}, {Oguri},
  {Oishi}, {Okabe}, {Okura}, {Ono}, {Onodera}, {Onoue}, {Osato}, {Ouchi},
  {Price}, {Pyo}, {Sako}, {Okamoto}, {Sawicki}, {Shibuya}, {Shimasaku},
  {Shimono}, {Shirasaki}, {Silverman}, {Simet}, {Speagle}, {Spergel},
  {Strauss}, {Sugahara}, {Sugiyama}, {Suto}, {Suyu}, {Suzuki}, {Tait},
  {Takata}, {Takada}, {Tamura}, {Tanaka}, {Tanaka}, {Tanaka}, {Tanaka},
  {Terai}, {Terashima}, {Toba}, {Toshikawa}, {Turner}, {Uchida}, {Uchiyama},
  {Umetsu}, {Uraguchi}, {Urata}, {Usuda}, {Utsumi}, {Wang}, {Wang}, {Wong},
  {Yabe}, {Yamada}, {Yamanoi}, {Yasuda}, {Yeh}, {Yonehara}, \&
  {Yuma}}]{2017arXiv170405858A}
{Aihara}, H., {Arimoto}, N., {Armstrong}, R., {et~al.} 2017{\natexlab{b}},
  ArXiv e-prints, arXiv:1704.05858

\bibitem[{{Akiyama} {et~al.}(2017){Akiyama}, {He}, {Ikeda}, {Niida}, {Nagao},
  {Bosch}, {Coupon}, {Enoki}, {Imanishi}, {Kashikawa}, {Kawaguchi}, {Komiyama},
  {Lee}, {Matsuoka}, {Miyazaki}, {Nishizawa}, {Oguri}, {Ono}, {Onoue}, {Ouchi},
  {Schulze}, {Silverman}, {Tanaka}, {Tanaka}, {Terashima}, {Toba}, \&
  {Ueda}}]{2017arXiv170405996A}
{Akiyama}, M., {He}, W., {Ikeda}, H., {et~al.} 2017, ArXiv e-prints,
  arXiv:1704.05996

\bibitem[{{Atek} {et~al.}(2015){Atek}, {Richard}, {Jauzac}, {Kneib},
  {Natarajan}, {Limousin}, {Schaerer}, {Jullo}, {Ebeling}, {Egami}, \&
  {Clement}}]{2015ApJ...814...69A}
{Atek}, H., {Richard}, J., {Jauzac}, M., {et~al.} 2015, \apj, 814, 69

\bibitem[{{Axelrod} {et~al.}(2010){Axelrod}, {Kantor}, {Lupton}, \&
  {Pierfederici}}]{2010SPIE.7740E..15A}
{Axelrod}, T., {Kantor}, J., {Lupton}, R.~H., \& {Pierfederici}, F. 2010, in
  \procspie, Vol. 7740, Software and Cyberinfrastructure for Astronomy, 774015

\bibitem[{{Ba{\~n}ados} {et~al.}(2016){Ba{\~n}ados}, {Venemans}, {Decarli},
  {Farina}, {Mazzucchelli}, {Walter}, {Fan}, {Stern}, {Schlafly}, {Chambers},
  {Rix}, {Jiang}, {McGreer}, {Simcoe}, {Wang}, {Yang}, {Morganson}, {De Rosa},
  {Greiner}, {Balokovi{\'c}}, {Burgett}, {Cooper}, {Draper}, {Flewelling},
  {Hodapp}, {Jun}, {Kaiser}, {Kudritzki}, {Magnier}, {Metcalfe}, {Miller},
  {Schindler}, {Tonry}, {Wainscoat}, {Waters}, \& {Yang}}]{2016ApJS..227...11B}
{Ba{\~n}ados}, E., {Venemans}, B.~P., {Decarli}, R., {et~al.} 2016, \apjs, 227,
  11

\bibitem[{{Barone-Nugent} {et~al.}(2015){Barone-Nugent}, {Wyithe}, {Trenti},
  {Treu}, {Oesch}, {Bouwens}, {Illingworth}, \&
  {Schmidt}}]{2015MNRAS.450.1224B}
{Barone-Nugent}, R.~L., {Wyithe}, J.~S.~B., {Trenti}, M., {et~al.} 2015,
  \mnras, 450, 1224

\bibitem[{{Benson} {et~al.}(2003){Benson}, {Bower}, {Frenk}, {Lacey}, {Baugh},
  \& {Cole}}]{2003ApJ...599...38B}
{Benson}, A.~J., {Bower}, R.~G., {Frenk}, C.~S., {et~al.} 2003, \apj, 599, 38

\bibitem[{{Bian} {et~al.}(2013){Bian}, {Fan}, {Jiang}, {McGreer}, {Dey},
  {Green}, {Maiolino}, {Walter}, {Lee}, \& {Dav{\'e}}}]{2013ApJ...774...28B}
{Bian}, F., {Fan}, X., {Jiang}, L., {et~al.} 2013, \apj, 774, 28

\bibitem[{{Binney}(1977)}]{1977ApJ...215..483B}
{Binney}, J. 1977, \apj, 215, 483

\bibitem[{{Binney}(2004)}]{2004MNRAS.347.1093B}
---. 2004, \mnras, 347, 1093

\bibitem[{{Bosch} {et~al.}(2017){Bosch}, {Armstrong}, {Bickerton}, {Furusawa},
  {Ikeda}, {Koike}, {Lupton}, {Mineo}, {Price}, {Takata}, {Tanaka}, {Yasuda},
  {AlSayyad}, {Becker}, {Coulton}, {Coupon}, {Garmilla}, {Huang}, {Krughoff},
  {Lang}, {Leauthaud}, {Lim}, {Lust}, {MacArthur}, {Mandelbaum}, {Miyatake},
  {Miyazaki}, {Murata}, {More}, {Okura}, {Owen}, {Swinbank}, {Strauss},
  {Yamada}, \& {Yamanoi}}]{2017arXiv170506766B}
{Bosch}, J., {Armstrong}, R., {Bickerton}, S., {et~al.} 2017, ArXiv e-prints,
  arXiv:1705.06766

\bibitem[{{Bouwens} {et~al.}(2008){Bouwens}, {Illingworth}, {Franx}, \&
  {Ford}}]{2008ApJ...686..230B}
{Bouwens}, R.~J., {Illingworth}, G.~D., {Franx}, M., \& {Ford}, H. 2008, \apj,
  686, 230

\bibitem[{{Bouwens} {et~al.}(2017{\natexlab{a}}){Bouwens}, {Illingworth},
  {Oesch}, {Atek}, {Lam}, \& {Stefanon}}]{2017ApJ...843...41B}
{Bouwens}, R.~J., {Illingworth}, G.~D., {Oesch}, P.~A., {et~al.}
  2017{\natexlab{a}}, \apj, 843, 41

\bibitem[{{Bouwens} {et~al.}(2017{\natexlab{b}}){Bouwens}, {Oesch},
  {Illingworth}, {Ellis}, \& {Stefanon}}]{2017ApJ...843..129B}
{Bouwens}, R.~J., {Oesch}, P.~A., {Illingworth}, G.~D., {Ellis}, R.~S., \&
  {Stefanon}, M. 2017{\natexlab{b}}, \apj, 843, 129

\bibitem[{{Bouwens} {et~al.}(2014){Bouwens}, {Illingworth}, {Oesch},
  {Labb{\'e}}, {van Dokkum}, {Trenti}, {Franx}, {Smit}, {Gonzalez}, \&
  {Magee}}]{2014ApJ...793..115B}
{Bouwens}, R.~J., {Illingworth}, G.~D., {Oesch}, P.~A., {et~al.} 2014, \apj,
  793, 115

\bibitem[{{Bouwens} {et~al.}(2015){Bouwens}, {Illingworth}, {Oesch}, {Trenti},
  {Labb{\'e}}, {Bradley}, {Carollo}, {van Dokkum}, {Gonzalez}, {Holwerda},
  {Franx}, {Spitler}, {Smit}, \& {Magee}}]{2015ApJ...803...34B}
---. 2015, \apj, 803, 34

\bibitem[{{Bower} {et~al.}(2006){Bower}, {Benson}, {Malbon}, {Helly}, {Frenk},
  {Baugh}, {Cole}, \& {Lacey}}]{2006MNRAS.370..645B}
{Bower}, R.~G., {Benson}, A.~J., {Malbon}, R., {et~al.} 2006, \mnras, 370, 645

\bibitem[{{Bowler} {et~al.}(2017){Bowler}, {Dunlop}, {McLure}, \&
  {McLeod}}]{2017MNRAS.466.3612B}
{Bowler}, R.~A.~A., {Dunlop}, J.~S., {McLure}, R.~J., \& {McLeod}, D.~J. 2017,
  \mnras, 466, 3612

\bibitem[{{Bowler} {et~al.}(2012){Bowler}, {Dunlop}, {McLure}, {McCracken},
  {Milvang-Jensen}, {Furusawa}, {Fynbo}, {Le F{\`e}vre}, {Holt}, {Ideue},
  {Ihara}, {Rogers}, \& {Taniguchi}}]{2012MNRAS.426.2772B}
{Bowler}, R.~A.~A., {Dunlop}, J.~S., {McLure}, R.~J., {et~al.} 2012, \mnras,
  426, 2772

\bibitem[{{Bowler} {et~al.}(2015){Bowler}, {Dunlop}, {McLure}, {McCracken},
  {Milvang-Jensen}, {Furusawa}, {Taniguchi}, {Le F{\`e}vre}, {Fynbo}, {Jarvis},
  \& {H{\"a}u{\ss}ler}}]{2015MNRAS.452.1817B}
---. 2015, \mnras, 452, 1817

\bibitem[{{Bruzual} \& {Charlot}(2003)}]{2003MNRAS.344.1000B}
{Bruzual}, G., \& {Charlot}, S. 2003, \mnras, 344, 1000

\bibitem[{{Calzetti} {et~al.}(2000){Calzetti}, {Armus}, {Bohlin}, {Kinney},
  {Koornneef}, \& {Storchi-Bergmann}}]{2000ApJ...533..682C}
{Calzetti}, D., {Armus}, L., {Bohlin}, R.~C., {et~al.} 2000, \apj, 533, 682

\bibitem[{{Capak} {et~al.}(2007){Capak}, {Aussel}, {Ajiki}, {McCracken},
  {Mobasher}, {Scoville}, {Shopbell}, {Taniguchi}, {Thompson}, {Tribiano},
  {Sasaki}, {Blain}, {Brusa}, {Carilli}, {Comastri}, {Carollo}, {Cassata},
  {Colbert}, {Ellis}, {Elvis}, {Giavalisco}, {Green}, {Guzzo}, {Hasinger},
  {Ilbert}, {Impey}, {Jahnke}, {Kartaltepe}, {Kneib}, {Koda}, {Koekemoer},
  {Komiyama}, {Leauthaud}, {Le Fevre}, {Lilly}, {Liu}, {Massey}, {Miyazaki},
  {Murayama}, {Nagao}, {Peacock}, {Pickles}, {Porciani}, {Renzini}, {Rhodes},
  {Rich}, {Salvato}, {Sanders}, {Scarlata}, {Schiminovich}, {Schinnerer},
  {Scodeggio}, {Sheth}, {Shioya}, {Tasca}, {Taylor}, {Yan}, \&
  {Zamorani}}]{2007ApJS..172...99C}
{Capak}, P., {Aussel}, H., {Ajiki}, M., {et~al.} 2007, \apjs, 172, 99

\bibitem[{{Castellano} {et~al.}(2010){Castellano}, {Fontana}, {Boutsia},
  {Grazian}, {Pentericci}, {Bouwens}, {Dickinson}, {Giavalisco}, {Santini},
  {Cristiani}, {Fiore}, {Gallozzi}, {Giallongo}, {Maiolino}, {Mannucci},
  {Menci}, {Moorwood}, {Nonino}, {Paris}, {Renzini}, {Rosati}, {Salimbeni},
  {Testa}, \& {Vanzella}}]{2010A&A...511A..20C}
{Castellano}, M., {Fontana}, A., {Boutsia}, K., {et~al.} 2010, \aap, 511, A20

\bibitem[{{Castellano} {et~al.}(2016){Castellano}, {Yue}, {Ferrara}, {Merlin},
  {Fontana}, {Amor{\'{\i}}n}, {Grazian}, {M{\'a}rmol-Queralto},
  {Micha{\l}owski}, {Mortlock}, {Paris}, {Parsa}, {Pilo}, \&
  {Santini}}]{2016ApJ...823L..40C}
{Castellano}, M., {Yue}, B., {Ferrara}, A., {et~al.} 2016, \apjl, 823, L40

\bibitem[{{Coleman} {et~al.}(1980){Coleman}, {Wu}, \&
  {Weedman}}]{1980ApJS...43..393C}
{Coleman}, G.~D., {Wu}, C.-C., \& {Weedman}, D.~W. 1980, \apjs, 43, 393

\bibitem[{{Coupon} {et~al.}(2017){Coupon}, {Czakon}, {Bosch}, {Komiyama},
  {Medezinski}, {Miyazaki}, \& {Oguri}}]{2017arXiv170500622C}
{Coupon}, J., {Czakon}, N., {Bosch}, J., {et~al.} 2017, ArXiv e-prints,
  arXiv:1705.00622

\bibitem[{{Croton} {et~al.}(2006){Croton}, {Springel}, {White}, {De Lucia},
  {Frenk}, {Gao}, {Jenkins}, {Kauffmann}, {Navarro}, \&
  {Yoshida}}]{2006MNRAS.365...11C}
{Croton}, D.~J., {Springel}, V., {White}, S.~D.~M., {et~al.} 2006, \mnras, 365,
  11

\bibitem[{{Curtis-Lake} {et~al.}(2012){Curtis-Lake}, {McLure}, {Pearce},
  {Dunlop}, {Cirasuolo}, {Stark}, {Almaini}, {Bradshaw}, {Chuter}, {Foucaud},
  \& {Hartley}}]{2012MNRAS.422.1425C}
{Curtis-Lake}, E., {McLure}, R.~J., {Pearce}, H.~J., {et~al.} 2012, \mnras,
  422, 1425

\bibitem[{{Curtis-Lake} {et~al.}(2016){Curtis-Lake}, {McLure}, {Dunlop},
  {Rogers}, {Targett}, {Dekel}, {Ellis}, {Faber}, {Ferguson}, {Grogin},
  {Kocevski}, {Koekemoer}, {Lai}, {M{\'a}rmol-Queralt{\'o}}, \&
  {Robertson}}]{2016MNRAS.457..440C}
{Curtis-Lake}, E., {McLure}, R.~J., {Dunlop}, J.~S., {et~al.} 2016, \mnras,
  457, 440

\bibitem[{{Dressler} {et~al.}(2011){Dressler}, {Bigelow}, {Hare}, {Sutin},
  {Thompson}, {Burley}, {Epps}, {Oemler}, {Bagish}, {Birk}, {Clardy},
  {Gunnels}, {Kelson}, {Shectman}, \& {Osip}}]{2011PASP..123..288D}
{Dressler}, A., {Bigelow}, B., {Hare}, T., {et~al.} 2011, \pasp, 123, 288

\bibitem[{{Eddington}(1913)}]{1913MNRAS..73..359E}
{Eddington}, A.~S. 1913, \mnras, 73, 359

\bibitem[{{Ellis} {et~al.}(2013){Ellis}, {McLure}, {Dunlop}, {Robertson},
  {Ono}, {Schenker}, {Koekemoer}, {Bowler}, {Ouchi}, {Rogers}, {Curtis-Lake},
  {Schneider}, {Charlot}, {Stark}, {Furlanetto}, \&
  {Cirasuolo}}]{2013ApJ...763L...7E}
{Ellis}, R.~S., {McLure}, R.~J., {Dunlop}, J.~S., {et~al.} 2013, \apjl, 763, L7

\bibitem[{{Faber} \& {Jackson}(1976)}]{1976ApJ...204..668F}
{Faber}, S.~M., \& {Jackson}, R.~E. 1976, \apj, 204, 668

\bibitem[{{Finkelstein} {et~al.}(2015){Finkelstein}, {Ryan}, {Papovich},
  {Dickinson}, {Song}, {Somerville}, {Ferguson}, {Salmon}, {Giavalisco},
  {Koekemoer}, {Ashby}, {Behroozi}, {Castellano}, {Dunlop}, {Faber}, {Fazio},
  {Fontana}, {Grogin}, {Hathi}, {Jaacks}, {Kocevski}, {Livermore}, {McLure},
  {Merlin}, {Mobasher}, {Newman}, {Rafelski}, {Tilvi}, \&
  {Willner}}]{2015ApJ...810...71F}
{Finkelstein}, S.~L., {Ryan}, Jr., R.~E., {Papovich}, C., {et~al.} 2015, \apj,
  810, 71

\bibitem[{{Finlator} {et~al.}(2017){Finlator}, {Prescott}, {Oppenheimer},
  {Dav{\'e}}, {Zackrisson}, {Livermore}, {Finkelstein}, {Thompson}, \&
  {Huang}}]{2017MNRAS.464.1633F}
{Finlator}, K., {Prescott}, M.~K.~M., {Oppenheimer}, B.~D., {et~al.} 2017,
  \mnras, 464, 1633

\bibitem[{{Furusawa} {et~al.}(2008){Furusawa}, {Kosugi}, {Akiyama}, {Takata},
  {Sekiguchi}, {Tanaka}, {Iwata}, {Kajisawa}, {Yasuda}, {Doi}, {Ouchi},
  {Simpson}, {Shimasaku}, {Yamada}, {Furusawa}, {Morokuma}, {Ishida}, {Aoki},
  {Fuse}, {Imanishi}, {Iye}, {Karoji}, {Kobayashi}, {Kodama}, {Komiyama},
  {Maeda}, {Miyazaki}, {Mizumoto}, {Nakata}, {Noumaru}, {Ogasawara}, {Okamura},
  {Saito}, {Sasaki}, {Ueda}, \& {Yoshida}}]{2008ApJS..176....1F}
{Furusawa}, H., {Kosugi}, G., {Akiyama}, M., {et~al.} 2008, \apjs, 176, 1

\bibitem[{{Furusawa} {et~al.}(2017)}]{furusawa2017} 
{Furusawa}, H., et al. 2017, submitted to PASJ

\bibitem[{{Gehrels}(1986)}]{1986ApJ...303..336G}
{Gehrels}, N. 1986, \apj, 303, 336

\bibitem[{{Giallongo} {et~al.}(2015){Giallongo}, {Grazian}, {Fiore}, {Fontana},
  {Pentericci}, {Vanzella}, {Dickinson}, {Kocevski}, {Castellano}, {Cristiani},
  {Ferguson}, {Finkelstein}, {Grogin}, {Hathi}, {Koekemoer}, {Newman}, \&
  {Salvato}}]{2015A&A...578A..83G}
{Giallongo}, E., {Grazian}, A., {Fiore}, F., {et~al.} 2015, \aap, 578, A83

\bibitem[{{Giavalisco}(2002)}]{2002ARA&A..40..579G}
{Giavalisco}, M. 2002, \araa, 40, 579

\bibitem[{{Glikman} {et~al.}(2011){Glikman}, {Djorgovski}, {Stern}, {Dey},
  {Jannuzi}, \& {Lee}}]{2011ApJ...728L..26G}
{Glikman}, E., {Djorgovski}, S.~G., {Stern}, D., {et~al.} 2011, \apjl, 728, L26

\bibitem[{{Gnedin}(2016)}]{2016ApJ...825L..17G}
{Gnedin}, N.~Y. 2016, \apjl, 825, L17

\bibitem[{{Granato} {et~al.}(2004){Granato}, {De Zotti}, {Silva}, {Bressan}, \&
  {Danese}}]{2004ApJ...600..580G}
{Granato}, G.~L., {De Zotti}, G., {Silva}, L., {Bressan}, A., \& {Danese}, L.
  2004, \apj, 600, 580

\bibitem[{{Gunn} \& {Stryker}(1983)}]{1983ApJS...52..121G}
{Gunn}, J.~E., \& {Stryker}, L.~L. 1983, \apjs, 52, 121

\bibitem[{{Harikane} {et~al.}(2017){Harikane}, {Ouchi}, {Ono}, {Saito},
  {Behroozi}, {More}, {Shimasaku}, {Toshikawa}, {Lin}, {Akiyama}, {Coupon},
  {Komiyama}, {Konno}, {Lin}, {Miyazaki}, {Nishizawa}, {Shibuya}, \&
  {Silverman}}]{2017arXiv170406535H}
{Harikane}, Y., {Ouchi}, M., {Ono}, Y., {et~al.} 2017, ArXiv e-prints,
  arXiv:1704.06535

\bibitem[{{Hildebrandt} {et~al.}(2009){Hildebrandt}, {Pielorz}, {Erben}, {van
  Waerbeke}, {Simon}, \& {Capak}}]{2009A&A...498..725H}
{Hildebrandt}, H., {Pielorz}, J., {Erben}, T., {et~al.} 2009, \aap, 498, 725

\bibitem[{{Hogg}(1999)}]{1999astro.ph..5116H}
{Hogg}, D.~W. 1999, ArXiv Astrophysics e-prints, astro-ph/9905116

\bibitem[{{Hook} {et~al.}(2004){Hook}, {J{\o}rgensen}, {Allington-Smith},
  {Davies}, {Metcalfe}, {Murowinski}, \& {Crampton}}]{2004PASP..116..425H}
{Hook}, I.~M., {J{\o}rgensen}, I., {Allington-Smith}, J.~R., {et~al.} 2004,
  \pasp, 116, 425

\bibitem[{{Huang} {et~al.}(2017){Huang}, {Leauthaud}, {Murata}, {Bosch},
  {Price}, {Lupton}, {Mandelbaum}, {Lackner}, {Bickerton}, {Miyazaki},
  {Coupon}, \& {Tanaka}}]{2017arXiv170501599H}
{Huang}, S., {Leauthaud}, A., {Murata}, R., {et~al.} 2017, ArXiv e-prints,
  arXiv:1705.01599

\bibitem[{{Ikeda} {et~al.}(2012){Ikeda}, {Nagao}, {Matsuoka}, {Taniguchi},
  {Shioya}, {Kajisawa}, {Enoki}, {Capak}, {Civano}, {Koekemoer}, {Masters},
  {Morokuma}, {Salvato}, {Schinnerer}, \& {Scoville}}]{2012ApJ...756..160I}
{Ikeda}, H., {Nagao}, T., {Matsuoka}, K., {et~al.} 2012, \apj, 756, 160

\bibitem[{{Ishigaki} {et~al.}(2017){Ishigaki}, {Kawamata}, {Ouchi}, {Oguri}, \&
  {Shimasaku}}]{2017arXiv170204867I}
{Ishigaki}, M., {Kawamata}, R., {Ouchi}, M., {Oguri}, M., \& {Shimasaku}, K.
  2017, ArXiv e-prints, arXiv:1702.04867

\bibitem[{{Ishigaki} {et~al.}(2015){Ishigaki}, {Kawamata}, {Ouchi}, {Oguri},
  {Shimasaku}, \& {Ono}}]{2015ApJ...799...12I}
{Ishigaki}, M., {Kawamata}, R., {Ouchi}, M., {et~al.} 2015, \apj, 799, 12

\bibitem[{{Ivezic} {et~al.}(2008){Ivezic}, {Tyson}, {Abel}, {Acosta},
  {Allsman}, {AlSayyad}, {Anderson}, {Andrew}, {Angel}, {Angeli}, {Ansari},
  {Antilogus}, {Arndt}, {Astier}, {Aubourg}, {Axelrod}, {Bard}, {Barr},
  {Barrau}, {Bartlett}, {Bauman}, {Beaumont}, {Becker}, {Becla}, {Beldica},
  {Bellavia}, {Blanc}, {Blandford}, {Bloom}, {Bogart}, {Borne}, {Bosch},
  {Boutigny}, {Brandt}, {Brown}, {Bullock}, {Burchat}, {Burke}, {Cagnoli},
  {Calabrese}, {Chandrasekharan}, {Chesley}, {Cheu}, {Chiang}, {Claver},
  {Connolly}, {Cook}, {Cooray}, {Covey}, {Cribbs}, {Cui}, {Cutri}, {Daubard},
  {Daues}, {Delgado}, {Digel}, {Doherty}, {Dubois}, {Dubois-Felsmann},
  {Durech}, {Eracleous}, {Ferguson}, {Frank}, {Freemon}, {Gangler}, {Gawiser},
  {Geary}, {Gee}, {Geha}, {Gibson}, {Gilmore}, {Glanzman}, {Goodenow},
  {Gressler}, {Gris}, {Guyonnet}, {Hascall}, {Haupt}, {Hernandez}, {Hogan},
  {Huang}, {Huffer}, {Innes}, {Jacoby}, {Jain}, {Jee}, {Jernigan},
  {Jevremovic}, {Johns}, {Jones}, {Juramy-Gilles}, {Juric}, {Kahn}, {Kalirai},
  {Kallivayalil}, {Kalmbach}, {Kantor}, {Kasliwal}, {Kessler}, {Kirkby},
  {Knox}, {Kotov}, {Krabbendam}, {Krughoff}, {Kubanek}, {Kuczewski},
  {Kulkarni}, {Lambert}, {Le Guillou}, {Levine}, {Liang}, {Lim}, {Lintott},
  {Lupton}, {Mahabal}, {Marshall}, {Marshall}, {May}, {McKercher}, {Migliore},
  {Miller}, {Mills}, {Monet}, {Moniez}, {Neill}, {Nief}, {Nomerotski},
  {Nordby}, {O'Connor}, {Oliver}, {Olivier}, {Olsen}, {Ortiz}, {Owen}, {Pain},
  {Peterson}, {Petry}, {Pierfederici}, {Pietrowicz}, {Pike}, {Pinto}, {Plante},
  {Plate}, {Price}, {Prouza}, {Radeka}, {Rajagopal}, {Rasmussen}, {Regnault},
  {Ridgway}, {Ritz}, {Rosing}, {Roucelle}, {Rumore}, {Russo}, {Saha},
  {Sassolas}, {Schalk}, {Schindler}, {Schneider}, {Schumacher}, {Sebag},
  {Sembroski}, {Seppala}, {Shipsey}, {Silvestri}, {Smith}, {Smith}, {Strauss},
  {Stubbs}, {Sweeney}, {Szalay}, {Takacs}, {Thaler}, {Van Berg}, {Vanden Berk},
  {Vetter}, {Virieux}, {Xin}, {Walkowicz}, {Walter}, {Wang}, {Warner},
  {Willman}, {Wittman}, {Wolff}, {Wood-Vasey}, {Yoachim}, {Zhan}, \& {for the
  LSST Collaboration}}]{2008arXiv0805.2366I}
{Ivezic}, Z., {Tyson}, J.~A., {Abel}, B., {et~al.} 2008, ArXiv e-prints,
  arXiv:0805.2366

\bibitem[{{Iwata} {et~al.}(2007){Iwata}, {Ohta}, {Tamura}, {Akiyama}, {Aoki},
  {Ando}, {Kiuchi}, \& {Sawicki}}]{2007MNRAS.376.1557I}
{Iwata}, I., {Ohta}, K., {Tamura}, N., {et~al.} 2007, \mnras, 376, 1557

\bibitem[{{Jaacks} {et~al.}(2013){Jaacks}, {Thompson}, \&
  {Nagamine}}]{2013ApJ...766...94J}
{Jaacks}, J., {Thompson}, R., \& {Nagamine}, K. 2013, \apj, 766, 94

\bibitem[{{Jiang} {et~al.}(2016){Jiang}, {McGreer}, {Fan}, {Strauss},
  {Ba{\~n}ados}, {Becker}, {Bian}, {Farnsworth}, {Shen}, {Wang}, {Wang},
  {Wang}, {White}, {Wu}, {Wu}, {Yang}, \& {Yang}}]{2016ApJ...833..222J}
{Jiang}, L., {McGreer}, I.~D., {Fan}, X., {et~al.} 2016, \apj, 833, 222

\bibitem[{{Juri{\'c}} {et~al.}(2015){Juri{\'c}}, {Kantor}, {Lim}, {Lupton},
  {Dubois-Felsmann}, {Jenness}, {Axelrod}, {Aleksi{\'c}}, {Allsman},
  {AlSayyad}, {Alt}, {Armstrong}, {Basney}, {Becker}, {Becla}, {Bickerton},
  {Biswas}, {Bosch}, {Boutigny}, {Carrasco Kind}, {Ciardi}, {Connolly},
  {Daniel}, {Daues}, {Economou}, {Chiang}, {Fausti}, {Fisher-Levine},
  {Freemon}, {Gee}, {Gris}, {Hernandez}, {Hoblitt}, {Ivezi{\'c}}, {Jammes},
  {Jevremovi{\'c}}, {Jones}, {Bryce Kalmbach}, {Kasliwal}, {Krughoff}, {Lang},
  {Lurie}, {Lust}, {Mullally}, {MacArthur}, {Melchior}, {Moeyens}, {Nidever},
  {Owen}, {Parejko}, {Peterson}, {Petravick}, {Pietrowicz}, {Price}, {Reiss},
  {Shaw}, {Sick}, {Slater}, {Strauss}, {Sullivan}, {Swinbank}, {Van Dyk},
  {Vuj{\v c}i{\'c}}, {Withers}, {Yoachim}, \& {LSST
  Project}}]{2015arXiv151207914J}
{Juri{\'c}}, M., {Kantor}, J., {Lim}, K., {et~al.} 2015, ArXiv e-prints,
  arXiv:1512.07914

\bibitem[{{Kashikawa} {et~al.}(2002){Kashikawa}, {Aoki}, {Asai}, {Ebizuka},
  {Inata}, {Iye}, {Kawabata}, {Kosugi}, {Ohyama}, {Okita}, {Ozawa}, {Saito},
  {Sasaki}, {Sekiguchi}, {Shimizu}, {Taguchi}, {Takata}, {Yadoumaru}, \&
  {Yoshida}}]{2002PASJ...54..819K}
{Kashikawa}, N., {Aoki}, K., {Asai}, R., {et~al.} 2002, \pasj, 54, 819

\bibitem[{{Kashikawa} {et~al.}(2006){Kashikawa}, {Shimasaku}, {Malkan}, {Doi},
  {Matsuda}, {Ouchi}, {Taniguchi}, {Ly}, {Nagao}, {Iye}, {Motohara},
  {Murayama}, {Murozono}, {Nariai}, {Ohta}, {Okamura}, {Sasaki}, {Shioya}, \&
  {Umemura}}]{2006ApJ...648....7K}
{Kashikawa}, N., {Shimasaku}, K., {Malkan}, M.~A., {et~al.} 2006, \apj, 648, 7

\bibitem[{{Kashikawa} {et~al.}(2015){Kashikawa}, {Ishizaki}, {Willott},
  {Onoue}, {Im}, {Furusawa}, {Toshikawa}, {Ishikawa}, {Niino}, {Shimasaku},
  {Ouchi}, \& {Hibon}}]{2015ApJ...798...28K}
{Kashikawa}, N., {Ishizaki}, Y., {Willott}, C.~J., {et~al.} 2015, \apj, 798, 28

\bibitem[{{Kawamata} {et~al.}(2015){Kawamata}, {Ishigaki}, {Shimasaku},
  {Oguri}, \& {Ouchi}}]{2015ApJ...804..103K}
{Kawamata}, R., {Ishigaki}, M., {Shimasaku}, K., {Oguri}, M., \& {Ouchi}, M.
  2015, \apj, 804, 103

\bibitem[{{Kawamata} {et~al.}(2016){Kawamata}, {Oguri}, {Ishigaki},
  {Shimasaku}, \& {Ouchi}}]{2016ApJ...819..114K}
{Kawamata}, R., {Oguri}, M., {Ishigaki}, M., {Shimasaku}, K., \& {Ouchi}, M.
  2016, \apj, 819, 114

\bibitem[{{Kawanomoto} {et~al.}(2017)}]{kawanomoto2017} 
{Kawanomoto}, S., et al. 2017, to be submitted to PASJ

\bibitem[{{Kelvin} {et~al.}(2014){Kelvin}, {Driver}, {Robotham}, {Graham},
  {Phillipps}, {Agius}, {Alpaslan}, {Baldry}, {Bamford}, {Bland-Hawthorn},
  {Brough}, {Brown}, {Colless}, {Conselice}, {Hopkins}, {Liske}, {Loveday},
  {Norberg}, {Pimbblet}, {Popescu}, {Prescott}, {Taylor}, \&
  {Tuffs}}]{2014MNRAS.439.1245K}
{Kelvin}, L.~S., {Driver}, S.~P., {Robotham}, A.~S.~G., {et~al.} 2014, \mnras,
  439, 1245

\bibitem[{{Knapp} {et~al.}(2004){Knapp}, {Leggett}, {Fan}, {Marley}, {Geballe},
  {Golimowski}, {Finkbeiner}, {Gunn}, {Hennawi}, {Ivezi{\'c}}, {Lupton},
  {Schlegel}, {Strauss}, {Tsvetanov}, {Chiu}, {Hoversten}, {Glazebrook},
  {Zheng}, {Hendrickson}, {Williams}, {Uomoto}, {Vrba}, {Henden}, {Luginbuhl},
  {Guetter}, {Munn}, {Canzian}, {Schneider}, \&
  {Brinkmann}}]{2004AJ....127.3553K}
{Knapp}, G.~R., {Leggett}, S.~K., {Fan}, X., {et~al.} 2004, \aj, 127, 3553

\bibitem[{{Komiyama} {et~al.}(2017)}]{komiyama2017} 
{Komiyama}, Y., et al. 2017, submitted to PASJ

\bibitem[{{Konno} {et~al.}(2017){Konno}, {Ouchi}, {Shibuya}, {Ono},
  {Shimasaku}, {Taniguchi}, {Nagao}, {Kobayashi}, {Kajisawa}, {Kashikawa},
  {Inoue}, {Oguri}, {Furusawa}, {Goto}, {Harikane}, {Higuchi}, {Komiyama},
  {Kusakabe}, {Miyazaki}, {Nakajima}, \& {Wang}}]{2017arXiv170501222K}
{Konno}, A., {Ouchi}, M., {Shibuya}, T., {et~al.} 2017, ArXiv e-prints,
  arXiv:1705.01222

\bibitem[{{Kriek} {et~al.}(2015){Kriek}, {Shapley}, {Reddy}, {Siana}, {Coil},
  {Mobasher}, {Freeman}, {de Groot}, {Price}, {Sanders}, {Shivaei}, {Brammer},
  {Momcheva}, {Skelton}, {van Dokkum}, {Whitaker}, {Aird}, {Azadi}, {Kassis},
  {Bullock}, {Conroy}, {Dav{\'e}}, {Kere{\v s}}, \&
  {Krumholz}}]{2015ApJS..218...15K}
{Kriek}, M., {Shapley}, A.~E., {Reddy}, N.~A., {et~al.} 2015, \apjs, 218, 15

\bibitem[{{Krumholz} \& {Dekel}(2012)}]{2012ApJ...753...16K}
{Krumholz}, M.~R., \& {Dekel}, A. 2012, \apj, 753, 16

\bibitem[{{Kuhlen} {et~al.}(2013){Kuhlen}, {Madau}, \&
  {Krumholz}}]{2013ApJ...776...34K}
{Kuhlen}, M., {Madau}, P., \& {Krumholz}, M.~R. 2013, \apj, 776, 34

\bibitem[{{Le F{\`e}vre} {et~al.}(2013){Le F{\`e}vre}, {Cassata}, {Cucciati},
  {Garilli}, {Ilbert}, {Le Brun}, {Maccagni}, {Moreau}, {Scodeggio}, {Tresse},
  {Zamorani}, {Adami}, {Arnouts}, {Bardelli}, {Bolzonella}, {Bondi},
  {Bongiorno}, {Bottini}, {Cappi}, {Charlot}, {Ciliegi}, {Contini}, {de la
  Torre}, {Foucaud}, {Franzetti}, {Gavignaud}, {Guzzo}, {Iovino}, {Lemaux},
  {L{\'o}pez-Sanjuan}, {McCracken}, {Marano}, {Marinoni}, {Mazure}, {Mellier},
  {Merighi}, {Merluzzi}, {Paltani}, {Pell{\`o}}, {Pollo}, {Pozzetti},
  {Scaramella}, {Tasca}, {Vergani}, {Vettolani}, {Zanichelli}, \&
  {Zucca}}]{2013A&A...559A..14L}
{Le F{\`e}vre}, O., {Cassata}, P., {Cucciati}, O., {et~al.} 2013, \aap, 559,
  A14

\bibitem[{{Lilly} {et~al.}(2009){Lilly}, {Le Brun}, {Maier}, {Mainieri},
  {Mignoli}, {Scodeggio}, {Zamorani}, {Carollo}, {Contini}, {Kneib}, {Le
  F{\`e}vre}, {Renzini}, {Bardelli}, {Bolzonella}, {Bongiorno}, {Caputi},
  {Coppa}, {Cucciati}, {de la Torre}, {de Ravel}, {Franzetti}, {Garilli},
  {Iovino}, {Kampczyk}, {Kovac}, {Knobel}, {Lamareille}, {Le Borgne}, {Pello},
  {Peng}, {P{\'e}rez-Montero}, {Ricciardelli}, {Silverman}, {Tanaka}, {Tasca},
  {Tresse}, {Vergani}, {Zucca}, {Ilbert}, {Salvato}, {Oesch}, {Abbas},
  {Bottini}, {Capak}, {Cappi}, {Cassata}, {Cimatti}, {Elvis}, {Fumana},
  {Guzzo}, {Hasinger}, {Koekemoer}, {Leauthaud}, {Maccagni}, {Marinoni},
  {McCracken}, {Memeo}, {Meneux}, {Porciani}, {Pozzetti}, {Sanders},
  {Scaramella}, {Scarlata}, {Scoville}, {Shopbell}, \&
  {Taniguchi}}]{2009ApJS..184..218L}
{Lilly}, S.~J., {Le Brun}, V., {Maier}, C., {et~al.} 2009, \apjs, 184, 218

\bibitem[{{Liu} {et~al.}(2016){Liu}, {Mutch}, {Angel}, {Duffy}, {Geil},
  {Poole}, {Mesinger}, \& {Wyithe}}]{2016MNRAS.462..235L}
{Liu}, C., {Mutch}, S.~J., {Angel}, P.~W., {et~al.} 2016, \mnras, 462, 235

\bibitem[{{Livermore} {et~al.}(2017){Livermore}, {Finkelstein}, \&
  {Lotz}}]{2017ApJ...835..113L}
{Livermore}, R.~C., {Finkelstein}, S.~L., \& {Lotz}, J.~M. 2017, \apj, 835, 113

\bibitem[{{Loveday} {et~al.}(2012){Loveday}, {Norberg}, {Baldry}, {Driver},
  {Hopkins}, {Peacock}, {Bamford}, {Liske}, {Bland-Hawthorn}, {Brough},
  {Brown}, {Cameron}, {Conselice}, {Croom}, {Frenk}, {Gunawardhana}, {Hill},
  {Jones}, {Kelvin}, {Kuijken}, {Nichol}, {Parkinson}, {Phillipps}, {Pimbblet},
  {Popescu}, {Prescott}, {Robotham}, {Sharp}, {Sutherland}, {Taylor}, {Thomas},
  {Tuffs}, {van Kampen}, \& {Wijesinghe}}]{2012MNRAS.420.1239L}
{Loveday}, J., {Norberg}, P., {Baldry}, I.~K., {et~al.} 2012, \mnras, 420, 1239

\bibitem[{{Madau}(1995)}]{1995ApJ...441...18M}
{Madau}, P. 1995, \apj, 441, 18

\bibitem[{{Magnier} {et~al.}(2013){Magnier}, {Schlafly}, {Finkbeiner}, {Juric},
  {Tonry}, {Burgett}, {Chambers}, {Flewelling}, {Kaiser}, {Kudritzki},
  {Morgan}, {Price}, {Sweeney}, \& {Stubbs}}]{2013ApJS..205...20M}
{Magnier}, E.~A., {Schlafly}, E., {Finkbeiner}, D., {et~al.} 2013, \apjs, 205,
  20

\bibitem[{{Mallery} {et~al.}(2012){Mallery}, {Mobasher}, {Capak}, {Kakazu},
  {Masters}, {Ilbert}, {Hemmati}, {Scarlata}, {Salvato}, {McCracken},
  {LeFevre}, \& {Scoville}}]{2012ApJ...760..128M}
{Mallery}, R.~P., {Mobasher}, B., {Capak}, P., {et~al.} 2012, \apj, 760, 128

\bibitem[{{Martin} {et~al.}(2005){Martin}, {Seibert}, {Buat},
  {Iglesias-P{\'a}ramo}, {Barlow}, {Bianchi}, {Byun}, {Donas}, {Forster},
  {Friedman}, {Heckman}, {Jelinsky}, {Lee}, {Madore}, {Malina}, {Milliard},
  {Morrissey}, {Neff}, {Rich}, {Schiminovich}, {Siegmund}, {Small}, {Szalay},
  {Welsh}, \& {Wyder}}]{2005ApJ...619L..59M}
{Martin}, D.~C., {Seibert}, M., {Buat}, V., {et~al.} 2005, \apjl, 619, L59

\bibitem[{{Mason} {et~al.}(2015){Mason}, {Treu}, {Schmidt}, {Collett},
  {Trenti}, {Marshall}, {Barone-Nugent}, {Bradley}, {Stiavelli}, \&
  {Wyithe}}]{2015ApJ...805...79M}
{Mason}, C.~A., {Treu}, T., {Schmidt}, K.~B., {et~al.} 2015, \apj, 805, 79

\bibitem[{{Masters} {et~al.}(2012){Masters}, {Capak}, {Salvato}, {Civano},
  {Mobasher}, {Siana}, {Hasinger}, {Impey}, {Nagao}, {Trump}, {Ikeda}, {Elvis},
  \& {Scoville}}]{2012ApJ...755..169M}
{Masters}, D., {Capak}, P., {Salvato}, M., {et~al.} 2012, \apj, 755, 169

\bibitem[{{Masters} {et~al.}(2017){Masters}, {Stern}, {Cohen}, {Capak},
  {Rhodes}, {Castander}, \& {Paltani}}]{2017ApJ...841..111M}
{Masters}, D.~C., {Stern}, D.~K., {Cohen}, J.~G., {et~al.} 2017, \apj, 841, 111

\bibitem[{{Matsuoka} {et~al.}(2016){Matsuoka}, {Onoue}, {Kashikawa}, {Iwasawa},
  {Strauss}, {Nagao}, {Imanishi}, {Niida}, {Toba}, {Akiyama}, {Asami}, {Bosch},
  {Foucaud}, {Furusawa}, {Goto}, {Gunn}, {Harikane}, {Ikeda}, {Kawaguchi},
  {Kikuta}, {Komiyama}, {Lupton}, {Minezaki}, {Miyazaki}, {Morokuma},
  {Murayama}, {Nishizawa}, {Ono}, {Ouchi}, {Price}, {Sameshima}, {Silverman},
  {Sugiyama}, {Tait}, {Takada}, {Takata}, {Tanaka}, {Tang}, \&
  {Utsumi}}]{2016ApJ...828...26M}
{Matsuoka}, Y., {Onoue}, M., {Kashikawa}, N., {et~al.} 2016, \apj, 828, 26

\bibitem[{{Matsuoka} {et~al.}(2017){Matsuoka}, {Onoue}, {Kashikawa}, {Iwasawa},
  {Strauss}, {Nagao}, {Imanishi}, {Lee}, {Akiyama}, {Asami}, {Bosch},
  {Foucaud}, {Furusawa}, {Goto}, {Gunn}, {Harikane}, {Ikeda}, {Izumi},
  {Kawaguchi}, {Kikuta}, {Kohno}, {Komiyama}, {Lupton}, {Minezaki}, {Miyazaki},
  {Morokuma}, {Murayama}, {Niida}, {Nishizawa}, {Oguri}, {Ono}, {Ouchi},
  {Price}, {Sameshima}, {Schulze}, {Shirakata}, {Silverman}, {Sugiyama},
  {Tait}, {Takada}, {Takata}, {Tanaka}, {Tang}, {Toba}, {Utsumi}, \&
  {Wang}}]{2017arXiv170405854M}
---. 2017, ArXiv e-prints, arXiv:1704.05854

\bibitem[{{McGreer} {et~al.}(2013){McGreer}, {Jiang}, {Fan}, {Richards},
  {Strauss}, {Ross}, {White}, {Shen}, {Schneider}, {Myers}, {Brandt}, {DeGraf},
  {Glikman}, {Ge}, \& {Streblyanska}}]{2013ApJ...768..105M}
{McGreer}, I.~D., {Jiang}, L., {Fan}, X., {et~al.} 2013, \apj, 768, 105

\bibitem[{{McLeod} {et~al.}(2016){McLeod}, {McLure}, \&
  {Dunlop}}]{2016MNRAS.459.3812M}
{McLeod}, D.~J., {McLure}, R.~J., \& {Dunlop}, J.~S. 2016, \mnras, 459, 3812

\bibitem[{{McLure} {et~al.}(2009){McLure}, {Cirasuolo}, {Dunlop}, {Foucaud}, \&
  {Almaini}}]{2009MNRAS.395.2196M}
{McLure}, R.~J., {Cirasuolo}, M., {Dunlop}, J.~S., {Foucaud}, S., \& {Almaini},
  O. 2009, \mnras, 395, 2196

\bibitem[{{McLure} {et~al.}(2013){McLure}, {Dunlop}, {Bowler}, {Curtis-Lake},
  {Schenker}, {Ellis}, {Robertson}, {Koekemoer}, {Rogers}, {Ono}, {Ouchi},
  {Charlot}, {Wild}, {Stark}, {Furlanetto}, {Cirasuolo}, \&
  {Targett}}]{2013MNRAS.432.2696M}
{McLure}, R.~J., {Dunlop}, J.~S., {Bowler}, R.~A.~A., {et~al.} 2013, \mnras,
  432, 2696

\bibitem[{{Miyazaki} {et~al.}(2012){Miyazaki}, {Komiyama}, {Nakaya}, {Kamata},
  {Doi}, {Hamana}, {Karoji}, {Furusawa}, {Kawanomoto}, {Morokuma}, {Ishizuka},
  {Nariai}, {Tanaka}, {Uraguchi}, {Utsumi}, {Obuchi}, {Okura}, {Oguri},
  {Takata}, {Tomono}, {Kurakami}, {Namikawa}, {Usuda}, {Yamanoi}, {Terai},
  {Uekiyo}, {Yamada}, {Koike}, {Aihara}, {Fujimori}, {Mineo}, {Miyatake},
  {Yasuda}, {Nishizawa}, {Saito}, {Tanaka}, {Uchida}, {Katayama}, {Wang},
  {Chen}, {Lupton}, {Loomis}, {Bickerton}, {Price}, {Gunn}, {Suzuki},
  {Miyazaki}, {Muramatsu}, {Yamamoto}, {Endo}, {Ezaki}, {Itoh}, {Miwa},
  {Yokota}, {Matsuda}, {Ebinuma}, \& {Takeshi}}]{2012SPIE.8446E..0ZM}
{Miyazaki}, S., {Komiyama}, Y., {Nakaya}, H., {et~al.} 2012, in \procspie, Vol.
  8446, Ground-based and Airborne Instrumentation for Astronomy IV, 84460Z

\bibitem[{{Miyazaki} {et~al.}(2017)}]{miyazaki2017} 
{Miyazaki}, S., et al. 2017, PASJ in press

\bibitem[{{Momcheva} {et~al.}(2016){Momcheva}, {Brammer}, {van Dokkum},
  {Skelton}, {Whitaker}, {Nelson}, {Fumagalli}, {Maseda}, {Leja}, {Franx},
  {Rix}, {Bezanson}, {Da Cunha}, {Dickey}, {F{\"o}rster Schreiber},
  {Illingworth}, {Kriek}, {Labb{\'e}}, {Ulf Lange}, {Lundgren}, {Magee},
  {Marchesini}, {Oesch}, {Pacifici}, {Patel}, {Price}, {Tal}, {Wake}, {van der
  Wel}, \& {Wuyts}}]{2016ApJS..225...27M}
{Momcheva}, I.~G., {Brammer}, G.~B., {van Dokkum}, P.~G., {et~al.} 2016, \apjs,
  225, 27

\bibitem[{{Mosleh} {et~al.}(2012){Mosleh}, {Williams}, {Franx}, {Gonzalez},
  {Bouwens}, {Oesch}, {Labbe}, {Illingworth}, \&
  {Trenti}}]{2012ApJ...756L..12M}
{Mosleh}, M., {Williams}, R.~J., {Franx}, M., {et~al.} 2012, \apjl, 756, L12

\bibitem[{{Mu{\~n}oz} \& {Loeb}(2011)}]{2011ApJ...729...99M}
{Mu{\~n}oz}, J.~A., \& {Loeb}, A. 2011, \apj, 729, 99

\bibitem[{{Murata} {et~al.}(2017)}]{murata2017} 
{Murata}, R., et al. 2017, to be submitted to PASJ

\bibitem[{{Murayama} {et~al.}(2007){Murayama}, {Taniguchi}, {Scoville},
  {Ajiki}, {Sanders}, {Mobasher}, {Aussel}, {Capak}, {Koekemoer}, {Shioya},
  {Nagao}, {Carilli}, {Ellis}, {Garilli}, {Giavalisco}, {Kitzbichler}, {Le
  F{\`e}vre}, {Maccagni}, {Schinnerer}, {Smol{\v c}i{\'c}}, {Tribiano},
  {Cimatti}, {Komiyama}, {Miyazaki}, {Sasaki}, {Koda}, \&
  {Karoji}}]{2007ApJS..172..523M}
{Murayama}, T., {Taniguchi}, Y., {Scoville}, N.~Z., {et~al.} 2007, \apjs, 172,
  523

\bibitem[{{Niida} {et~al.}(2016){Niida}, {Nagao}, {Ikeda}, {Matsuoka},
  {Kobayashi}, {Toba}, \& {Taniguchi}}]{2016ApJ...832..208N}
{Niida}, M., {Nagao}, T., {Ikeda}, H., {et~al.} 2016, \apj, 832, 208

\bibitem[{{Ocvirk} {et~al.}(2016){Ocvirk}, {Gillet}, {Shapiro}, {Aubert},
  {Iliev}, {Teyssier}, {Yepes}, {Choi}, {Sullivan}, {Knebe}, {Gottl{\"o}ber},
  {D'Aloisio}, {Park}, {Hoffman}, \& {Stranex}}]{2016MNRAS.463.1462O}
{Ocvirk}, P., {Gillet}, N., {Shapiro}, P.~R., {et~al.} 2016, \mnras, 463, 1462

\bibitem[{{Oesch} {et~al.}(2010){Oesch}, {Bouwens}, {Carollo}, {Illingworth},
  {Trenti}, {Stiavelli}, {Magee}, {Labb{\'e}}, \&
  {Franx}}]{2010ApJ...709L..21O}
{Oesch}, P.~A., {Bouwens}, R.~J., {Carollo}, C.~M., {et~al.} 2010, \apjl, 709,
  L21

\bibitem[{{Oke} \& {Gunn}(1983)}]{1983ApJ...266..713O}
{Oke}, J.~B., \& {Gunn}, J.~E. 1983, \apj, 266, 713

\bibitem[{{Ono} {et~al.}(2013){Ono}, {Ouchi}, {Curtis-Lake}, {Schenker},
  {Ellis}, {McLure}, {Dunlop}, {Robertson}, {Koekemoer}, {Bowler}, {Rogers},
  {Schneider}, {Charlot}, {Stark}, {Shimasaku}, {Furlanetto}, \&
  {Cirasuolo}}]{2013ApJ...777..155O}
{Ono}, Y., {Ouchi}, M., {Curtis-Lake}, E., {et~al.} 2013, \apj, 777, 155

\bibitem[{{O'Shea} {et~al.}(2015){O'Shea}, {Wise}, {Xu}, \&
  {Norman}}]{2015ApJ...807L..12O}
{O'Shea}, B.~W., {Wise}, J.~H., {Xu}, H., \& {Norman}, M.~L. 2015, \apjl, 807,
  L12

\bibitem[{{Ouchi} {et~al.}(2004){Ouchi}, {Shimasaku}, {Okamura}, {Furusawa},
  {Kashikawa}, {Ota}, {Doi}, {Hamabe}, {Kimura}, {Komiyama}, {Miyazaki},
  {Miyazaki}, {Nakata}, {Sekiguchi}, {Yagi}, \& {Yasuda}}]{2004ApJ...611..660O}
{Ouchi}, M., {Shimasaku}, K., {Okamura}, S., {et~al.} 2004, \apj, 611, 660

\bibitem[{{Ouchi} {et~al.}(2008){Ouchi}, {Shimasaku}, {Akiyama}, {Simpson},
  {Saito}, {Ueda}, {Furusawa}, {Sekiguchi}, {Yamada}, {Kodama}, {Kashikawa},
  {Okamura}, {Iye}, {Takata}, {Yoshida}, \& {Yoshida}}]{2008ApJS..176..301O}
{Ouchi}, M., {Shimasaku}, K., {Akiyama}, M., {et~al.} 2008, \apjs, 176, 301

\bibitem[{{Ouchi} {et~al.}(2009){Ouchi}, {Mobasher}, {Shimasaku}, {Ferguson},
  {Fall}, {Ono}, {Kashikawa}, {Morokuma}, {Nakajima}, {Okamura}, {Dickinson},
  {Giavalisco}, \& {Ohta}}]{2009ApJ...706.1136O}
{Ouchi}, M., {Mobasher}, B., {Shimasaku}, K., {et~al.} 2009, \apj, 706, 1136

\bibitem[{{Ouchi} {et~al.}(2010){Ouchi}, {Shimasaku}, {Furusawa}, {Saito},
  {Yoshida}, {Akiyama}, {Ono}, {Yamada}, {Ota}, {Kashikawa}, {Iye}, {Kodama},
  {Okamura}, {Simpson}, \& {Yoshida}}]{2010ApJ...723..869O}
{Ouchi}, M., {Shimasaku}, K., {Furusawa}, H., {et~al.} 2010, \apj, 723, 869

\bibitem[{{Ouchi} {et~al.}(2017){Ouchi}, {Harikane}, {Shibuya}, {Shimasaku},
  {Taniguchi}, {Konno}, {Kobayashi}, {Kajisawa}, {Nagao}, {Ono}, {Inoue},
  {Umemura}, {Mori}, {Hasegawa}, {Higuchi}, {Komiyama}, {Matsuda}, {Nakajima},
  {Saito}, \& {Wang}}]{2017arXiv170407455O}
{Ouchi}, M., {Harikane}, Y., {Shibuya}, T., {et~al.} 2017, ArXiv e-prints,
  arXiv:1704.07455

\bibitem[{{P{\^a}ris} {et~al.}(2017){P{\^a}ris}, {Petitjean}, {Ross}, {Myers},
  {Aubourg}, {Streblyanska}, {Bailey}, {Armengaud}, {Palanque-Delabrouille},
  {Y{\`e}che}, {Hamann}, {Strauss}, {Albareti}, {Bovy}, {Bizyaev}, {Niel
  Brandt}, {Brusa}, {Buchner}, {Comparat}, {Croft}, {Dwelly}, {Fan},
  {Font-Ribera}, {Ge}, {Georgakakis}, {Hall}, {Jiang}, {Kinemuchi},
  {Malanushenko}, {Malanushenko}, {McMahon}, {Menzel}, {Merloni}, {Nandra},
  {Noterdaeme}, {Oravetz}, {Pan}, {Pieri}, {Prada}, {Salvato}, {Schlegel},
  {Schneider}, {Simmons}, {Viel}, {Weinberg}, \& {Zhu}}]{2017A&A...597A..79P}
{P{\^a}ris}, I., {Petitjean}, P., {Ross}, N.~P., {et~al.} 2017, \aap, 597, A79

\bibitem[{{Parsa} {et~al.}(2017){Parsa}, {Dunlop}, \&
  {McLure}}]{2017arXiv170407750P}
{Parsa}, S., {Dunlop}, J.~S., \& {McLure}, R.~J. 2017, ArXiv e-prints,
  arXiv:1704.07750

\bibitem[{{Ravindranath} {et~al.}(2006){Ravindranath}, {Giavalisco},
  {Ferguson}, {Conselice}, {Katz}, {Weinberg}, {Lotz}, {Dickinson}, {Fall},
  {Mobasher}, \& {Papovich}}]{2006ApJ...652..963R}
{Ravindranath}, S., {Giavalisco}, M., {Ferguson}, H.~C., {et~al.} 2006, \apj,
  652, 963

\bibitem[{{Reddy} {et~al.}(2008){Reddy}, {Steidel}, {Pettini}, {Adelberger},
  {Shapley}, {Erb}, \& {Dickinson}}]{2008ApJS..175...48R}
{Reddy}, N.~A., {Steidel}, C.~C., {Pettini}, M., {et~al.} 2008, \apjs, 175, 48

\bibitem[{{Rees} \& {Ostriker}(1977)}]{1977MNRAS.179..541R}
{Rees}, M.~J., \& {Ostriker}, J.~P. 1977, \mnras, 179, 541

\bibitem[{{Ribeiro} {et~al.}(2016){Ribeiro}, {Le F{\`e}vre}, {Tasca}, {Lemaux},
  {Cassata}, {Garilli}, {Maccagni}, {Zamorani}, {Zucca}, {Amor{\'{\i}}n},
  {Bardelli}, {Fontana}, {Giavalisco}, {Hathi}, {Koekemoer}, {Pforr}, {Tresse},
  \& {Dunlop}}]{2016A&A...593A..22R}
{Ribeiro}, B., {Le F{\`e}vre}, O., {Tasca}, L.~A.~M., {et~al.} 2016, \aap, 593,
  A22

\bibitem[{{Rowe} {et~al.}(2015){Rowe}, {Jarvis}, {Mandelbaum}, {Bernstein},
  {Bosch}, {Simet}, {Meyers}, {Kacprzak}, {Nakajima}, {Zuntz}, {Miyatake},
  {Dietrich}, {Armstrong}, {Melchior}, \& {Gill}}]{2015A&C....10..121R}
{Rowe}, B.~T.~P., {Jarvis}, M., {Mandelbaum}, R., {et~al.} 2015, Astronomy and
  Computing, 10, 121

\bibitem[{{Saito} {et~al.}(2008){Saito}, {Shimasaku}, {Okamura}, {Ouchi},
  {Akiyama}, {Yoshida}, \& {Ueda}}]{2008ApJ...675.1076S}
{Saito}, T., {Shimasaku}, K., {Okamura}, S., {et~al.} 2008, \apj, 675, 1076

\bibitem[{{Salpeter}(1955)}]{1955ApJ...121..161S}
{Salpeter}, E.~E. 1955, \apj, 121, 161

\bibitem[{{Sawicki} \& {Thompson}(2006)}]{2006ApJ...642..653S}
{Sawicki}, M., \& {Thompson}, D. 2006, \apj, 642, 653

\bibitem[{{Scannapieco} \& {Oh}(2004)}]{2004ApJ...608...62S}
{Scannapieco}, E., \& {Oh}, S.~P. 2004, \apj, 608, 62

\bibitem[{{Schechter}(1976)}]{1976ApJ...203..297S}
{Schechter}, P. 1976, \apj, 203, 297

\bibitem[{{Schenker} {et~al.}(2013){Schenker}, {Robertson}, {Ellis}, {Ono},
  {McLure}, {Dunlop}, {Koekemoer}, {Bowler}, {Ouchi}, {Curtis-Lake}, {Rogers},
  {Schneider}, {Charlot}, {Stark}, {Furlanetto}, \&
  {Cirasuolo}}]{2013ApJ...768..196S}
{Schenker}, M.~A., {Robertson}, B.~E., {Ellis}, R.~S., {et~al.} 2013, \apj,
  768, 196

\bibitem[{{Schlafly} {et~al.}(2012){Schlafly}, {Finkbeiner}, {Juri{\'c}},
  {Magnier}, {Burgett}, {Chambers}, {Grav}, {Hodapp}, {Kaiser}, {Kudritzki},
  {Martin}, {Morgan}, {Price}, {Rix}, {Stubbs}, {Tonry}, \&
  {Wainscoat}}]{2012ApJ...756..158S}
{Schlafly}, E.~F., {Finkbeiner}, D.~P., {Juri{\'c}}, M., {et~al.} 2012, \apj,
  756, 158

\bibitem[{{Schlegel} {et~al.}(1998){Schlegel}, {Finkbeiner}, \&
  {Davis}}]{1998ApJ...500..525S}
{Schlegel}, D.~J., {Finkbeiner}, D.~P., \& {Davis}, M. 1998, \apj, 500, 525

\bibitem[{{Shibuya} {et~al.}(2015){Shibuya}, {Ouchi}, \&
  {Harikane}}]{2015ApJS..219...15S}
{Shibuya}, T., {Ouchi}, M., \& {Harikane}, Y. 2015, \apjs, 219, 15

\bibitem[{{Shibuya} {et~al.}(2017{\natexlab{a}}){Shibuya}, {Ouchi}, {Konno},
  {Higuchi}, {Harikane}, {Ono}, {Shimasaku}, {Taniguchi}, {Kobayashi},
  {Kajisawa}, {Nagao}, {Furusawa}, {Goto}, {Kashikawa}, {Komiyama}, {Kusakabe},
  {Lee}, {Momose}, {Nakajima}, {Tanaka}, {Wang}, \&
  {Yuma}}]{2017arXiv170408140S}
{Shibuya}, T., {Ouchi}, M., {Konno}, A., {et~al.} 2017{\natexlab{a}}, ArXiv
  e-prints, arXiv:1704.08140

\bibitem[{{Shibuya} {et~al.}(2017{\natexlab{b}}){Shibuya}, {Ouchi}, {Harikane},
  {Rauch}, {Ono}, {Mukae}, {Higuchi}, {Kojima}, {Yuma}, {Lee}, {Furusawa},
  {Konno}, {Martin}, {Shimasaku}, {Taniguchi}, {Kobayashi}, {Kajisawa},
  {Nagao}, {Goto}, {Kashikawa}, {Komiyama}, {Kusakabe}, {Momose}, {Nakajima},
  {Tanaka}, \& {Wang}}]{2017arXiv170500733S}
{Shibuya}, T., {Ouchi}, M., {Harikane}, Y., {et~al.} 2017{\natexlab{b}}, ArXiv
  e-prints, arXiv:1705.00733

\bibitem[{{Shimasaku} {et~al.}(2005){Shimasaku}, {Ouchi}, {Furusawa},
  {Yoshida}, {Kashikawa}, \& {Okamura}}]{2005PASJ...57..447S}
{Shimasaku}, K., {Ouchi}, M., {Furusawa}, H., {et~al.} 2005, \pasj, 57, 447

\bibitem[{{Shimasaku} {et~al.}(2006){Shimasaku}, {Kashikawa}, {Doi}, {Ly},
  {Malkan}, {Matsuda}, {Ouchi}, {Hayashino}, {Iye}, {Motohara}, {Murayama},
  {Nagao}, {Ohta}, {Okamura}, {Sasaki}, {Shioya}, \&
  {Taniguchi}}]{2006PASJ...58..313S}
{Shimasaku}, K., {Kashikawa}, N., {Doi}, M., {et~al.} 2006, \pasj, 58, 313

\bibitem[{{Shioya} {et~al.}(2009){Shioya}, {Taniguchi}, {Sasaki}, {Nagao},
  {Murayama}, {Saito}, {Ideue}, {Nakajima}, {Matsuoka}, {Trump}, {Scoville},
  {Sanders}, {Mobasher}, {Aussel}, {Capak}, {Kartaltepe}, {Koekemoer},
  {Carilli}, {Ellis}, {Garilli}, {Giavalisco}, {Kitzbichler}, {Impey},
  {LeFevre}, {Schinnerer}, \& {Smolcic}}]{2009ApJ...696..546S}
{Shioya}, Y., {Taniguchi}, Y., {Sasaki}, S.~S., {et~al.} 2009, \apj, 696, 546

\bibitem[{{Silk}(1977)}]{1977ApJ...211..638S}
{Silk}, J. 1977, \apj, 211, 638

\bibitem[{{Stark}(2016)}]{2016ARA&A..54..761S}
{Stark}, D.~P. 2016, \araa, 54, 761

\bibitem[{{Stefanon} {et~al.}(2017){Stefanon}, {Labb{\'e}}, {Bouwens},
  {Brammer}, {Oesch}, {Franx}, {Fynbo}, {Milvang-Jensen}, {Muzzin},
  {Illingworth}, {Le F{\`e}vre}, {Caputi}, {Holwerda}, {McCracken}, {Smit}, \&
  {Magee}}]{2017arXiv170604613S}
{Stefanon}, M., {Labb{\'e}}, I., {Bouwens}, R.~J., {et~al.} 2017, ArXiv
  e-prints, arXiv:1706.04613

\bibitem[{{Steidel} {et~al.}(1999){Steidel}, {Adelberger}, {Giavalisco},
  {Dickinson}, \& {Pettini}}]{1999ApJ...519....1S}
{Steidel}, C.~C., {Adelberger}, K.~L., {Giavalisco}, M., {Dickinson}, M., \&
  {Pettini}, M. 1999, \apj, 519, 1

\bibitem[{{Steidel} {et~al.}(1996){Steidel}, {Giavalisco}, {Pettini},
  {Dickinson}, \& {Adelberger}}]{1996ApJ...462L..17S}
{Steidel}, C.~C., {Giavalisco}, M., {Pettini}, M., {Dickinson}, M., \&
  {Adelberger}, K.~L. 1996, \apjl, 462, L17

\bibitem[{{Takahashi} {et~al.}(2011){Takahashi}, {Oguri}, {Sato}, \&
  {Hamana}}]{2011ApJ...742...15T}
{Takahashi}, R., {Oguri}, M., {Sato}, M., \& {Hamana}, T. 2011, \apj, 742, 15

\bibitem[{{Takata} {et~al.}(2017)}]{takata2017} 
{Takata}, T., et al. 2017, to be submitted to PASJ

\bibitem[{{Tasca} {et~al.}(2017){Tasca}, {Le F{\`e}vre}, {Ribeiro}, {Thomas},
  {Moreau}, {Cassata}, {Garilli}, {Le Brun}, {Lemaux}, {Maccagni},
  {Pentericci}, {Schaerer}, {Vanzella}, {Zamorani}, {Zucca}, {Amorin},
  {Bardelli}, {Cassar{\`a}}, {Castellano}, {Cimatti}, {Cucciati}, {Durkalec},
  {Fontana}, {Giavalisco}, {Grazian}, {Hathi}, {Ilbert}, {Paltani}, {Pforr},
  {Scodeggio}, {Sommariva}, {Talia}, {Tresse}, {Vergani}, {Capak}, {Charlot},
  {Contini}, {de la Torre}, {Dunlop}, {Fotopoulou}, {Guaita}, {Koekemoer},
  {L{\'o}pez-Sanjuan}, {Mellier}, {Salvato}, {Scoville}, {Taniguchi}, \&
  {Wang}}]{2017A&A...600A.110T}
{Tasca}, L.~A.~M., {Le F{\`e}vre}, O., {Ribeiro}, B., {et~al.} 2017, \aap, 600,
  A110

\bibitem[{{Tonry} {et~al.}(2012){Tonry}, {Stubbs}, {Lykke}, {Doherty},
  {Shivvers}, {Burgett}, {Chambers}, {Hodapp}, {Kaiser}, {Kudritzki},
  {Magnier}, {Morgan}, {Price}, \& {Wainscoat}}]{2012ApJ...750...99T}
{Tonry}, J.~L., {Stubbs}, C.~W., {Lykke}, K.~R., {et~al.} 2012, \apj, 750, 99

\bibitem[{{Toshikawa} {et~al.}(2016){Toshikawa}, {Kashikawa}, {Overzier},
  {Malkan}, {Furusawa}, {Ishikawa}, {Onoue}, {Ota}, {Tanaka}, {Niino}, \&
  {Uchiyama}}]{2016ApJ...826..114T}
{Toshikawa}, J., {Kashikawa}, N., {Overzier}, R., {et~al.} 2016, \apj, 826, 114

\bibitem[{{Toshikawa} {et~al.}(2017)}]{toshikawa2017} 
{Toshikawa}, J., et al. 2017, PASJ in press

\bibitem[{{Trenti} \& {Stiavelli}(2008)}]{2008ApJ...676..767T}
{Trenti}, M., \& {Stiavelli}, M. 2008, \apj, 676, 767

\bibitem[{{van der Burg} {et~al.}(2010){van der Burg}, {Hildebrandt}, \&
  {Erben}}]{2010A&A...523A..74V}
{van der Burg}, R.~F.~J., {Hildebrandt}, H., \& {Erben}, T. 2010, \aap, 523,
  A74

\bibitem[{{Wang} \& {Heckman}(1996)}]{1996ApJ...457..645W}
{Wang}, B., \& {Heckman}, T.~M. 1996, \apj, 457, 645

\bibitem[{{Wang} {et~al.}(2016){Wang}, {Wu}, {Fan}, {Yang}, {Yi}, {Bian},
  {McGreer}, {Yang}, {Ai}, {Dong}, {Zuo}, {Jiang}, {Green}, {Wang}, {Cai},
  {Wang}, \& {Yue}}]{2016ApJ...819...24W}
{Wang}, F., {Wu}, X.-B., {Fan}, X., {et~al.} 2016, \apj, 819, 24

\bibitem[{{Willott} {et~al.}(2010{\natexlab{a}}){Willott}, {Albert},
  {Arzoumanian}, {Bergeron}, {Crampton}, {Delorme}, {Hutchings}, {Omont},
  {Reyl{\'e}}, \& {Schade}}]{2010AJ....140..546W}
{Willott}, C.~J., {Albert}, L., {Arzoumanian}, D., {et~al.} 2010{\natexlab{a}},
  \aj, 140, 546

\bibitem[{{Willott} {et~al.}(2010{\natexlab{b}}){Willott}, {Delorme},
  {Reyl{\'e}}, {Albert}, {Bergeron}, {Crampton}, {Delfosse}, {Forveille},
  {Hutchings}, {McLure}, {Omont}, \& {Schade}}]{2010AJ....139..906W}
{Willott}, C.~J., {Delorme}, P., {Reyl{\'e}}, C., {et~al.} 2010{\natexlab{b}},
  \aj, 139, 906

\bibitem[{{Willott} {et~al.}(2013){Willott}, {McLure}, {Hibon}, {Bielby},
  {McCracken}, {Kneib}, {Ilbert}, {Bonfield}, {Bruce}, \&
  {Jarvis}}]{2013AJ....145....4W}
{Willott}, C.~J., {McLure}, R.~J., {Hibon}, P., {et~al.} 2013, \aj, 145, 4

\bibitem[{{Wise} {et~al.}(2014){Wise}, {Demchenko}, {Halicek}, {Norman},
  {Turk}, {Abel}, \& {Smith}}]{2014MNRAS.442.2560W}
{Wise}, J.~H., {Demchenko}, V.~G., {Halicek}, M.~T., {et~al.} 2014, \mnras,
  442, 2560

\bibitem[{{Wyithe} {et~al.}(2011){Wyithe}, {Yan}, {Windhorst}, \&
  {Mao}}]{2011Natur.469..181W}
{Wyithe}, J.~S.~B., {Yan}, H., {Windhorst}, R.~A., \& {Mao}, S. 2011, \nat,
  469, 181

\bibitem[{{Yang} {et~al.}(2017){Yang}, {Fan}, {Wu}, {Wang}, {Bian}, {Yang},
  {McGreer}, {Yi}, {Jiang}, {Green}, {Yue}, {Wang}, {Li}, {Ding}, {Dye}, \&
  {Lawrence}}]{2017AJ....153..184Y}
{Yang}, J., {Fan}, X., {Wu}, X.-B., {et~al.} 2017, \aj, 153, 184

\bibitem[{{Yoshida} {et~al.}(2006){Yoshida}, {Shimasaku}, {Kashikawa}, {Ouchi},
  {Okamura}, {Ajiki}, {Akiyama}, {Ando}, {Aoki}, {Doi}, {Furusawa},
  {Hayashino}, {Iwamuro}, {Iye}, {Karoji}, {Kobayashi}, {Kodaira}, {Kodama},
  {Komiyama}, {Malkan}, {Matsuda}, {Miyazaki}, {Mizumoto}, {Morokuma},
  {Motohara}, {Murayama}, {Nagao}, {Nariai}, {Ohta}, {Sasaki}, {Sato},
  {Sekiguchi}, {Shioya}, {Tamura}, {Taniguchi}, {Umemura}, {Yamada}, \&
  {Yasuda}}]{2006ApJ...653..988Y}
{Yoshida}, M., {Shimasaku}, K., {Kashikawa}, N., {et~al.} 2006, \apj, 653, 988

\end{thebibliography}

%%%%%%%%%%%%%%%%%%%%%%%%%%%%%%%%%%%%%%%%%%%%%%%%%%%%%%%%%%%%%%%%%
%%%%%%%%%%%%%%%%%%%%%%%%%%%%%%%%%%%%%%%%%%%%%%%%%%%%%%%%%%%%%%%%%

\appendix

\section{The UV Luminosity Function Results for Different Layers}

\begin{figure*}
 \begin{center}
  \includegraphics[width=15cm]{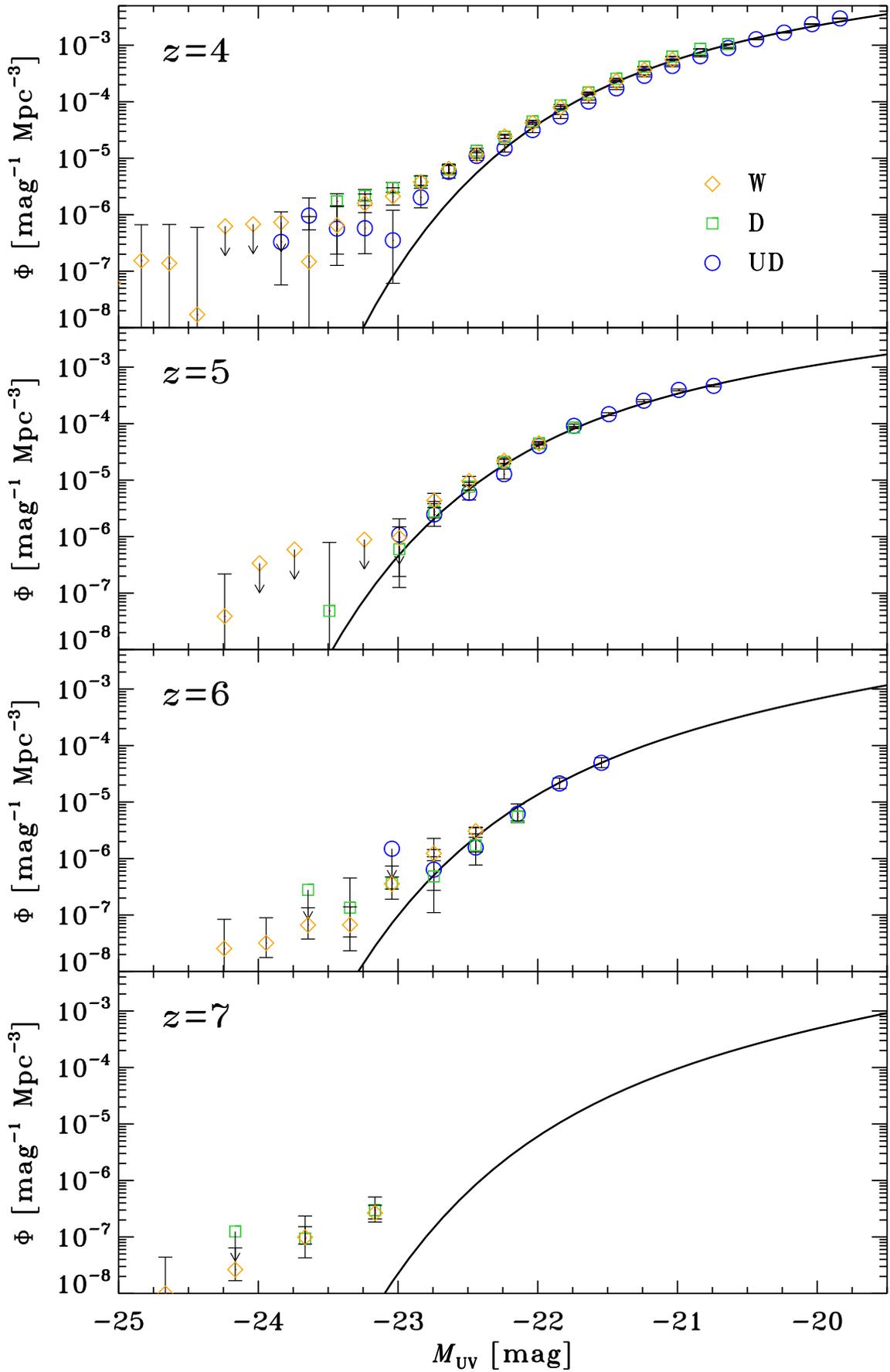} 
 \end{center}
\caption{
Rest-frame UV luminosity functions of dropouts at $z \sim 4$, $z \sim 5$, $z \sim 6$, and $z \sim 7$ 
derived for our dropout samples in the 
W (orange diamonds), 
D (green squares), and 
UD (blue circles) 
layers separately,  
where the quasar contamination correction is not taken into account. 
The symbols with a downward arrow 
denote the $1\sigma$ upper limits. 
The solid lines denote the best-fit Schechter functions 
shown in Figure \ref{fig:UVLF_selected}.
}\label{fig:LF_for_each_field}
\end{figure*}

In this appendix, 
we present the UV LF determination results for the W, D, and UD layers separately. 
The obtained UV LFs of $z \sim 4-7$ dropouts for each of the three layers is shown in Figure \ref{fig:LF_for_each_field}. 
We confirm that 
our LF results for the different layers are consistent with each other
mostly within a factor of $1.5$. 
Although
our $z \sim 4$ LF results 
between these layers show larger differences
in the bright magnitude range from $M_{\rm UV} = -24$ mag to $-23$ mag, 
the significances of the differences are still 
$\lesssim 2 \sigma$ due to the large uncertainties.

%%%%%%%%%%%%%%%%%%%%%%%%%%%%%%%%%%%%%%
%%%%%%%%%%%%%%%%%%%%%%%%%%%%%%%%%%%%%%%

\end{document}